\documentclass{article}

\usepackage{arxiv}

\usepackage[utf8]{inputenc} 
\usepackage[T1]{fontenc}    
\usepackage{hyperref}       
\usepackage{url}            
\usepackage{booktabs}       
\usepackage{amsfonts}       
\usepackage{nicefrac}       
\usepackage{microtype}      
\usepackage{lipsum}
\usepackage{graphicx}
\graphicspath{ {./images/} }
\usepackage{graphicx} 
\usepackage{tabularx}
\graphicspath{{./figures}}
\usepackage{longtable}
\usepackage{float} 
\usepackage{pifont} 
\newcommand{\cmark}{\ding{51}}  
\newcommand{\xmark}{\ding{55}}  

\usepackage{array} 
\newcolumntype{L}[1]{>{\raggedright\arraybackslash}p{#1}} 
\newcolumntype{C}[1]{>{\centering\arraybackslash}p{#1}}   
\newcolumntype{R}[1]{>{\raggedleft\arraybackslash}p{#1}}  
\usepackage{url}

\title{Large Language Models for Security Operations Centers: A Comprehensive Survey}

\author{
 Ali Habibzadeh \\
  Department of Computer Engineering\\
  University of Guilan\\
  Rasht, Iran \\
  \texttt{ali78@webmail.guilan.ac.ir} \\
   \And
 Farid Feyzi \\
  Department of Computer Engineering\\
  University of Guilan\\
  Rasht, Iran \\
  \texttt{feizi@guilan.ac.ir} \\
  \And
 Reza Ebrahimi Atani \\
  Department of Computer Engineering\\
  University of Guilan\\
  Rasht, Iran \\
  \texttt{rebrahimi@guilan.ac.ir} \\
}

\begin{document}
\maketitle
\begin{abstract}
Large Language Models (LLMs) have emerged as powerful tools capable of understanding and generating human-like text, offering transformative potential across diverse domains. The Security Operations Center (SOC), responsible for safeguarding digital infrastructure, represents one of these domains. SOCs serve as the frontline of defense in cybersecurity, tasked with continuous monitoring, detection, and response to incidents. However, SOCs face persistent challenges such as high alert volumes, limited resources, high demand for experts with advanced knowledge, delayed response times, and difficulties in leveraging threat intelligence effectively. In this context, LLMs can offer promising solutions by automating log analysis, streamlining triage, improving detection accuracy, and providing the required knowledge in less time. This survey systematically explores the integration of generative AI and more specifically LLMs into SOC workflow, providing a structured perspective on its capabilities, challenges, and future directions. We believe that this survey offers researchers and SOC managers a broad overview of the current state of LLM integration within academic study. To the best of our knowledge, this is the first comprehensive study to examine LLM applications in SOCs in details.
\end{abstract}


\section{Introduction}\label{sec:intro}
Large Language Models (LLMs) have significantly transformed the landscape of automated tasks and content generation across numerous fields. These models represent a major advancement in artificial intelligence, characterized as large-scale, pre-trained, statistical language models built upon neural networks, capable of comprehending and generating human-like text by learning from vast and diverse linguistic datasets \cite{minaee2024large}. Moreover, their powerful comprehension for downstream tasks and ability to handle various natural language processing applications have positioned them as versatile problem-solving tools \cite{kumar2024large}. These efforts have intensified in recent years, accompanied by a sudden surge in research articles encompassing various fields such as telecommunications \cite{zhou2024large}, agriculture \cite{gong2025application}, the transportation industry \cite{KARIM2025100004}, and healthcare \cite{wang2025healthq}. Nevertheless, researchers are still discovering new capabilities of LLMs in technology-related disciplines. One of these areas is cybersecurity.

Cybersecurity is a critical domain focused on safeguarding digital systems, including computer systems, networks, and data, from unauthorized access, interruption, or any compromise. Its importance continues to grow as technology integrates into nearly every aspect of life, leading to an expanding scope of cyber threats \cite{hassanin2024comprehensive}. Moreover, the complexity and sophistication of cyberattacks have risen significantly, posing substantial risks to businesses and critical infrastructure. The global cost of cybercrime was estimated to be around \$8.4 trillion in 2022 and is predicted to reach \$23.84 trillion annually by 2027 \cite{125_smet}.

In response to the escalating sophistication and frequency of cyber threats, Security Operations Center (SOC) functions as the central hub for an organization’s cybersecurity operations \cite{khayat2025empowering}. They tasked with continuous monitoring and defending to maintain an organization’s overall security posture \cite{vielberth2020security}. SOC's processes are typically structured around the incident lifecycle, encompassing detection, analysis, and response \cite{vielberth2021security_soc}. Log analysis, threat detection, incident response, cyber threat intelligence (CTI) analysis, and vulnerability management are some key operational tasks in it which collectively enable coordinated defense against cyber threats.

Despite their use of advanced tools, modern SOCs face significant challenges that impede their effectiveness. They are often overwhelmed by the sheer volume, velocity, and variety of IT infrastructure and security data, leading to an over-reliance on manual analysis by human security analysts \cite{gupta2019automated}. This manual effort results in substantial backlogs and slower resolution times for critical security events. These conditions give rise to alert fatigue, driven by the massive influx of alarms, wherein analysts struggle to distinguish genuine threats from false positives \cite{tariq2025alert}. This human-centric model is becoming more unsustainable given the rapid growth of data in today’s world, contributing to analyst burnout and a general lack of confidence in identifying threats. What further complicates this issue is the fact that 80\% of the SOC budget is allocated to labor costs, highlighting the financial burden \cite{48_attackqa}. Global shortage of skilled cybersecurity professionals is another notable challenge, with industry estimates indicating a gap of 1.8 to 4 million unfilled positions worldwide, making it difficult for organizations to adequately staff their SOCs with experienced analysts \cite{furnell2021cybersecurity}. This severe lack of expertise is a frequently cited barrier to achieving excellence in SOCs, contributing to overworked teams, increased burnout rates, and a diminished ability to effectively detect and respond to increasingly sophisticated cyber threats \cite{crowley2019common}. 

The revolutionary potential of LLMs provides a promising approach to address or significantly mitigate many of the challenges currently encountered by SOCs. LLMs can automate labor-intensive tasks and streamline workflows, thereby reducing human workload and alleviating alert fatigue. Their ability to efficiently process large, diverse knowledge sources allows for the automation of organization-specific threat intelligence extraction, enhancing threat identification and response capabilities. Furthermore, LLMs can provide natural language explanations for alerts, process vast volumes of log data to detect anomalies, and offer contextual and explainable mappings for network intrusion detection system (NIDS) rules. They can also improve threat assessment, support secure code generation and detection, and automate routine compliance checks and regulatory reporting, ultimately empowering security analysts and enabling faster, more efficient decision-making \cite{gupta2023chatgpt, kshetri2025transforming}.

\begin{figure}[H]
    \centering
    \includegraphics[width=0.8\linewidth]{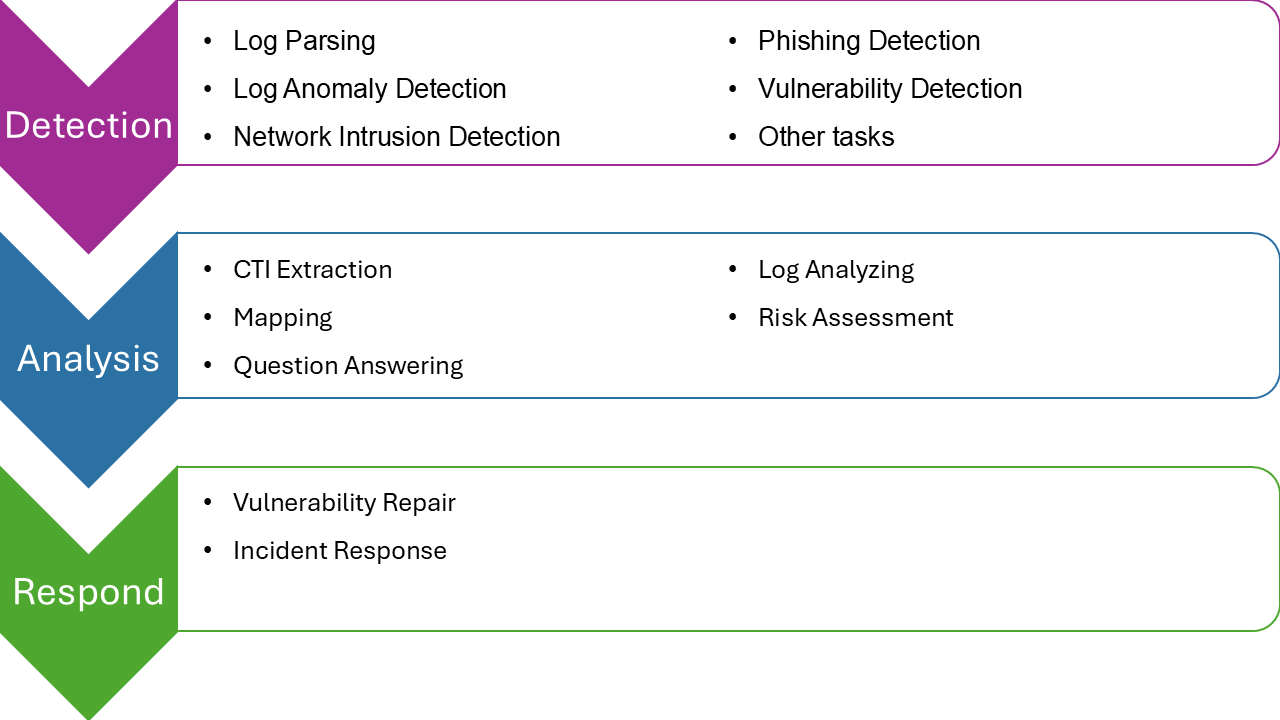}
    \caption{Overall workflow of this survey.}
    \label{fig:workflow}
\end{figure}

In this survey, we examine studies that leverage LLMs to enhance cybersecurity defenses, with a particular focus on SOCs. The goal of this survey is not to provide an exhaustive review of all existing literature, but rather to offer a structured overview that informs managers and professionals about the current state of LLMs regarding SOCs. This perspective enables developers to address emerging needs when designing security solutions, supports managers in making informed decisions to strengthen organizational defenses, and helps employees reduce engagement with repetitive and labor-intensive tasks. To structure our analysis, we categorize SOC-related operations into three distinct phases and, within each phase, highlight and evaluate the most significant tasks discussed in the academic literature. This workflow is demonstrated in Figure \ref{fig:workflow}.
Compared to other similar works, this review offers several unique contributions, as outlined below:

\begin{itemize}
    \item In this survey, we provide an overview of the fundamental concepts of SOCs, including the classification of employee tiers, the types of tools utilized, the different phases of workflow, and the challenges associated with their operation.
    \item This review highlights the datasets and LLMs most frequently utilized in the development of models for SOC-related tasks. By consolidating this information, it provides a reference framework that enables future researchers to reduce time spent on preliminary exploration and focus more on advancing their own models.
    \item This research provides a comprehensive and precise comparison of proposed models in the SOC domain. On the one hand, It encompasses a range of tasks related to logs, code, and networks. On the other hand, it also analyzes their underlying methodologies in detail, enabling the identification of recurring patterns across the literature.
    \item Lastly, our research highlights the challenges of fully integrating LLM-based models into real-world SOC operations and identifies future research directions informed by trends observed in the reviewed literature.
\end{itemize}

The rest of the survey is organized as follows: In section \ref{sec:pre}, the Preliminaries concepts are introduced and related studies are reviewed. Section \ref{sec:method} presents the methodology employed in this review. In sections \ref{sec:rq1} and \ref{sec:rq2}, the most widely adopted LLMs and datasets reported in the literature are presented, respectively. Sections \ref{sec:rq3} to \ref{sec:rq5} examines the applications of LLMs and the adopted approaches in SOC tasks categorized to detection, analysis, and response phases. In the end, the key challenges are outlined and potential future research directions are discussed in section \ref{sec:rq6}.

\section{Preliminaries and Related studies} \label{sec:pre}
This section introduces the fundamental concepts of LLMs and SOCs, followed by a review of related survey articles, with a discussion of how the present work distinguishes itself from them.

\subsection{Large Language models}
Large Language Models (LLMs) are computational models that possess the capability to understand and generate human language. These models are characterized by their massive parameter sizes, often ranging from billions to trillions, and their exceptional learning capabilities. LLMs are typically transformer-based neural networks that undergo pre-training on extensive and diverse datasets, including web pages, literature, conversations, and code. This extensive training equips them with general-purpose language comprehension and generation capabilities, allowing them to estimate the probability of word sequences or generate new text conditioned on a given input. A key phenomenon observed with larger models is the emergence of unpredictable abilities, such as instruction following, in-context learning, and multi-step reasoning, which are typically absent in smaller ones \cite{chang2024survey}.

Building upon this foundational understanding of LLMs, their diverse capabilities stem from variations in their underlying architectures. LLMs are generally categorized into three primary architectural types: encoder-only models, decoder-only models, and encoder-decoder models. Encoder-only LLMs, exemplified by models like BERT, primarily focus on understanding and encoding input information, making them ideal for generating contextualized embeddings for data such as source code, network traffic, or system logs, which are then used in downstream tasks like classification. Decoder-only LLMs, such as GPT and LLaMA, operate using an autoregressive approach, generating outputs token-by-token. This makes them highly suitable for tasks requiring detailed generation, including security analyses, advisories, and code generation. In recent years, decoder-only architectures have become the predominant choice in research due to their powerful text comprehension, reasoning capabilities, and open-ended generation. Encoder-decoder LLMs, like T5, offer a hybrid approach, capable of accommodating both language understanding and generation tasks. Different research fields may show a preference for specific pre-trained models, with GPT-based and BERT-based models being widely adopted \cite{jiang2024survey}. 

To fully leverage the capabilities of these diverse LLM architectures and tailor them for particular applications, several key methods are employed:

\begin{itemize}
    \item \textbf{Pre-training from scratch: }This involves training models from the ground up on large, specialized corpora to enable them to acquire domain-specific knowledge, such as programming syntax and semantics. This process requires massive datasets, often comprising billions or trillions of tokens, and is computationally intensive. However, it lays a foundational understanding crucial for the model's capabilities.
    \item \textbf{Fine-tuning: }This widely utilized technique involves updating the parameters of pre-trained LLMs using smaller, task-relevant datasets to adapt them to specific downstream tasks. It encompasses several approaches, including continual pre-training on large domain-specific datasets, instruction fine-tuning with labeled examples, reinforcement learning with human feedback (RLHF) to align models with desired objectives, and knowledge distillation to transfer knowledge from larger models to smaller ones while maintaining performance. Fine-tuning also can be comprehensive (full-parameter) or more resource-efficient (parameter-efficient fine-tuning). Consequently, these diverse fine-tuning strategies are critical for specializing LLMs, enabling them to reliably address the intricate and dynamic demands of various real-world applications, particularly within sensitive domains like cybersecurity.
    \item \textbf{Prompt Engineering: }This method focuses on designing effective prompts to guide LLMs toward desired outputs. It is the most widely used domain technique for applying LLMs to security tasks in the past few years. Recent studies have highlighted performance variations between simple and advanced prompts when LLMs are applied to NLP tasks \cite{peng2023towards, kocmi2023large, wang2024investigating, yao2025comparing}. Poorly formulated prompts can compromise output quality, leading to inaccurate results or the generation of irrelevant content. Prompt engineering leverages the LLM's in-context learning abilities, often by providing examples within the prompt itself. Techniques such as Chain-of-Thought (CoT) and Tree-of-Thought (ToT) are also employed to improve the LLMs' ability to perform complex reasoning by generating intermediate steps.
    \item \textbf{Retrieval-Augmented Generation (RAG): }This technique combines LLMs with external knowledge repositories. It helps LLMs overcome limitations like generating plausible but incorrect information (hallucinations) and lacking domain expertise. RAG enables LLMs to retrieve and summarize knowledge from global and local databases, thereby providing more accurate and contextually relevant outputs for tasks such as threat intelligence.
    \item \textbf{Agents: }This approach involves equipping LLMs with external tools to expand their capabilities and enable them to perform complex tasks. LLM agents can serve as coordinators for intricate operations, and experiments have shown that they can achieve performance comparable to or even surpass that of extensively trained agents in sequential decision-making tasks.

\end{itemize}

Leveraging these architectural types and adaptation methodologies, LLMs are increasingly being applied to solve complex problems across a multitude of industries. For instance, in healthcare, LLMs assist with clinical decision-making \cite{zhang2025multimodal}, patient care \cite{cascella2023evaluating}, medical query answering \cite{lee2024development}, and even transcribing medical dictations into electronic health records \cite{onitilo2023evaluating}. Within education, LLMs assist educators with various tasks such as assignment assessment \cite{hsiao2023developing}, providing feedback \cite{guo2024resist, chhetri2024exploring}, and generating questions \cite{elkins2023useful}. They also empower students by improving their writing skills \cite{han2023llm}, facilitating comprehension of complex concepts \cite{dai2023can}, and expediting information delivery \cite{nelson2025sensai}. Furthermore, Within software engineering, LLMs are utilized for generating source code \cite{liu2024no, zheng2023codegeex}, Software testing \cite{schafer2023empirical}, bug fixing \cite{zhao2025codinggenie, ahmad2024hardware}, and program repair \cite{bouzenia2024repairagent, luo2024fine}. In the further sections of this review, we examine the potential applications of LLMs in cybersecurity, with a particular focus on SOCs.

\subsection{Security Operation Centers}
\begin{figure}[H]
    \centering
    \includegraphics[width=0.8\linewidth]{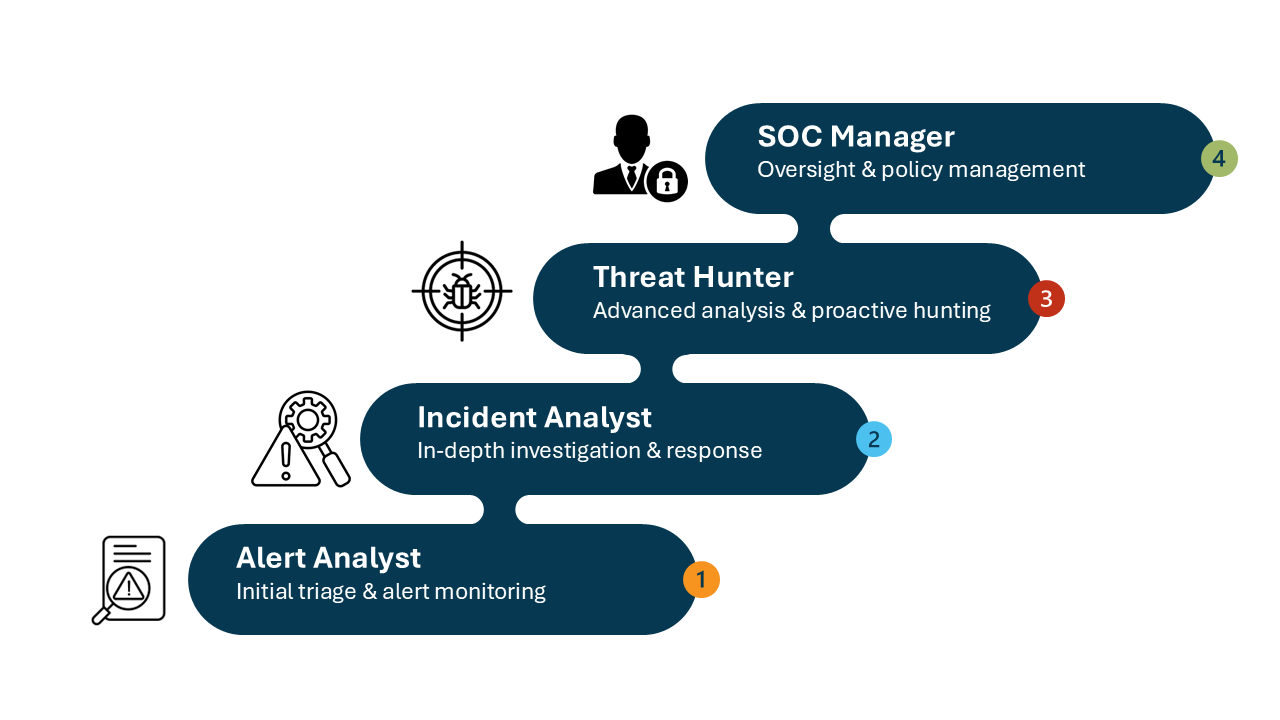}
    \caption{SOC Analyst Tiers. Analysts are typically organized into four tiers based on their expertise, responsibilities, and the complexity of alerts they handle.}
    \label{fig:soctiers}
\end{figure}

SOC is an organizational unit that integrates people, processes, and technologies to manage and strengthen an organization's overall security posture \cite{vielberth2021security_soc}. SOCs function as centralized hubs for cybersecurity, continuously monitoring, detecting, analyzing, and responding to security incidents. Their primary objectives are to protect sensitive data, ensure operational continuity, and safeguard the organization’s reputation. Analysts within SOCs are typically structured into Tiers 1 through 4 according to their experience and responsibilities, as illustrated in Figure \ref{fig:soctiers}. Tier 1 analysts, often junior, focusing on initial alert triage, classification, and prioritization of security events to identify potential incidents. More complex or higher-severity incidents are escalated to Tier 2 analysts for in-depth investigation and remediation efforts, while the most experienced Tier 3 analysts typically perform proactive threat hunting and handle major incidents. Managing a SOC involves coordinating complex processes to mitigate risks, comply with regulatory requirements, and enhance security effectiveness, which is the duty of a Tier 4 analyst. This tiered structure ensures efficient handling of security events and proper escalation. Moreover, SOC activities encompass both reactive measures, such as incident response, and proactive strategies, including threat hunting, and vulnerability assessment \cite{decusatis2019design, vielberth2020security}.

Within a SOC, the processes are necessary for effective operations and are typically aligned with the incident lifecycle, which broadly consists of three main phases: Detection, Analysis, and Response. This structured approach is essential for handling the escalating volume, velocity, and variety of security data and increasingly sophisticated cyber threats \cite{vielberth2021security_soc}.

\begin{figure}[H]
    \centering
    \includegraphics[width=1\linewidth]{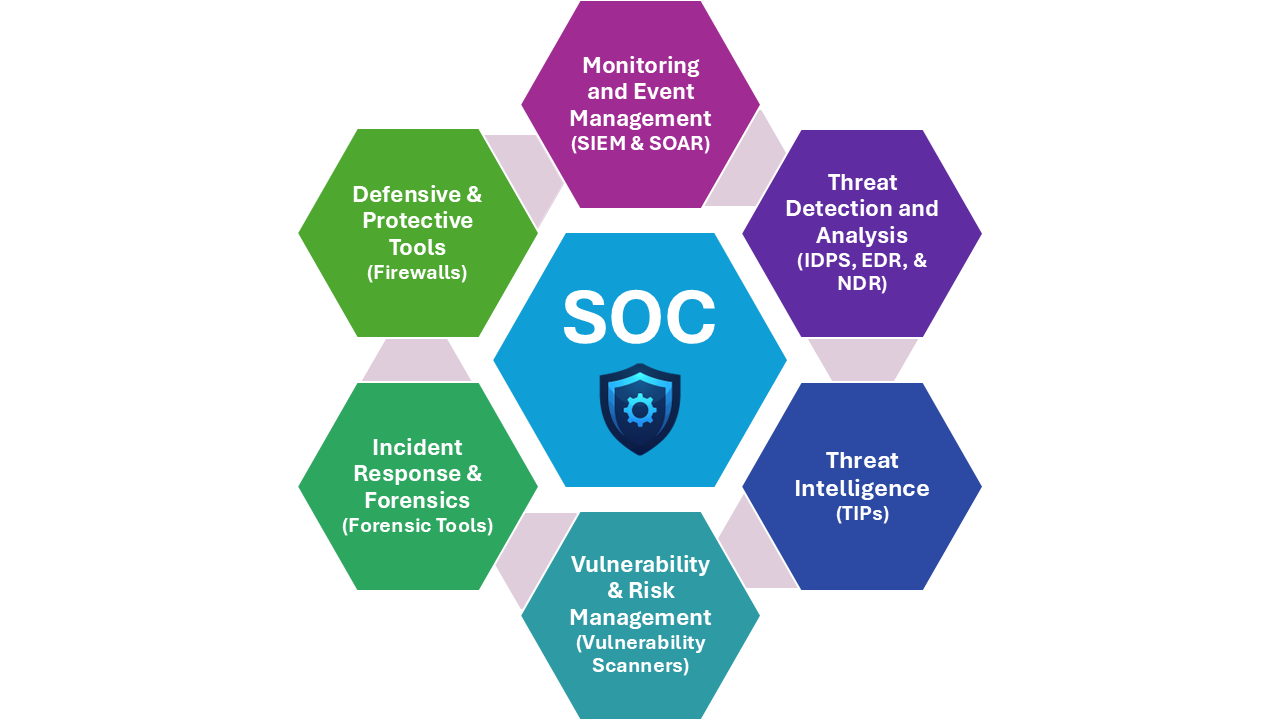}
    \caption{Main categories of SOC tools based on their purpose. }
    \label{fig:soctools}
\end{figure}

The Detection phase involves the generation of alerts indicating suspicious activities, often from various sources such as Network Intrusion Detection Systems (NIDS), Firewalls, Operating Systems (OSs), and application programs. This phase identifies anomalous patterns and potential threats that may be part of an unfolding attack. Furthermore, automation in this phase is vital, as the immense volume of data would otherwise overwhelm human analysts. As shown in Figure \ref{fig:soctools}, to ensure a robust security posture, SOCs employ a range of integrated tools that operate in correlation to provide visibility, threat detection, incident response, and overall risk management. Intrusion Detection and Prevention Systems (IDPS) are mandatory components in the detection phase of a SOC, designed to monitor network traffic and system activities to identify potentially malicious activity and abnormal patterns \cite{griffith2024network}. These systems, which can flag known attacks through signature detection or identify anomalies by modeling normal behavior, include tools such as Snort \cite{snort2025}, Suricata \cite{suricata2025}, and Zeek \cite{zeek2025}. Among these open-source offerings, Snort and Suricata predominantly rely on signature-based detection engines by default, whereas Zeek uniquely incorporates anomaly-based detection capabilities. Furthermore, while Zeek is particularly valuable for detecting zero-day attacks due to its anomaly-based approach, its practical adoption is often limited by a smaller default rule set and a less active community for rule generation, in contrast to Snort and Suricata which benefit from extensive rule bases and robust community support \cite{waleed2022open}.

The Analysis phase begins with alert triage, where SOC analysts assess the validity, severity, and potential impact of generated alerts. This involves correlating the alert with other contextual information, such as threat intelligence feeds, asset criticality, and vulnerability data. The primary goal is to filter out false positives and prioritize genuine incidents requiring immediate attention \cite{kersten2023give}. Analysts review each alert to classify it as interesting, indicating a potential incident that requires further investigation, or not-interesting, representing false positives. If an alert is deemed interesting, it is then escalated to higher-tier analysts. Threat Intelligence Platforms (TIPs) and Security Information and Event Management (SIEM) are two widely used technological components in the analysis phase. TIPs, such as Malware Information Sharing Platform (MISP) \cite{wagner2016misp, misp_project2025}, are open-source Cyber Threat Intelligence (CTI) sharing tools used by thousands of organizations worldwide, enabling the timely and efficient collection, analysis, and exchange of CTI. These platforms are requisite for mitigating ongoing attacks, preventing potential threats, and accelerating the identification and understanding of threats, threat actors, and their tactics, techniques, and procedures (TTPs) \cite{stojkovski2021s}. Moreover, SIEM tools like Wazuh \cite{wazuh2025}, IBM QRadar \cite{ibm_qradar2025}, and Cisco Umbrella \cite{cisco_umbrella2025} are central to collecting, normalizing, aggregating, and analyzing event logs and data from various IT infrastructure sources \cite{vielberth2021security_siem}. Utilized in both the Detection and Analysis phases, these tools Primarily relies on correlation rules, thresholds, and signatures to detect known suspicious behaviors across logs in near real-time. They enrich data with contextual information, and generate prioritized alerts and reports \cite{bassey2024building}.

Finally, the Response phase focuses on mitigating the identified threats. Based on the findings from the analysis, the SOC executes an incident response plan to contain the threat, eradicate the attacker's presence, and recover affected systems and data. This can involve informing relevant stakeholders and taking appropriate measures to prevent further damage \cite{freitas2025ai}. Technologies such as Security Orchestration, Automation, and Response (SOAR) platforms \cite{bridges2023testing} and Endpoint Detection and Response (EDR) solutions \cite{arfeen2021endpoint} are employed to streamline and automate response actions through predefined playbooks. Following an incident, post-incident activities are conducted to review the effectiveness of the response, identify gaps in security controls, and update incident response plans and playbooks for continuous improvement.

Despite these structured processes, modern SOCs face significant operational challenges, primarily stemming from the ever-increasing volume, velocity, and variety of security data, leading to phenomena such as alert fatigue and an overwhelming number of false positives \cite{10_triageprocess}. This deluge of information often conceals genuine threats, making the detection of actual attacker activities akin to "finding a needle in a haystack" \cite{vielberth2020security}. Coupled with a global skill shortage in cybersecurity and the monotonous, high-pressure nature of Tier 1 analyst tasks, this leads to analyst burnout and high turnover rates, compromising overall SOC effectiveness and increasing mean time to resolution for critical events \cite{gupta2019automated}. Furthermore, the increasing sophistication and stealth of modern cyber threats, including Advanced Persistent Threats (APTs) and zero-day exploits, often evade traditional rule-based detection systems, demanding more advanced analytical capabilities \cite{waleed2022open, griffith2024network}. Artificial Intelligence (AI), particularly LLMs, are poised to mitigate these challenges by automating key aspects of the SOC workflow. These technologies can significantly improve the accuracy and efficiency of alert classification and prioritization, reducing false positive rates and freeing human analysts from repetitive tasks to focus on complex investigations and threat hunting. By leveraging AI to analyze vast and heterogeneous datasets, detect anomalous behaviors, and provide contextual insights, SOCs can enhance their ability to proactively identify and respond to sophisticated threats, thereby reducing analyst cognitive load and improving overall organizational security posture.

\subsection{Related Surveys}
Since the emergence of LLMs, researchers have increasingly investigated their applications across diverse domains, including cybersecurity and Software Engineering. This has resulted in a growing body of survey articles that review, categorize, and compare existing studies in these fields. Several works specifically focus on providing a comprehensive overview of the literature related to either the offensive or defensive aspects of LLMs in cybersecurity. These studies provides a holistic perspective of the integration of LLMs into the cybersecurity domain \cite{yao2024survey, zhang2025llms}.
In this regard, a recent work \cite{yao2024survey} explores the intersection of LLMs with security and privacy. The authors categorized 281 papers into three distinct groups: "The Good" for security-beneficial applications, "The Bad" for offensive applications that could adversely impact security, and "The Ugly" for security vulnerabilities within LLMs and their defense mechanisms. Key findings from the survey reveal that LLMs have a predominantly positive impact on the security community, contributing significantly to both code security and data security and privacy. They also state user-level attacks are the most common type among offensive applications. User-level attacks are malicious activities where LLMs leverage their human-like reasoning abilities to create remarkably convincing but ultimately deceptive content.
Although these type of surveys address various aspects of cybersecurity, their coverage often remains high-level and fragmented. They primarily mention the applications of LLMs but lack a structured and rigorous comparison of the reviewed methodologies, therefore limiting the ability to discern strengths and weaknesses across studies. As a result, common patterns, prevailing practices, and critical research gaps are not sufficiently highlighted in the existing literature.

A substantial body of research has focused on more specific tasks and application domains. Notably, software vulnerability detection and rapair stands out as a particularly active and well-reviewed area within the broader intersection of LLMs and cybersecurity, reflecting its critical importance in safeguarding software systems. Dedicated systematic literature reviews have emerged to provide in-depth analyses of LLM applications, techniques, and datasets specifically for this crucial domain and its related task of program repair \cite{taghavi2025large, tamberg2025harnessing, zhou2025large}. For instance, a recent work \cite{zhou2025large} presents a systematic literature review focusing specifically on the utilization of LLMs for software vulnerability detection and repair. The authors analyzed 58 primary studies published from 2018 to 2024, encompassing research from leading Software Engineering, Artificial Intelligence, and Security venues. Their methodology aimed to summarize the types of LLMs employed, categorize adaptation techniques for both detection and repair, and identify limitations of existing studies.
In addition, several systematic reviews offer broader perspectives on the use of LLMs across other aspects of cybersecurity, including network operations \cite{liu2025large} and threat detection \cite{chen2024survey}. These review articles contribute to a more comprehensive understanding of how LLMs are being applied across different domains of security practice. In a recent study \cite{chen2024survey}, they mention the growing challenges in cyber threat detection due to the escalating complexity of cyber threats and the limitations of traditional models.  This survey aims to bridge that gap by examining how LLMs can effectively enhance detection and monitoring tasks, including cyber threat intelligence (CTI), phishing email detection, and threat prediction. It specifically focuses on the defender's perspective, exploring how LLMs optimize different stages of various security tasks and providing insights into their suitability for practical security scenarios.
Although these literature reviews examine the application of LLMs and their methodologies in specific downstream tasks within cybersecurity, they often fail to provide a comprehensive perspective on the broader operational context and the distinct requirements of SOC environments. This broader scope means these studies may not comprehensively reflect the nuanced integration and capabilities of LLMs within various SOC workflows. Consequently, the information provided often falls short of offering sufficiently specific and actionable insights for SOC managers to effectively guide strategic decisions and enhance team performance.

Finally, it is noteworthy that only a limited number of studies have been published recently on SOCs, despite their critical role as central points of defense against cyberattacks. Among the existing work, two studies \cite{feng2017user, oesch2020assessment} explored the use of machine learning to enhance SOC operations, Nevertheless, both were published prior to 2021. This indicates a clear gap in contemporary research, particularly in the integration of emerging AI techniques such as LLMs within SOC environments. Addressing this gap is essential for advancing both the practical capabilities of SOCs and the theoretical understanding of intelligent cybersecurity operations. Consequently, more systematic and up-to-date investigations are required to guide future developments and inform best practices in this rapidly evolving field.

\section{Methodology}\label{sec:method}
Our survey methodology follows the systematic approach proposed by Kitchenham’s guidelines \cite{kitchenham2007guidelines} for performing Systematic Literature Reviews (SLR) in software engineering. This approach involves three major activities: planning, conducting, and reporting the review. In the planning phase, we formulated the research questions that guided this study and developed the search queries for identifying relevant publications. During the conducting phase, we searched across selected digital libraries and retrieved the initial set of studies. We then applied inclusion and exclusion criteria to refine this set to the final group of studies included in our analysis. To ensure relevance to our research questions, we screened the selected studies based on their titles, abstracts, and conclusions. Full-text reviews were conducted when necessary to resolve ambiguities and confirm eligibility. Finally, in the reporting phase, we synthesized and organized the extracted data to systematically highlight key findings.

\subsection{Research Questions}
To understand the current state of LLM-based models in SOC-related tasks, it is necessary to create appropriate and effective research questions.
Through this SLR, six research questions are defined as follows, which are answered in sections \ref{sec:rq1} to \ref{sec:rq6}.
\begin{itemize}
    \item \textbf{RQ1: }What LLMs have been utilized? 
    \item \textbf{RQ2: }Which datasets are utilized in developing LLM-based models within the context of SOC?
    \item \textbf{RQ3: }How are LLMs adapted for the detection phase of SOCs?
    \item \textbf{RQ4: }How are LLMs adapted for the analysis phase of SOCs?
    \item \textbf{RQ5: }How are LLMs adapted for the response phase of SOCs?
    \item \textbf{RQ6: }What are the challenges and future research directions for applying LLMs in SOC domain?
\end{itemize}
Within the research questions, RQ1 seeks to identify specific patterns between the reviewed papers and the LLMs employed, enabling researchers to make more informed choices regarding the LLM most appropriate for their downstream tasks. RQ2 highlights the most commonly used datasets in studies focused on developing appropriate models for various SOC tasks. Moreover, RQ3 to RQ5 focus on methodologies for employing LLMs and comparing their performance across different phases of SOC operations. Lastly, RQ6 outlines the challenges in this research area and identifies potential future directions for the field.

To address RQ3–RQ5 and provide a clearer understanding of how Generative AI is being integrated into SOC workflows, this survey adopts the Technology Readiness Level (TRL) framework. TRL is a rigorous maturity scale that was originally developed by NASA \cite{NASA2012TRL} and later standardized internationally, including through ISO 16290 \cite{ISO16290}. The framework consists of nine levels that track the progression of a technology from the observation of basic principles to the deployment of fully operational systems. Theses levels are started with Basic principles observed and ends with Actual system proven in operational environment. In this survey, we use the TRL framework to systematically assess the maturity of LLM applications within SOC contexts based on evidence gathered from the reviewed literature.
However, for the purposes of this study, these nine levels are consolidated into four broader categories to facilitate comparison across different SOC tasks:
\begin{enumerate}
    \item \textbf{Emerging stage: }Concept development and proof of concept
    \item \textbf{Performance validation: } Laboratory and pilot validation
    \item \textbf{Operational adaptation: } System prototype demonstrated in operational environments
    \item \textbf{Full implementation: }Complete, qualified, and proven systems ready for deployment
\end{enumerate}

\subsection{Search Strategy}
A search strategy is a structured approach to identifying relevant studies in order to provide a comprehensive and unbiased overview of the literature.  This study primarily utilized five major scientific databases: ScienceDirect, SpringerLink, IEEE Xplore, ACM Digital Library, and ArXiv. To supplement these sources, Google Scholar was employed to capture additional relevant publications that may not have been indexed in the primary databases. We conducted the initial search using broad queries to capture a comprehensive set of papers in the field, followed by a manual screening process to eliminate irrelevant studies. The initial search queries focused on keywords related to SOCs, cybersecurity, and the application of LLMs in threat detection, analysis, and response. We employed the following search string to identify all primary studies relevant to the topic: \texttt{(Large Language Model} \texttt{\textbf{ OR }} \texttt{LLM} 
\texttt{\textbf{ OR }} \texttt{Generative AI} \texttt{\textbf{ OR }} 
\texttt{BERT} \texttt{\textbf{ OR }} \texttt{GPT} \texttt{\textbf{ OR }} 
\texttt{LLama)} \texttt{\textbf{ AND }} 
\texttt{(Cybersecurity} \texttt{\textbf{ OR }} 
\texttt{Security Operation Center} \texttt{\textbf{ OR }} 
\texttt{SOC} \texttt{\textbf{ OR }} \texttt{Log} 
\texttt{\textbf{ OR }} \texttt{Vulnerability} \texttt{\textbf{ OR }} 
\texttt{Network} \texttt{\textbf{ OR }} \texttt{Cyber Threat} 
\texttt{\textbf{ OR }} \texttt{Cyber Threat Intelligence)}.

\subsection{Study selection}
Study selection involves applying predefined inclusion and exclusion criteria to determine which papers are relevant for the review. This process ensures that only studies directly addressing the research questions are analyzed, improving the rigor and focus of the survey. 
We restricted our search to papers published between 2021 and 2025, ensuring the study captures the latest advancements in the field. Additionally, studies were included only if they directly employed LLMs in their architecture and their downstream task aligns with SOCs. We excluded duplicate studies, non-English publications, and documents that were not full-length papers, including books, short papers, and tutorials.
In the next step, each paper was screened based on its abstract and conclusion to evaluate its relevance, novelty, and currency. When ambiguities remained after this step, the full text was reviewed to determine eligibility. Through this process, we ensured that only relevant, novel, and methodologically significant studies were included in the final review.

\subsection{Data extraction}
\begin{figure}[H]
    \centering
    \includegraphics[width=0.8\linewidth]{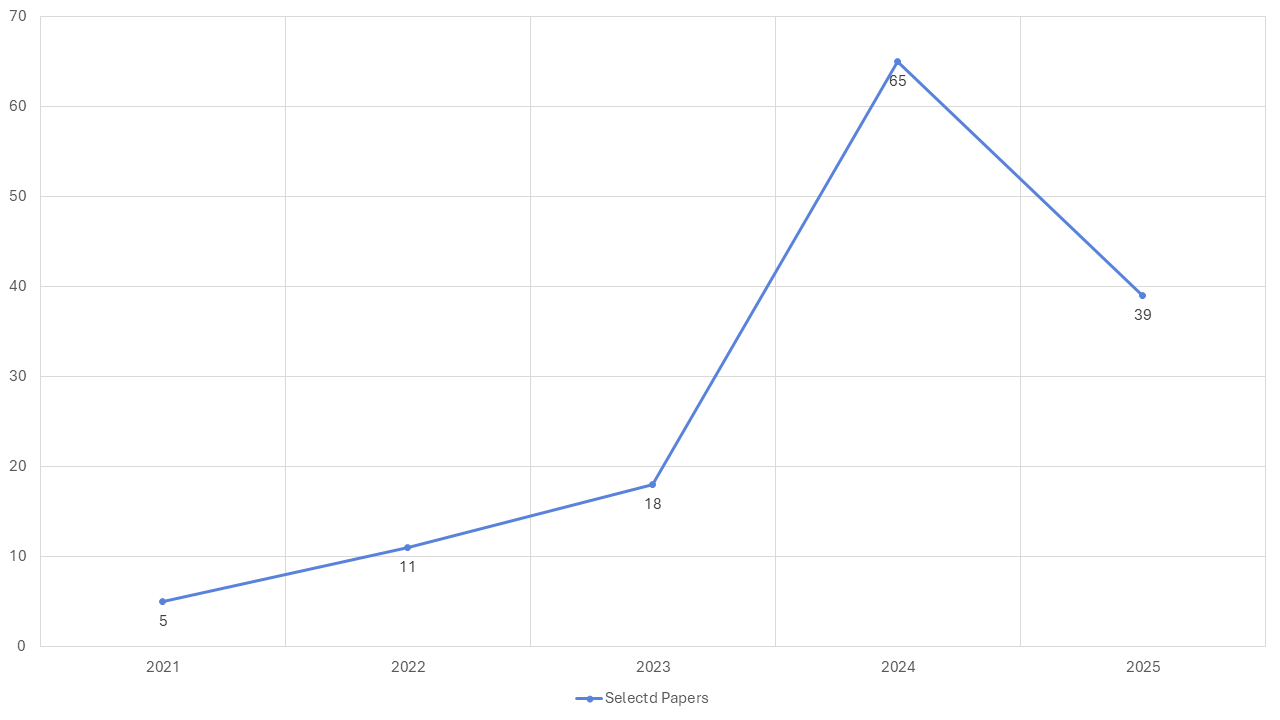}
    \caption{Statistical distribution of reviewed papers in each year.}
    \label{fig:paperyears}
\end{figure}

\begin{figure}[H]
    \centering
    \includegraphics[width=0.8\linewidth]{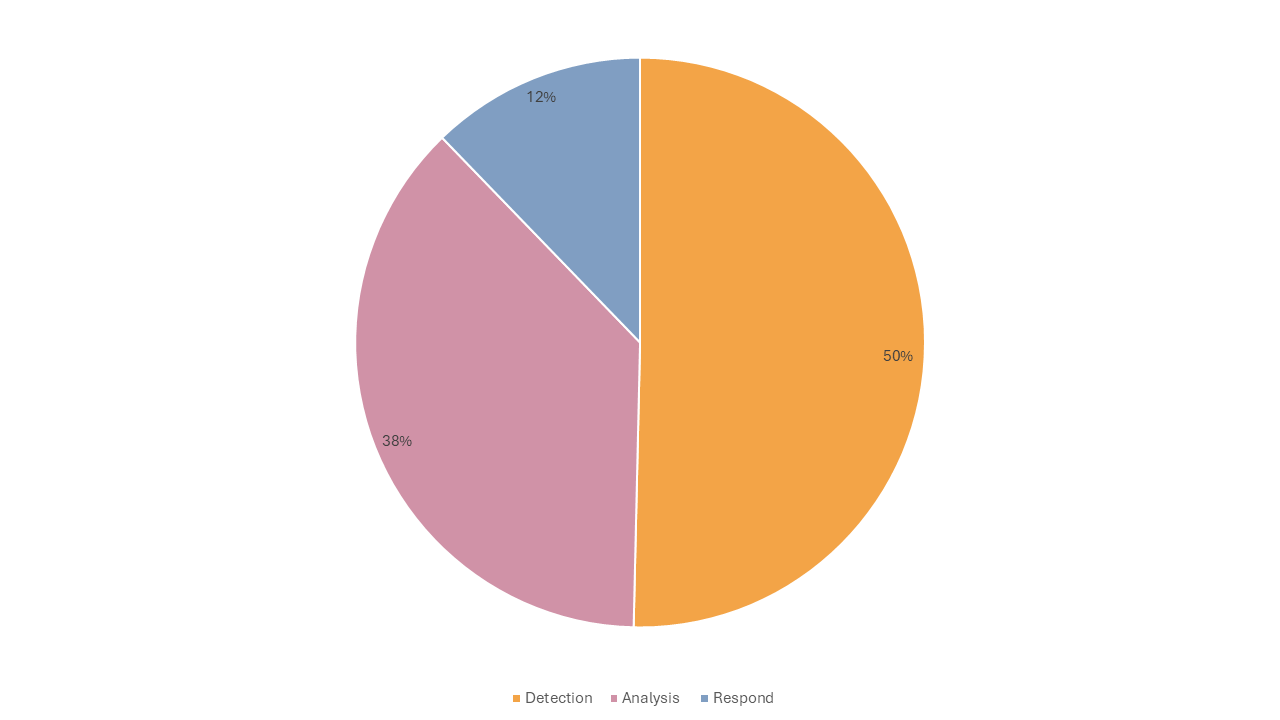}
    \caption{Statistical distribution of reviewed papers in SOC workflow.}
    \label{fig:distribution}
\end{figure}

Data extraction and categorization systematically collect and organize key information from selected studies to enable structured analysis and meaningful comparisons across the literature. Following the above processes, we identified 138 articles, and their distribution by published year is illustrated in Figure \ref{fig:paperyears}. The chart indicates a sharp increase in the introduction of LLM-based models within SOC-related domains, a trend that is likely to continue as LLMs advance further. Subsequently, the selected papers were categorized according to the workflow of SOCs, allowing for a structured analysis of the role of LLMs across detection, analysis, and response tasks. 
The distribution of papers across each phase, shown in Figure \ref{fig:distribution}, reveals a clear decline in the number of studies from Detection to Response. This suggests a significant research gap in the response phase, where LLMs could enhance SOC operations by automating generating actionable insights and assisting in decision-making, ultimately improving both efficiency and effectiveness.

\section{RQ1: What LLMs and methods have been utilized?}
\label{sec:rq1}

\begin{figure}[H]
    \centering
    \includegraphics[width=0.6\linewidth]{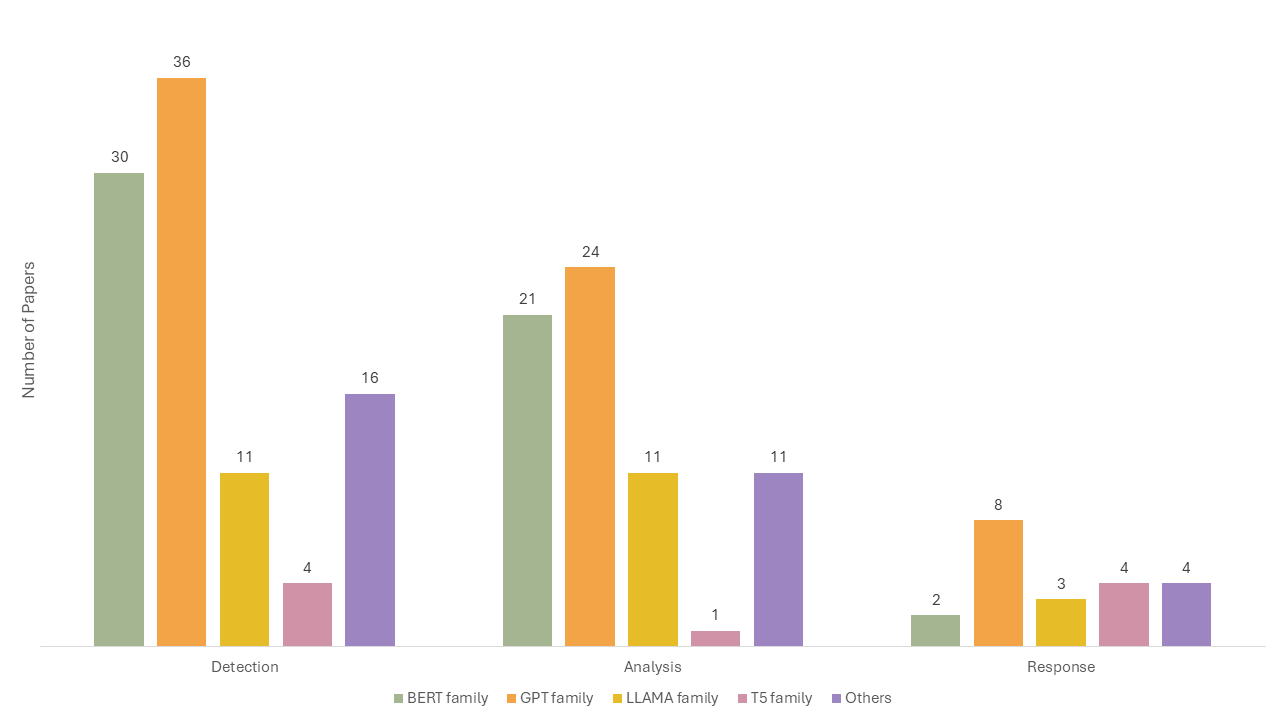}
    \caption{Statistical distribution of LLM types across reviewed papers classified by SOC workflow.}
    \label{fig:LLM_types}
\end{figure}

\begin{figure}[H]
    \centering
    \includegraphics[width=0.6\linewidth]{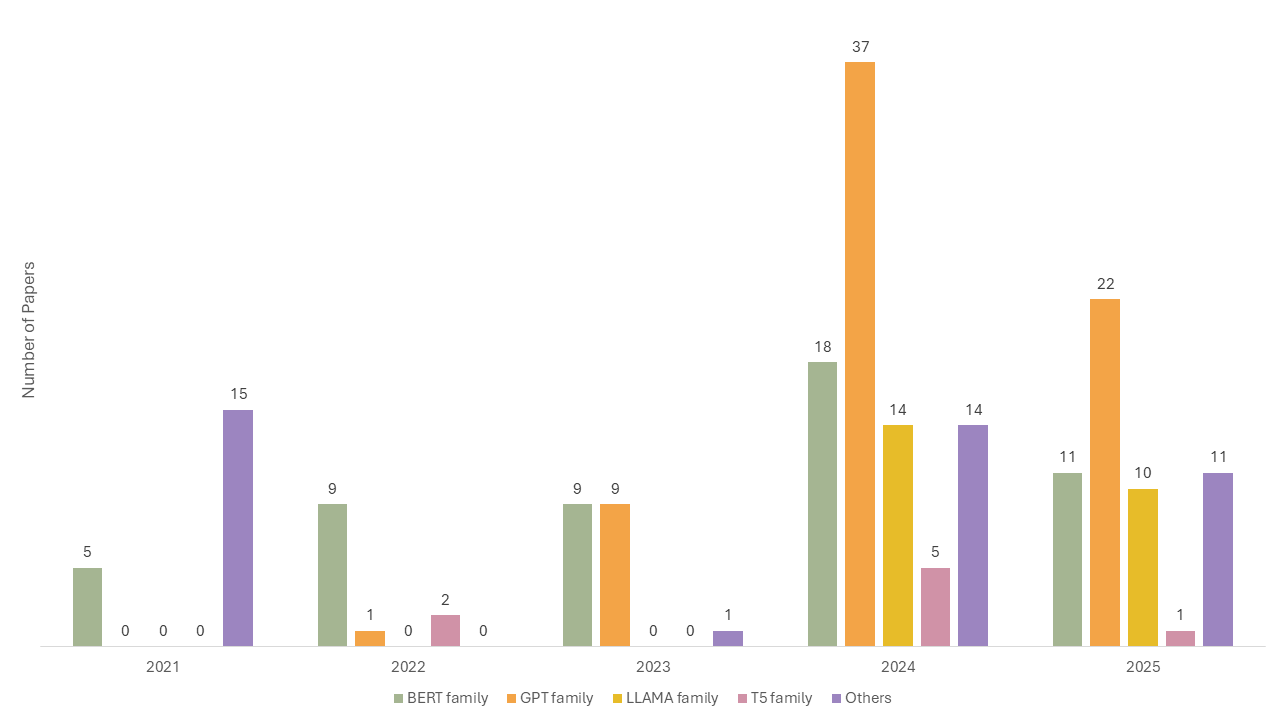}
    \caption{Statistical distribution of LLM types across reviewed papers classified by year of publication.}
    \label{fig:LLM_types_years}
\end{figure}

 Figures \ref{fig:LLM_types} and \ref{fig:LLM_types_years} illustrate the most frequently used LLMs among the studies reviewed in this paper. From these two figures, it can be observed that the BERT and the GPT families are the most widely adopted models in SOC-related tasks. Their popularity, however, stems from different factors. On the one hand, BERT family models are particularly attractive due to their open-source nature, which allows researchers to fine-tune them extensively or even pre-train them from scratch. Additionally, their bidirectional encoding architecture enhances representation learning by enabling the model to capture contextual information more effectively. Further advantages include lower computational demands, a reduced number of parameters, a compact and lightweight design, open availability, and a wide variety of pre-trained variants that enable adaptation to diverse tasks. On the other hand, the GPT family is particularly well-suited for exploiting the in-context learning capabilities of LLMs and is frequently employed in prompt engineering tasks. These models demonstrate strong performance in generating coherent and contextually relevant text, supporting zero-shot and few-shot learning across diverse domains, and adapting to various tasks without requiring task-specific fine-tuning. Its strong generalization capability allows GPT models to be applied effectively to a variety of security tasks, from alert analysis to threat intelligence. Although this family is not open-source to the public, its features can be accessed via available APIs for a fee. Another important observation is the considerable rise in the adoption of the GPT family in recent years relative to other model families, a trend that can be attributed to its compatibility with emerging methods such as Agents and RAG, in addition to its generalizability and scalability.

\begin{figure}[H]
    \centering
    \includegraphics[width=0.6\linewidth]{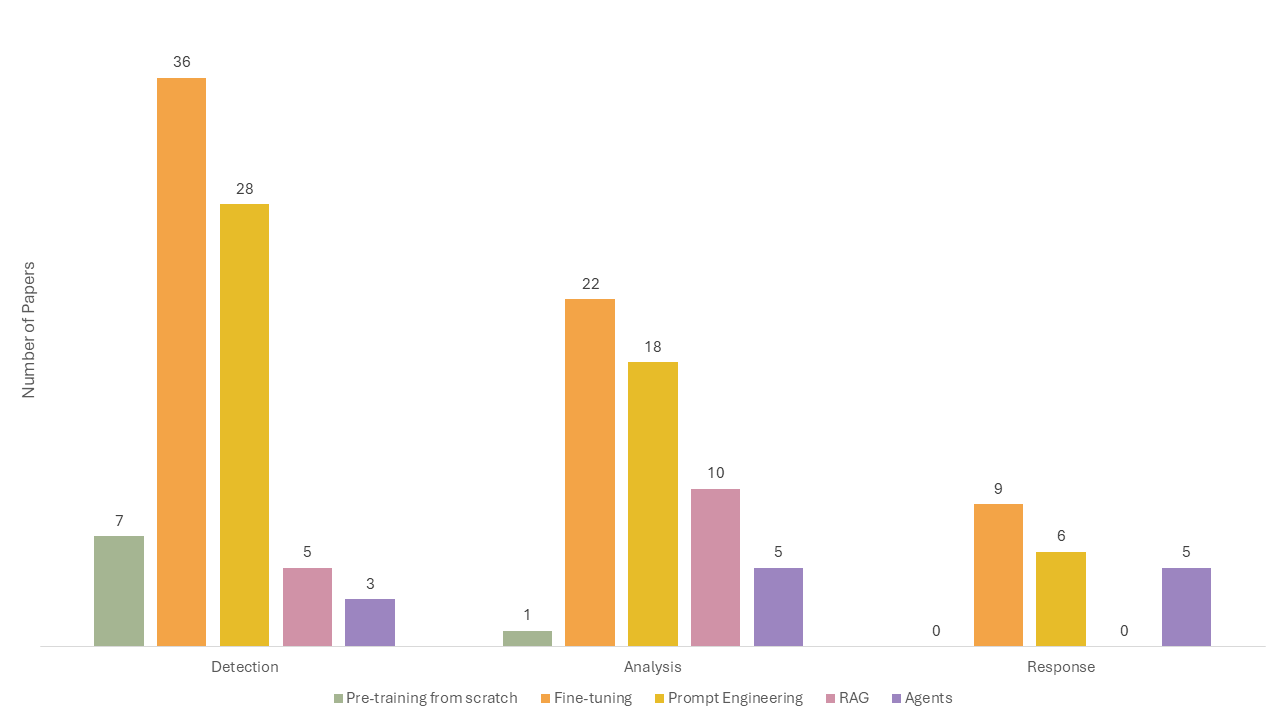}
    \caption{Statistical distribution of adopted methods across reviewed papers classified by SOC workflow.}
    \label{fig:LLM_methods}
\end{figure}

\begin{figure}[H]
    \centering
    \includegraphics[width=0.6\linewidth]{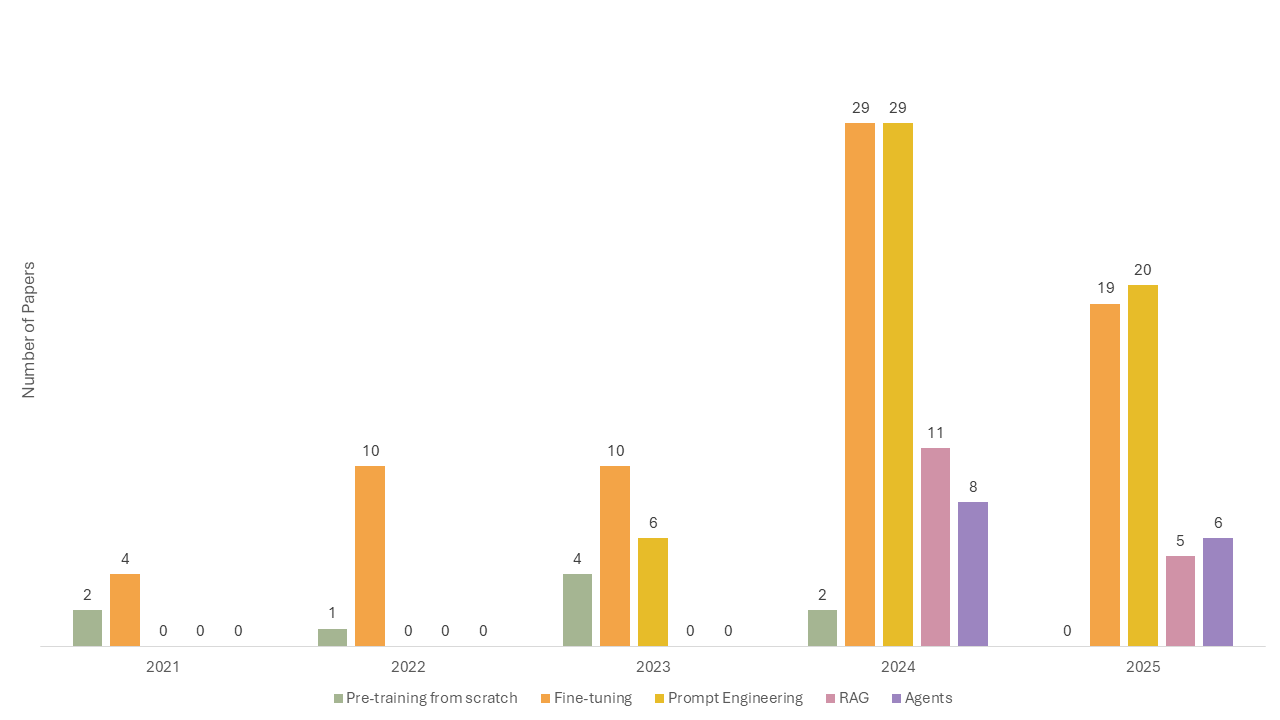}
    \caption{Statistical distribution of adopted methods across reviewed papers classified by year of publication.}
    \label{fig:LLM_methods_years}
\end{figure}

The statistical distribution of the methods adopted by the models analyzed in this study is presented in Figures \ref{fig:LLM_methods} and \ref{fig:LLM_methods_years}. As can be seen, Fine-tuning and prompt engineering are the most commonly used methods in the studies.  On one hand, The widespread use of fine-tuning is due to the need to transfer cybersecurity knowledge to language models for specialized tasks such as log anomaly detection, Network Intrusion Detection (NID), and vulnerability detection and remediation. In contrast, the capabilities of prompt engineering, including the flexibility and efficiency of zero-shot and few-shot learning, make it an ideal approach for more general tasks during analysis and response phases. These methods are crucial for harnessing the significant potential of LLMs to transform cybersecurity practices, enabling more automated, intelligent, and effective defenses against complex and evolving cyber threats. Ongoing research and development are essential to refine these techniques, address inherent limitations like biases and interpretability, and ensure the robust and reliable deployment of LLM-powered solutions in sensitive security environments. 

At the end, two key insights can be drawn from these observations. First, as the complexity of SOC workflows increases, the proportion of GPT-based model usage rises, whereas the use of BERT-based models declines. This trend may be attributed to the increasingly generative nature of tasks, which progress from detection to response. Second, the adoption of GPT models and prompt engineering techniques has grown substantially in recent years. This increase is likely driven by the widespread application of GPT in advanced approaches such as RAG and multi-agent systems, as well as researchers’ preference for methods that do not require extensive training or computational resources.

\subsection{Answer of RQ1: }BERT and GPT families are the most widely used models in SOC-related studies, primarily due to their trainability and in-context learning capabilities, respectively. Among the presented methods, fine-tuning and prompt engineering are the most commonly adopted approaches in the literature. This observation can also be attributed to the need for domain-specific knowledge in some tasks and the efficiency of zero-shot or few-shot task adaptation.

\section{RQ2: Which datasets are utilized in developing LLM-based models within the context of SOC?} \label{sec:rq2}

Datasets are an essential component in the development of effective LLM-based models. They can serve for training, evaluation, or as a knowledge base to address user requests. Moreover, drawing connections between different datasets can provide additional insights. The availability of numerous datasets and benchmarks within a specific domain can enhance trust, improve model performance, and facilitate broader generalization, whereas their absence may indicate limitations in model reliability and coverage. In the cybersecurity domain, particularly within SOCs, concerns regarding confidentiality, sensitivity of security data, and regulatory constraints have led organizations to refrain from sharing their datasets, resulting in a scarcity of publicly available resources; this scarcity in turn emphasizes the importance and value of the existing datasets. In this section, we provide a structured overview of the existing datasets in this domain. For clarity, these datasets are categorized into four groups: Log-related, Code-related, Network-related, and CTI-related.

\subsection{Log Related datasets}
Logs are fundamental to cybersecurity, serving as text files that record user activities, system events, and other vital data. They are essential for cybersecurity monitoring, forensics, and alert mechanisms. Logs provide  fault information and enable the detection, prevention, and response to cyber incidents by identifying anomalies and suspicious activity. The SIEM system actively collects, correlates, and analyzes these events and alerts from logs generated by network devices, servers, applications, and users. This comprehensive analysis enables SOCs to detect patterns, identify anomalies, and prioritize critical events efficiently. Log-related tasks in a SOC encompass a wide range of activities, including log parsing, anomaly detection, failure prediction, log compression, log summarization, root cause analysis, and incident response. Prominent benchmarks such as Loghub and Loghub-2k provide diverse log datasets from various systems, serving as fundamental resources for evaluating log analysis techniques \cite{madani2011log, karlsen2024benchmarking, balta2025cybersecurity}.

Loghub \cite{zhu2023loghub} is a widely used benchmark for various log analysis tasks such as anomaly detection, log parsing, failure prediction, log compression, and log summarization. This benchmark, particularly its Loghub-2k version, comprises 16 diverse log datasets including those from distributed systems, supercomputers, operating systems, mobile systems, server applications, and standalone softwares. For evaluation, 2,000 log messages from each system were manually labeled with ground-truth templates, and some datasets are also manually labeled for anomalies, making them suitable for anomaly detection tasks. HDFS, BGL, and Thunderbird are among the most frequently used datasets from Loghub for anomaly detection.
Moreover, Loghub-2.0 \cite{jiang2024large} is another updated and corrected version of Loghub, primarily used for log parsing evaluation. It contains ground-truth templates for 14 log datasets from Loghub, with an average of 3.6 million log messages and approximately 3,500 unique log templates across the collection, offering a more comprehensive evaluation environment.
{
\scriptsize
\begin{longtable}{p{0.75cm} p{1.5cm} p{1.5cm} p{2cm} p{1cm} p{2.25cm} p{1cm} p{1cm}}
\caption{Log Analysis datasets. The numbers in parentheses indicate the number of citations in this survey.} \label{tab:logdataset} \\
\hline
Ref. & Name & Type & Source & Size & Use Cases & Availability & Citations \\
\hline
\endfirsthead

\multicolumn{8}{c}{{\bfseries Table \thetable\ (continued)}} \\
\hline
Ref. & Name & Type & Source & Size & Use Cases & Availability & Citations \\
\hline
\endhead

\hline \multicolumn{8}{r}{{Continued on next page}} \\
\endfoot

\hline
\endlastfoot

\cite{oliner2007supercomputers} & Liberty (2007) & Individual Dataset & Super computers & 944 GB & Anomaly Detection & - & 812 (1) \\ \hline
\cite{zhu2019tools} & LogPAI (2019) & Dataset Collection & Multiple systems & - & Various Log Analysis tasks & \cite{logpai_github} & 600 (2) \\ \hline
\cite{le2020analyzing} & CERT (2020) & Individual Dataset & Organizational Activities & 20 GB & Insider Threat Detection & \cite{mansur_insider_threat_logon} & 190 (2) \\ \hline
\cite{laprade2020picodomain} & PicoDomain (2020) & Individual Dataset & Network & - & Anomaly Detection, and Intrusion Analysis & \cite{picoDomain2025} & 8 (2) \\ \hline
\cite{alsaheel2021atlas} & ATLAS (2021) & Individual Dataset & security event logs & 6.7 GB & Attack Investigation & \cite{alsaheel2021atlas_raw_logs} & 276(0) \\ \hline
\cite{jiang2023large} & LogPub (2023) & Dataset Collection & Multiple systems & 11 GB & Log Parsing & \cite{issta2024_log_parser_eval} & 16 (2)  \\ \hline
\cite{zhu2023loghub} & Loghub (2023) & Dataset Collection & Multiple systems & Over 100 GB & Various Log Analysis tasks & \cite{zhu2023loghub_github} & 174 (0) \\ \hline
\cite{zhu2023loghub} & HDFS (2023) & Individual Dataset & Distributed systems & 16 GB & Log Parsing, Anomaly Detection, and Failure Analysis & \cite{zhu2023loghub_github} & 174 (9) \\ \hline
\cite{zhu2023loghub} & BGL (2023) & Individual Dataset & Super computers & 0.7 GB & Log Parsing, Anomaly Detection, and Failure Analysis & \cite{zhu2023loghub_github} & 174 (14) \\ \hline
\cite{zhu2023loghub} & Thunderbird (2023) & Individual Dataset & Super computers & 30 GB & Log Parsing, Anomaly Detection, and Failure Analysis & \cite{zhu2023loghub_github} & 174 (10) \\ \hline
\cite{zhu2023loghub} & Spirit (2023) & Individual Dataset & Super computers & - & Log Parsing, Anomaly Detection, and Failure Analysis & \cite{zhu2023loghub_github} & 174 (3) \\ \hline
\cite{jiang2024large} & Loghub-2.0 (2024) & Dataset Collection & Multiple systems & 30 GB  & Log Parsing & \cite{jiang2024loghub2} & 43 (6) \\ \hline 
        
\end{longtable}
}

\subsection{Code Related datasets}
Code-related tasks protecting software systems against cyber threats by identifying and mitigating security flaws. Considering the severe societal and financial impacts of exploited vulnerabilities, timely and accurate detection and remediation are essential. However, manual approaches are often time-consuming, subjective, and inadequate to keep pace with the rapidly growing number and complexity of vulnerabilities. Consequently, automated techniques are indispensable for finding, localizing, classifying, and ultimately fixing these weaknesses in software \cite{wartschinski2019vudenc}.

The success of these automated approaches heavily relies on high-quality and diverse datasets. These datasets originate from real-world projects, synthetic examples, or a combination of both, and are collected at varying levels of granularity, including project, commit, file, function, slice, and line levels. As can be seen in Table \ref{tab:codedataset}, majority of the datasets are labeled at the function level, although this approach can limit their capacity to capture vulnerabilities that span across multiple functions or files. Another notable observation is that most datasets are based on the C programming language. This could be attributed to the widespread use of C/C++ in technological infrastructure, its susceptibility to critical memory and resource management vulnerabilities, and the greater availability of public datasets for these languages. Well-known datasets in this field include Vudenc for Python code, and Big-Vul, CVEfixes, and SARD which are predominantly for C/C++ vulnerabilities. Furthermore, the number of datasets focused on vulnerability repair is considerably smaller than those for vulnerability detection. Among the datasets listed in the table, only CVEfixes, D2A, FormAI, ContractTinker, PythonVulnDB, and PairVul are dedicated to repair tasks. Despite their growing availability in recent years, these datasets often encounter challenges such as label noise, data imbalance, duplication, limited scope, and the absence of associated test cases for repair tasks, all of which may hinder model performance and limit generalization to real-world scenarios \cite{ zhou2025large, kaniewski2025systematic}.

{
\scriptsize
\begin{longtable}{p{0.75cm} p{1.75cm} p{1cm} p{1.5cm} p{1.5cm} p{1.75cm} p{1.25cm} p{1cm}}
\caption{Vulnerability Detection and Repair datasets. The numbers in parentheses indicate the number of citations in this survey.} \label{tab:codedataset} \\
\hline
Ref. & Name & Type & Granularity & Languages & Vulnerability Types & Availability & Citations \\
\hline
\endfirsthead

\multicolumn{7}{c}{{\bfseries Table \thetable\ (continued)}} \\
\hline
Ref. & Name & Type & Granularity & Languages & Vulnerability Types & Availability & Citations \\
\hline
\endhead

\hline \multicolumn{7}{r}{{Continued on next page}} \\
\endfoot

\hline
\endlastfoot

    \cite{nist_sard} & SARD (2006) & Mixed & File & C/C++, Java, PHP, and C\# &  150 different CWEs & \cite{nist_sard} & - (1) \\ \hline
    \cite{nist_sard_juliet_cpp_1_3} & Juliet C/C++ (2017) & Synthetic & File & C/C++ &  91 different CWEs & \cite{nist_sard_juliet_cpp_1_3} & - (0) \\ \hline
    \cite{nist_sard_juliet_java_1_3} & Juliet Java (2017) & Synthetic & File & Java &  118 different CWEs & \cite{nist_sard_juliet_java_1_3} & - (0) \\ \hline
    \cite{zhou2019devign} & Devign (2019) & Real & Function & C &  - & \cite{Devign2019} & 1255 (3) \\ \hline
    \cite{fan2020ac} & Big-Vul (2020) & Real & Function, Line & C/C++ &  112 different CWEs & \cite{msr20codevuln} & 551 (8) \\ \hline
    \cite{ghaleb2020effective} & SolidiFI  (2020) & Synthetic & Function & Solidity  & 7 different CWEs & \cite{solidifibenchmark2025} & 244 (1) \\ \hline
    \cite{hazimeh2020magma} & Magma  (2020) & Synthetic & Function, Line & C/C++  & 138 different CVEs & \cite{magma2021} & 219 (1) \\ \hline
    \cite{li2021sysevr} & SeVC (2021) & Mixed & Function & C/C++ &  126 different CWEs & \cite{SySeVR2020} & 868 (2) \\ \hline
    \cite{chakraborty2021deep} & ReVeal (2021) & Real & Function & C/C++ &  - & \cite{reveal2020} & 806 (2) \\ \hline
    \cite{zheng2021d2a} & D2A (2021) & Real & Function, Line & C/C++  & - & \cite{d2a2021} & 196 (1) \\ \hline
    \cite{bhandari2021cvefixes} & CVEfixes (2021) & Real & Function, Commit, File & 27 languages  & 11873 CVEs in 272 different CWEs & \cite{moonen2024cvefixes} & 283 (4) \\ \hline
    \cite{nikitopoulos2021crossvul} & CrossVul (2021) & Real & File & 40+ languages  & 5131 CVEs in 168 different CWEs & \cite{nikitopoulos2021crossvulzenodo} & 105  (0)\\ \hline
    \cite{wartschinski2022vudenc} & VUDENC (2022) & Real & Line & Python  & 7 different CWEs & \cite{wartschinski2019vudenc} & 138 (0)\\ \hline
    \cite{chen2023diversevul} & DiverseVul (2023) & Real & Function & C/C++  & 150 different CWEs & \cite{diversevul2023} & 229 (3) \\ \hline
    \cite{russell2018automated} & Draper VDISC (2023) & Mixed & Function & C/C++  & 5 different CWEs & \cite{kim2018vd} & 20 (1) \\ \hline
    \cite{tihanyi2023formai} & FormAI (2023) & Synthetic & File, Function, Line & C/C++  & 41 different CWEs & \cite{formai2025} & 75 (1)\\ \hline
    \cite{ding2024vulnerability} & PRIMEVUL  (2024) & Real & Function & C/C++  & 140 different CWEs & \cite{ding2024primevul} & 67 (1) \\ \hline
    \cite{168_contracttinker} & ContractTinker (2024) & Real & Function & Solidity  & - & \cite{wang2025llm4smapr} & 16 (1) \\ \hline
    \cite{151_secureqwen} & PythonVulnDB (2025) & Mixed & Function & Python  & 15 different CWEs & - & 17 (1) \\ \hline
    \cite{147_vulrag} & PairVul  (2025) & Real & Function & C/C++  & 10 different CWEs & \cite{knowledgerag4llmvuld2024} & 68 (1) \\ \hline
         
\end{longtable}
}

\subsection{Network related datasets}
Network-related attacks remain one of the most pervasive and damaging categories of cyber threats, targeting the availability, integrity, and confidentiality of digital infrastructures. Such attacks encompass a wide spectrum of malicious activities, ranging from large-scale network intrusions to sophisticated phishing campaigns that exploit human and technical vulnerabilities. In the following sections, we examine the most commonly utilized datasets for network intrusion and phishing attack research in academic studies.

\subsubsection{Network Intrusion}
Network Intrusion Detection (NID) involves the analysis of network traffic and cybersecurity incidents to identify and respond to malicious activities. NIDS serve as a cornerstone of network security, actively monitoring traffic in real-time to detect potential threats and effectively safeguard data and systems. By automating the analysis of cybersecurity incidents, NIDS can significantly reduce the workload of SOC analysts, while also providing strategic recommendations and comprehensive reports for improved defense measures\cite{gupta2023chatgpt}.

Labeled datasets are necessary in this domain to train and evaluate machine learning-based NIDS as they allow models to learn from both normal and malicious network traffic patterns. However, many traditional datasets have limitations such as being outdated, lacking traffic diversity, containing anonymized packet information, or not accurately reflecting contemporary real-world network environments \cite{ring2019survey}. To address these shortcomings, new and comprehensive datasets like CICIDS2017, TON\_IoT, ACI-IoT-2023, CICIoT2023, UNSW-NB15, and Edge-IIoTset have been developed, providing diverse and realistic attack scenarios for robust NIDS research and deployment.

{
\scriptsize
\begin{longtable}{p{0.75cm} p{1.75cm} p{1.75cm} p{1.5cm} p{1.5cm} p{1.25cm} p{1.25cm} p{1.5cm}}
\caption{Network Intrusion detection datasets. The numbers in parentheses indicate the number of citations in this survey.} \label{tab:networkdataset} \\
\hline
Ref. & Name & Devices & Traffic Type & Attacks & Features & Availability & Citations \\
\hline
\endfirsthead

\multicolumn{7}{c}{{\bfseries Table \thetable\ (continued)}} \\
\hline
Ref. & Name & Devices & Traffic Type & Attacks & Features & Availability & Citations \\
\hline
\endhead

\hline \multicolumn{7}{r}{{Continued on next page}} \\
\endfoot

\hline
\endlastfoot

         \cite{ferrag2022edge} & Edge-IIoTset (2022) & IoT & Real & 5 types & Flow-based & \cite{ferrag2022edgeiiotset} & 789 (1) \\ \hline
         \cite{neto2023ciciot2023} & CICIoT2023 (2023) & IoT & Real & 7 types & Flow-based & \cite{neto2023ciciot2023r} & 610 (4) \\ \hline
         \cite{sharafaldin2019developing} & CICDDoS2019 (2019) & firewall, router, switches, and operating systems & Emulated & 11 types of DDoS attacks & Flow-based & \cite{cic_ddos2019} & 1306 (1) \\ \hline
         \cite{sharafaldin2018toward} & CICIDS2017 (2018) & firewall, router, switches, and operating systems & Emulated & 7 types & Flow-based & \cite{cic_ids2017} & 5058 (0) \\ \hline
         \cite{koroniotis2020new} & Bot-IoT (2019) & IoT & Real & 3 types & Flow-based & \cite{koroniotis2019botiot} & 270 (1)\\ \hline
         \cite{song2020vehicle} & Car Hacking Dataset (2018) & vehicle & Hybrid & 4 types & Packet-based & \cite{song2020carhacking} & 655 (0) \\ \hline
         \cite{nack2024aci} & ACI-IoT-2023 (2023) & IoT & Hybrid & 4 types & Flow-based and Packet-based & \cite{bastian2023aci_iot} & 2 (0) \\ \hline
         \cite{farzaneh2023dtl} & DTL-IDS (2023) & Various devices in testbed environment & Hybrid & 9 types of DDoS attacks & Flow-based & \cite{farzaneh2023dtlids5g} & 15 (1) \\ \hline
         \cite{herzalla2023tii} &  TII-SSRC-23 (2023) & Various devices in testbed environment & Emulated & 26 types & Flow-based and Packet-based & \cite{herzalla2023tiissrc23} & 32 (1) \\ \hline
         \cite{sarhan2020netflow} &  NF-UNSW-NB15 (2020) & Various devices in testbed environment & Hybrid & 20 types & Flow-based & \cite{moustafa2015unsw} & 463 (1) \\ \hline
        \cite{alsaedi2020ton_iot} &  TON\_IoT (2020) & IoT & Hybrid & 9 types & Flow-based and Packet-based & \cite{moustafa2021toniot} & 723 (0) \\ \hline

\end{longtable}
}

\subsubsection{Phishing attacks}
Phishing attacks are a pervasive cybercrime tactic that exploits human vulnerabilities through social engineering to fraudulently obtain sensitive information, such as login credentials, financial details, and personal identification \cite{chiew2018building}. These deceptions are crafted to trick individuals into disclosing sensitive data, posing a significant and continually evolving threat in the digital landscape. Phishing attacks commonly occur through various channels, with email phishing being a primary vector, where fraudulent messages mimic legitimate communications to manipulate users. A more sophisticated variant, spear phishing, targets specific individuals with highly personalized emails, relying on extensive research about the target to enhance the message's credibility \cite{nahmias2024prompted}. Another prevalent type is phishing websites, in which attackers design counterfeit webpages that closely resemble legitimate platforms, often replicating official logos and branding, to deceive users into submitting sensitive information. This includes brand-based phishing, targeting users by specifically imitating popular and trusted brands. Moreover, the  detection process typically involves identifying the brand that a webpage is attempting to impersonate and verifying whether its domain name corresponds to that of the legitimate brand \cite{tan2023hybrid}. Finally, this attacks can also occur via SMS, known as smishing, where malicious links are sent through text messages to mobile phone users \cite{do2022deep}.

Despite the prominent role of datasets in developing detection models in this field, yet the research community faces significant challenges due to the lack of a publicly accessible standard dataset, leading many researchers to use their own limited collections \cite{dephides}. Publicly available datasets for phishing detection are often suffer from issues such as lack of diversity, recency, and quality, frequently containing dead, duplicate, or incomplete links \cite{do2022deep}. To overcome these issues and ensure robust, high-performing detection systems, there is a recognized need for large-scale, balanced, and high-quality datasets that reflect current and diverse phishing patterns.

Studies on phishing detection frequently construct their datasets by augmenting their phishing datasets with legitimate websites from publicly available sources to enable a comprehensive and representative evaluation of detection systems. For phishing data, widely used resources include PhishTank, which provides verified phishing URLs, and email-based datasets such as the Nazario phishing corpus, IWSPA-AP training data, FRAUD corpus, and Kaggle’s fraudulent email corpus. Conversely, legitimate and benign data is often sourced from extensive collections like Alexa \cite{kaggle_top1m} and Tranco lists \cite{tranco2025} for URLs, or email archives such as Enron \cite{benston2002enron, shetty2004enron} and the benign portions of SpamAssassin datasets. Subsequently, we review phishing datasets in the table \ref{tab:phishingdataset}.

{
\scriptsize
\begin{longtable}{p{1.75cm} p{1.5cm} p{1.75cm} p{1.5cm} p{1.5cm} p{1.5cm} p{1cm}}
\caption{Phishing Detection datasets. The numbers in "Citation" column indicate the number of citations in this survey. Datasets without any published year indicate that they are being updated continuously.} \label{tab:phishingdataset} \\
\hline
Ref. & Name & Data Type & Collection Method & Number of Samples & Attacks type & Citations \\
\hline
\endfirsthead

\multicolumn{7}{c}{{\bfseries Table \thetable\ (continued)}} \\
\hline
Ref. & Name & Data Type & Collection Method & Number of Samples & Attacks type & Citations \\
\hline
\endhead

\hline \multicolumn{7}{r}{{Continued on next page}} \\
\endfoot

\hline
\endlastfoot

        \cite{rtatman_fraudulent_email_corpus} & Fraudulent E-mail Corpus (1998-2007) & Email-based & Manual collection & 4,098 phishing emails & Primarily Nigerian-style fraud emails & 1 \\ \hline
         \cite{naserabdullahalam_phishing_email_dataset} & Nazario (2000-2015) & Email-based & Manual collection & 5,000 phishing emails & Various phishing attacks  &  2 \\ \hline
         \cite{spamassassin_corpus} & SpamAssassin (2002-) & email-based & Real-world collection & 6,000 spam and ham emails & General spam emails  & 1 \\ \hline
         \cite{phishtank} & PhishTank (2006-) & URL-based & Crowd-sourced & Over 1.9 million & Various phishing attacks  & 2 \\ \hline
         \cite{lindsey98_phishpedia} & Phishpedia (2021) & Website HTML/visual & Web scraping & 1,820 phishing webpages & Brand impersonation  & 1 \\ \hline
         \cite{openphish} & OpenPhish & URL-based & Automated collection & - & Various phishing attacks  & 5 \\ \hline
         \cite{apwg} &  APWG & email-based, URL-based, and website HTML/visual & Crowd-sourced & Over 4.7 million phishing attacks & Business Email Compromise, social media phishing, and credential theft  & 1 \\ \hline

\end{longtable}
}

\subsection{Cyber Threat Intelligence datasets}

Cyber Threat Intelligence (CTI) knowledge bases are pivotal resources for SOC teams, enabling them to proactively defend against the increasingly sophisticated cyber threat landscape. SOC analyst teams heavily rely on comprehensive data sources such as the MITRE framework and the National Vulnerability Database (NVD). The MITRE framework plays a central role in SOC operations by providing a common language and structured knowledge base that enhances the capacity of security teams to understand, categorize, and respond to threats effectively, thereby strengthening overall threat intelligence and defense strategies. It offers a standardized lexicon of adversarial behaviors, organized into Tactics, Techniques, and Procedures (TTPs), which model attacker operations. The framework also maintains the Common Vulnerabilities and Exposures (CVE) list, documenting publicly known cybersecurity vulnerabilities, and the Common Weakness Enumeration (CWE), which details software and hardware weakness types that may lead to vulnerabilities. Additionally, the Common Attack Pattern Enumeration and Classification (CAPEC) describes cyber attack patterns. A significant portion of this intelligence is derived from Open Source Intelligence (OSINT), gathered from various unstructured formats like security blogs, social media, and news articles. By processing and analyzing these diverse CTI data sources, often with the aid of advanced NLP techniques and specifically LLMs, SOC teams gain actionable insights for informed decision-making, enhancing their security posture, refining incident response, identifying and mitigating vulnerabilities, and performing proactive threat hunting. This automated extraction and structuring of CTI helps bridge the gap between vast amounts of unstructured information and the timely, effective implementation of cyber defense actions, significantly reducing manual workload for analysts. In the following, We examine each knowledge base within MITRE with more details.

\subsubsection{Common Vulnerabilities and Exposures}
Common Vulnerabilities and Exposures (CVEs) represent a foundational component of cybersecurity knowledge, providing a standardized naming system for publicly known cybersecurity vulnerabilities. Each CVE entry is a brief, low-level description that uniquely identifies a vulnerability in a specific product or protocol, typically including an identification number, a textual description, and relevant references. The MITRE Corporation oversees and maintains the authoritative CVE list. Complementing MITRE's efforts, the NVD, managed by the National Institute of Standards and Technology (NIST), enriches these CVE entries with additional crucial information such as severity estimations (CVSS), impact metrics, affected products, and references to exploit information. While MITRE focuses on uniquely identifying and cataloging vulnerabilities, the NVD provides a comprehensive, machine-readable repository that supports vulnerability assessment, risk prioritization, and automated integration into security tools, making it especially valuable for operational cybersecurity and SOC environments. Furthermore, another key distinction lies in the completeness and hierarchical relationships of associated data; for instance, while NVD utilizes a subset of CWEs from MITRE, it may initially lack associated CWE mappings for some CVEs, often labeled as "NVD-CWE-noinfo", or provide limited information on mitigation strategies, which can be further elaborated by linking to the MITRE ATT\&CK framework. 
Despite these variations in detail, CVEs are leveraged by organizations to monitor newly discovered vulnerabilities, coordinate defense efforts, and enhance their security posture against evolving cyber threats \cite{31_hwrex, 125_smet}.

\subsubsection{Common Weakness Enumeration}
Common Weakness Enumeration (CWE), maintained by the MITRE Corporation, functions as a standardized and hierarchically organized dictionary of common software and hardware weakness types. These entries delineate fundamental flaws, faults, bugs, or errors in system design, architecture, code, or implementation that, if left unaddressed, can lead to exploitable security vulnerabilities. The key distinction between CVEs and CWEs lies in their scope: CWEs classify the underlying types of weaknesses, providing a conceptual framework for understanding root causes and guiding detection and mitigation strategies, whereas CVEs identify specific, actionable instances of these vulnerabilities. For example, CVE-2004-0366 denotes a concrete vulnerability involving arbitrary SQL statements in a particular product, while CWE-89: SQL Injection describes the broader class of flaw, outlining its inherent properties and informing effective countermeasures \cite{1_v2wbert}. Accordingly, CVEs alert practitioners to concrete threats, whereas CWEs offer foundational knowledge essential for comprehensive vulnerability management and the design of robust defensive measures \cite{31_hwrex}.

\subsubsection{Adversarial Tactics, Techniques, and Common Knowledge}

The MITRE Adversarial Tactics, Techniques, and Common Knowledge (ATT\&CK) framework is a globally acknowledged, comprehensive knowledge base that details adversarial behavior observed in the real world. It provides a standardized lexicon and hierarchical structure of Tactics, Techniques, and Procedures (TTPs). In the hierarchical TTPs matrix, tactics define why an adversarial action is performed, techniques specify how it is carried out, and procedures represent the concrete, real-world implementations of those techniques\cite{li2022attackg}. For SOCs and cybersecurity professionals, ATT\&CK is an essential resource that significantly enhances their capabilities by offering actionable insights into threats. It enables teams to proactively anticipate and design effective defense strategies, facilitates threat hunting activities by providing a structured understanding of attack patterns, and supports incident response by improving situational awareness and enabling swift reactions to detected attacks. Furthermore, ATT\&CK helps bridge the gap between vulnerability information (like CVEs) and attack actions, allowing analysts to understand how vulnerabilities are exploited and to prioritize mitigation measures. Its widespread adoption across government, private sectors, and cybersecurity products makes it a valuable tool for understanding threat actor behaviors and creating effective cyber defense strategies \cite{43_shah2024ai, ampel2021linking}.

\subsubsection{Common Attack Pattern Enumeration and Classification}

The Common Attack Pattern Enumeration and Classification (CAPEC) is a comprehensive database maintained by the MITRE Corporation, serving as a publicly available catalog that meticulously documents and categorizes a wide array of attack patterns utilized by cyber adversaries. This repository provides detailed descriptions of these attack patterns, including their underlying mechanisms, necessary prerequisites, and typical consequences. CAPEC's structured taxonomy is also integral to connecting various cybersecurity knowledge bases, acting as a critical bridge linking MITRE ATT\&CK Tactics and Techniques with CWE and CVE.
For SOC teams and cybersecurity professionals, CAPEC is an indispensable resource. It significantly enhances threat modeling and security assessments, enabling teams to anticipate potential threats and design more effective defensive measures, thereby contributing to a more resilient cybersecurity posture. By providing a structured understanding of how attacks are executed, CAPEC fosters clearer communication regarding security issues and supports systematic analysis and response to cyber threats \cite{14_bron, 31_hwrex}.

\subsection{Answer of RQ2: }Among the reviewed studies, BGL, BIG-VUL, CICIOT2023, and OpenPhish are the most commonly used datasets for log analysis, vulnerability analysis, network analysis, and phishing detection, respectively. Furthermore, MITRE and NVD serve as the primary repositories supporting models developed to perform CTI analysis. 

\section{RQ3: How are LLMs adapted for the detection phase of SOCs?}\label{sec:rq3}

The detection phase constitutes a central stage in the SOC workflow, dedicated to identifying potential security incidents. This phase is crucial to the organization's defense architecture, involving continuous monitoring of IT infrastructure and security controls to detect threats. Incidents are identified through both automated and manual procedures, determining whether collected data indicates a security incident. While automated systems handle the sheer volume of data, human analysts are essential for detecting advanced or unknown attacks that technology might miss. Advanced solutions, including LLMs, are being explored to enhance threat detection by capturing semantic information from unstructured logs, facilitating pattern recognition, and providing explainable outcomes to bridge the "semantic gap" that analysts often encounter with traditional ML outputs. This evolution aims to make the detection phase more efficient, accurate, and interpretable, enabling proactive defense strategies against evolving cyber threats.

\subsection{Log anomaly detection}
System logs are records generated by logging statements in software source code that capture system events, status, and activities at runtime. They serve as a primary information resource for monitoring and maintaining the reliability, stability, and security of computer systems, often appearing as semi-structured or unstructured text \cite{hadadi2024anomaly}. These logs provide valuable insights for troubleshooting and security analysis, and play a key role in cybersecurity monitoring. Furthermore, modern SOC platforms such as Splunk \cite{splunk} and ElasticStack \cite{elastic_stack} depend on collecting and analyzing event logs from numerous sources \cite{llmtd}. Understanding the information contained in system logs constitutes a essential element of system Operation and Maintenance (O\&M) and timely detection of anomalous events represents a fundamental step in protecting online computer systems against malicious attacks or operational malfunctions \cite{105_loglm, liu2023log}.

Log-based anomaly detection serves a vital role by identifying patterns that deviate from established norms of system behavior, thereby providing an effective technique for monitoring system activities and detecting potentially suspicious actions. Anomalies detected within log data often indicate the presence of potential system malfunctions, security breaches, or operational failures. Furthermore, the projected expansion of interconnected Internet of Things (IoT) devices, estimated to reach 55.7 billion units by 2025 and expected to generate approximately 80 zettabytes of data, highlights the increasing importance of efficient log data management, analysis, and interpretation in maintaining the reliability and security of networked systems \cite{lou2010mining, sarlog}. Therfore, the effective detection of abnormal events in online computer systems is essential for preserving both their security and operational reliability. But before examining the applications of LLMs in log anomaly detection, it is necessary to first provide a brief overview of log parsing, a critical component of log data preprocessing.

\subsubsection{Parsing}
The first step in log analysis tasks is log parsing, which is the conversion of raw log messages into structured log messages \cite{he2021survey}. When a standardized log template is derived from a raw log entry, complex data are simplified through preprocessing, allowing them to be represented by a limited number of distinct events. For example, 4,747,964 log messages generated from the BGL system are converted to 376 events through the Drain parser \cite{drain}. This allows security tools to more effectively perform tasks such as anomaly detection and event analysis across systems in real time. Raw log messages are typically composed of two components that log parsers are responsible to extracting them: 1) log template: the fixed, unchangeable part of the logging statements, and 2) log parameters: dynamic parts that can change across executions.

Manual log parsing is impractical and challenging due to the large volume and complexity of logs generated by modern software, necessitating automated log parsing approaches \cite{manuallogparsing}. Before the advent of LLMs, syntax-based log parsers such as Drain \cite{drain} and AEL \cite{ael} were the most commonly used log parsers due to their acceptable performance and high efficiency. But these models faced critical problems: Firstly, they have low accuracy in distinguishing log templates from parameters when there are not many instances of that log message. Secondly, traditional parsers often lack robustness across different types of logs and datasets due to different logging formats and behaviours. Thirdly, syntax-based log parsers rely heavily on rules designed by experts, and their performance can degrade significantly when log messages deviates from those established rules. To address these challenges, recent models have moved toward designing semantic-based log parsers, where LLMs play a pivotal role \cite{khan2022guidelines}.

LLMs are highly effective for log parsing due to their powerful semantic understanding, which facilitates accurate distinction between static keywords and dynamic parameters. Additionally, their inherent generalizability and adaptability minimize manual intervention and improve robustness across diverse and evolving log formats. One approach in semantic-based models is to further train or fine-tune a language model \cite{logppt, llmparser}. For example, LogPPT \cite{logppt} treats log parsing as a token classification task and utilizes prompt-based few-shot learning to effectively capture the patterns of templates and parameters in a log message. LogPPT initially creates a small and diverse set of log messages with their labels and then, during training, the model is tasked with predicting a virtual label token ("PARAM") at the positions of parameters while keeping template tokens unchanged. LogPPT eliminates the requirement for manual preprocessing, enhancing its robustness and significantly outperforming traditional methods such as Drain and AEL.

Nevertheless, the training-based approach often also comes with challenges such as limited generalization to unseen and evolving logs, high computational cost, and low efficiency, which has led recent log parsers to consider utilizing the in-context learning capabilities of LLMs instead of further training them \cite{logparser-llm}. The learning strategies employed by these log parsers can be categorized into supervised and unsupervised methods. The supervised models, which includes Divlog \cite{divlog}, LILAC \cite{lilac}, and LogParser-LLM \cite{logparser-llm} require some labeled datasets. Specifically, these parsers select a few examples of ground truth data as demonstrations in the prompt to guide the LLM in predicting the template for the requested log message. The primary distinction among these models are candidate sampling and demonstration selection employed to identify the most suitable examples for the LLM's ICL process. Candidate sampling is an initial, often offline, process that selects a small, diverse, and representative subset of log messages from a larger, typically unlabeled, log dataset. This step aims to minimize the labor-intensive effort of manual data labeling by focusing annotation on a compact yet informative set of examples. Subsequently, demonstration selection operates during the online parsing phase, where for each new log message to be parsed, a few most relevant examples are dynamically retrieved from this pre-sampled and labeled candidate set. Moreover, LILAC, and LogParser-LLM enhance their efficiency by strategically minimizing expensive LLM invocations; LILAC achieves this through an adaptive parsing cache that stores and refines templates to prevent duplicate queries and mitigate the overhead of LLM inference, while LogParser-LLM integrates a prefix tree structure to efficiently cluster logs and employs a stringent matching paradigm that reduces the necessity for repetitive LLM calls for individual log messages. This strategic reduction in LLM queries makes both models significantly more efficient and practical for real-world deployment in large-scale log analysis systems.

The unsupervised category encompasses log parsers that achieve optimal performance without the need for any manually labeled examples. This category of parsers more accurately captures the characteristics of real-world operational environments and demonstrates strong alignment with them. LUNAR \cite{lunar}, AdaParser \cite{adaparser}, and LEMUR \cite{lemur} fall into this category. The core methodology of LUNAR \cite{lunar} centers on leveraging the comparative analysis capability of LLMs. It does this by identifying and presenting the LLM with Log Contrastive Units (LCUs), which are groups of log messages that are similar enough to imply a common underlying structure but vary in specific segments, thereby highlighting the dynamic parameters. To efficiently find these LCUs within vast log data, LUNAR utilizes a hierarchical sharder that groups logs by length and top-k tokens to select suitable comparison sets within these groups. Extensive evaluation on large-scale datasets demonstrates that LUNAR achieves comparable accuracy and efficiency to state-of-the-art label-dependent methods like LILAC. The novelty of AdaParser \cite{adaparser} also lies in the use of the self-correction property of LLMs. This model compares the template generated by GPT-3 with a set of predefined rules and, if they are not satisfied, asks the model to correct the log template. The self-correction process iterates until the template either passes verification or a predefined number of iterations is reached. AdaParser exhibits remarkable adaptability to evolving logs, showing significantly less performance degradation than other LLM-based parsers when historical data is limited or absent.

An analysis of the evolution of the presented models reveals a clear trend: as we approach more recent developments, LLM-based log parsers have progressively advanced toward greater adaptability to real-world operational environments. Initially, these models eliminated the dependency on extensive training and substantial computational resources, subsequently removed the dependence on labeled data, and ultimately focused on maximizing efficiency.
Moreover, as shown in Table \ref{tab:parser_table}, the performance of the majority of LLM-based log parsers is assessed by Group Accuracy (GA) and Parsing Accuracy (PA).  On the one hand, GA measures the ratio of logs that are correctly grouped and belong to the same template, compared to the total number of logs, regardless of whether the individual constants and variables are correctly identified. On the other hand, PA is a more strict log-level metric. It measures the ratio of logs that all their tokens parsed correctly to the total number of log messages. However, these commonly used log parsing metrics may yield misleading evaluations because they are influenced more by the frequency of log messages than by the diversity of unique templates. Consequently, a parser might achieve a high accuracy score by effectively handling a large number of repetitive and insignificant messages, while demonstrating poor performance in accurately identifying a wide range of templates \cite{khan2022guidelines}.
Efficiency, or how long it takes to parse new logs in inference mode, is another metric that can be helpful in assessing how well log parsers perform. To improve efficiency, most LLM-based parsers utilize a parse tree to store templates, and when a new log message arrives, it is first traverse through this tree to find the matching template. This component is effective because the number of log templates is several times smaller than the number of log messages in real-world systems. For example, the Loghub-2.0 dataset contains over 50 million log messages, yet the total number of log templates is less than 3500 \cite{50million}. 
The parse tree significantly reduces the number of queries to the LLM, making the efficiency of these log parsers comparable to syntactic-based ones such as Drain.

{
\scriptsize
\begin{longtable}{p{1.25cm} p{1.5cm} p{1.75cm} p{1.25cm} p{1.25cm} p{1.5cm} p{1.5cm}}
\caption{Overview of LLM-based Log Parsing models} \label{tab:parser_table} \\
\hline
Ref. & Name & Training method & LLM & Datasets & Grouping Accuracy & Parsing Accuracy \\
\hline
\endfirsthead

\multicolumn{7}{c}{{\bfseries Table \thetable\ (continued)}} \\
\hline
Ref. & Name & Training method & LLM & Datasets & Grouping Accuracy & Parsing Accuracy \\
\hline
\endhead

\hline \multicolumn{7}{r}{{Continued on next page}} \\
\endfoot

\hline
\endlastfoot

        \cite{logppt} & LogPPT (2023) & Prompt Tuning (Supervised) & RoBERTa & LogPAI & 0.92 & 0.92  \\ \hline
         \cite{llmparser} & LLMParser (2024) & Fine-Tuning (Supervised) & Various models & Loghub-2.0 & 0.88 & 0.96  \\ \hline
         \cite{divlog} & DivLog (2024) & Prompt Engineering (Supervised) & GPT-3 & LogPAI & - & 0.98  \\ \hline
         \cite{lilac} & LILAC (2024) & Prompt Engineering (Supervised) & GPT-3.5 & Loghub-2.0 & 0.93 & 0.84  \\ \hline
         \cite{lunar} & LUNAR (2024) & Prompt Engineering (Unsupervised) & GPT-3.5 & Loghub-2.0 & 0.93 & 0.88  \\ \hline
         \cite{logparser-llm} & LogParser-LLM (2024) & Prompt Engineering (Supervised) & GPT-4 & LogPub & 0.91 & 0.90  \\ \hline
         \cite{adaparser} & AdaParser (2025) & Prompt Engineering (Unsupervised) & GPT-3.5 & Loghub-2.0 & 0.96 & 0.94  \\ \hline
         \cite{lemur} & LEMUR (2025) & Prompt Engineering (Unsupervised) & GPT-4 & Loghub & 0.99 & -  \\ \hline

\end{longtable}
}

\subsubsection{Anomaly Detection}
Ensuring the reliability of software systems is a critical task, especially for large-scale cloud environments providing 24/7 services globally, as even brief downtime or failures can result in significant financial losses, service interruptions, data problems, and even threats to human safety. Nevertheless, contemporary systems generate log data at an enormous scale, with individual service instances capable of producing tens of gigabytes of logs per day \cite{liu2023log, mi2013toward}.
Manually analyzing this vast, complex, and continuously evolving log data is impractical, labor-intensive, prone to errors and bias, and cannot provide the timely detection needed \cite{lanobert, adaparser}. Therefore, automated log-based anomaly detection has become essential. Its importance lies in timely and accurate identification of log messages that indicate suspicious system behaviors or potential problems, thereby maintaining system reliability and security with minimal human intervention. Practical anomaly detection methods must be accurate, lightweight, and adaptive to handle the scale, limited resources, and evolving nature of real-world log data.

Traditional log-based anomaly detection methods face several challenges, including their reliance on manual feature engineering, imbalanced training data, difficulty adapting to diverse and evolving log data frequently seen in systems due to software updates, performance degradation when training data is scarce, and lack of interpretability in their analysis results \cite{qiloggpt, hanloggpt, logprompt}. Specifically, many traditional deep learning methods struggle to effectively capture the semantic and temporal information embedded within natural language-based logs and depend on syntactic-based log parsers, which can lead to the loss of crucial information and reduced adaptability \cite{bertlog, logfit}. LLMs advance this field by leveraging their pre-training on vast datasets to gain a robust understanding of diverse patterns and contextual information, helping to mitigate the issue of data insufficiency in evolving environments. They can comprehend the semantic and temporal meaning of logs and offer capabilities such as adaptability to evolving or unstable logs, enabling the development of parser-free methods, and development of expert systems. Furthermore, LLMs can provide explanations and justifications for detected anomalies, significantly enhancing interpretability.

The LLM-based models proposed in this field can be divided into two categories: Training-based methods and Prompt Engineering-based methods. One approach in training-based methods involves training an LLM only on normal data. Reconstruction-based models can predict the normal behavior of the system well and detect anomalies using the difference between the predicted log and the log generated by the system. In this regard, most models \cite{logbert, hanloggpt, lanobert} map log sentences to template IDs referred as log key. LogBERT \cite{logbert} is a self-supervised framework that is trained on normal data based on the masked log key prediction (MLKP) and Volume of Hypersphere Minimization (VHM) tasks and, at inference time, detects anomalies by comparing the actual log key of randomly masked tokens with top k predictions. Due to differing correlations among log keys between anomalous and normal sequences, LogBERT struggles to accurately predict masked log keys within anomalous sequences, enabling it to effectively distinguish between normal and anomalous log sequences after training. The value of the hyperparameter K could have a large impact on the performance of these kind of models and determines the strictness. A smaller K usually leads to high recall and low precision, and a larger K results in high precision and low recall. LogGPT \cite{hanloggpt} addresses the sensitivity and randomness associated with mask ratios and the existing gap between language modeling objectives and anomaly detection inference in previous model.  
LogGPT is initially pre-trained to predict the next log entry based on the preceding sequence. Subsequently, it is fine-tuned using reinforcement learning, employing a reward mechanism that evaluates whether the observed log entry falls within the Top-K predicted entries generated by the log language model. Moreover, LanoBERT \cite{lanobert} eliminates the dependency on log parsing and instead uses regular expressions. In this model, during inference, all log keys are masked once in the log sequence. For a given log sequence, after calculating a predictive probability for each masked log key, the model aggregates these values and the abnormal score is then obtained based on the average of the top-k predictive probability values. A low abnormal score indicates uncertainty of the model and is therefore considered more likely to be abnormal.

The primary distinction of LogFiT \cite{logfit} compared to prior models lies in its use of raw log sentences rather than mapping logs to template IDs, coupled with its exploitation of the pre-trained knowledge embedded within LLMs. Approaches relying on log keys exhibit semantic limitations, as they fail to capture the nuanced meanings within log sentences. In addition, they struggle to accommodate variability in log content and alterations in the lexical structure of logs over time. Unseen or slightly modified log sentences often fail to align with existing templates, which can lead to failures or degraded performance \cite{le2021log}. LogFiT incorporates a heuristic to choose between RoBERTa and Longformer based on log sequence lengths and during the inference, it identifies anomalies based on the model's top-k token prediction accuracy for masked tokens: if accuracy falls below a certain threshold, the log is flagged as anomalous. Experimental results indicate that LogFiT attains higher F1 scores than baseline models such as DeepLog \cite{du2017deeplog} and LogBERT across the HDFS, BGL, and Thunderbird datasets. Importantly, LogFiT retains its performance despite variability in log data content, a condition under which traditional parser-dependent models typically suffer substantial declines, underscoring its robustness and adaptability to evolving log structures.
Reconstruction-based models have the benefit of not requiring any anomaly data, but they also have the drawback of requiring constant training on normal data to become used to new emerging patterns.

Binary-based classification models is another approach among training based methods. These models consist of two part: First part involves utilizing a LLM for extracting semantic vectors from log messages and second part involves a deep learning model for binary classification which could be MLP \cite{bertlog}, custom Siamese network \cite{sarlog}, or another Language model \cite{guan2025logllm}. SaRLog employs a dual-strategy framework to enhance log anomaly detection, focusing on generating semantically rich representations and minimizing dependence on large labeled datasets. It first incorporates SecureBERT, a domain-specific fine-tuned BERT model, to capture context-aware semantics of log messages, thereby overcoming the limitations of general purpose LLMs in handling specialized system log terminology. Building upon this, SaRLog integrates a Siamese network with contrastive loss to improve detection of rare and subtle anomalies. This design minimizes embedding distances for similar (normal) log pairs and maximizes separation for dissimilar (anomalous) pairs, enabling effective few-shot learning.
Moreover, a recent study \cite{guan2025logllm} introduced LogLLM, which integrates BERT to extract semantic embeddings from log messages and employs Llama, to classify log sequences. LogLLM is designed to mitigate out-of-memory issues prevalent in LLM-based models by employing BERT to summarize individual log messages \cite{working-llm-limiti}. 
Although Binary-based classification models demonstrates strong performance, their large number of parameters results in substantial computational overhead and prolonged processing time, rendering it impractical for operational deployment. Additionally, challenges such as training data imbalance and limited limited availability of anomaly samples have led to a lack of interest among researchers.

Several recent studies \cite{105_loglm, 100_luk, 187-llmlade} have shown that training a language model on multiple related tasks, rather than focusing solely on a single task, can enhance its performance on that specific task. In this context, LogLM \cite{105_loglm} adopts an instruction-based training paradigm, converting multiple log analysis tasks into instruction–response pairs to train a unified model. This study also highlights the successful real-world deployment of LogLM as a central log analysis component within Huawei's O\&M platform, demonstrating its robust log analysis and adaptive instruction-following abilities by processing over 30,000 queries in six months. Similarly, LUK \cite{100_luk} serves as a knowledge enhancement framework that first extracts expert knowledge from LLMs, incorporates this knowledge into the log pre-training phase of a smaller model, and subsequently fine-tunes the enriched model for downstream log analysis tasks. The experimental findings of the study demonstrate that LUK achieves state-of-the-art performance across various log analysis tasks, exhibiting robustness and strong generalization capabilities in low-resource and unstable log data environments. Additionally, LUK offers substantially improved inference efficiency compared to direct utilization of LLMs, highlighting its practicality for real-world applications.

Prompt engineering-based methods leverage the language understanding capabilities of LLMs to identify anomalous patterns in log data and involves crafting specific prompts that include task descriptions, desired output formats, and sometimes examples, enabling LLMs to perform the detection without requiring extensive in-domain training data. LogGPT \cite{qiloggpt} and LogPrompt \cite{logprompt} are both exploring capacity of GPT-3.5 on log analyzing tasks by examining different prompt engineering techniques like self-prompt, Chain-of-thought, and in-context learning. Although these models generally underperform compared to training-based approaches, they offer the advantages of not requiring training data and exhibiting strong generalizability. They also demonstrates acceptable interpretability, providing reports and preventive suggestions for detected anomalies, which can aid administrators in troubleshooting. 
Audit-LLM \cite{auditllm} is a multi-agent framework designed for log-based insider threat detection (ITD). The framework comprises three collaborative agents working together to tackle the complexities of ITD. Furthermore, to address faithfulness hallucination \cite{hallucinationsurvey} and increase accuracy, Audit-LLM incorporates a pair-wise Evidence-based Multi-agent Debate (EMAD) mechanism, where two independent Executors iteratively refine their conclusions through reasoning exchange to reach a more reliable consensus. Experimental results indicate that Audit-LLM outperforms existing baselines in log-based insider threat detection across the CERT r4.2, CERT r5.2, and PicoDomain datasets, achieving accuracies of 0.961, 0.959, and 0.931, respectively. These results reflect substantial improvements, including accuracy gains of up to 21.8\% over DeepLog and 3.5\% over LogGPT on CERT r4.2.

RAGLog \cite{raglog} and RAPID \cite{rapid} are retrieval-based models that involves finding known normal log sequences corresponding to new test logs and judging them based on the comparison between them. These models aim to address the limitations of other LLM-based methods, including constraints related to token capacity, memory retention, hallucination, and real-time applicability. 
RAPID \cite{rapid} reframes anomaly detection by treating logs as natural language and comparing test logs to a knowledge base of known normal sequences. The innovation of this model lies in its focus on token-level information, enabling the detection of subtle anomalies that are often overlooked by sequence-level methods. In this regard, anomaly scores are computed by comparing each test log to its nearest neighbor in the embedding space using a token-level distance metric, enhancing sensitivity to nuanced deviations. To ensure real-time applicability and reduce computational overhead, RAPID employs a core set technique for efficient comparison. It achieves this by first quickly identifying a small subset of the most relevant normal log sequences using overall log summaries (CLS tokens), and then performing the more detailed, intensive comparisons only within this reduced core set. Experimental results demonstrate its competitive performance with existing models, achieving an average F1-score of 97.38\% on three representative benchmarks without any log-specific training. The method’s capability to function without prior training, combined with its efficiency and robustness in handling unseen logs, makes it particularly well-suited for real-time anomaly detection in dynamic environments.

Despite the widespread adoption of LLMs among researchers in this field, several challenges needs to be addressed before real world deployment. This challenges involves context length limitations that restrict processing of extensive log files, contradictory outputs, and high computational costs that impede real-time deployment. Additionally, LLMs often struggle with domain-specific terminology found in system logs.
Future research aims to mitigate these limitations by reducing hallucinations, enhancing result consistency, and advancing real-time response capabilities through the development of lightweight models capable of efficiently handling large-scale log analysis \cite{raglog}.
Table \ref{tab:anomaly_table} presents log anomaly detection models that incorporate LLMs in their architecture. Drain is the main parser chosen by almost all the models that have a log parser in their structures.

{
\scriptsize
\begin{longtable}{p{1cm} p{1cm} p{1.5cm} p{1cm} p{1cm} p{1cm} p{1.5cm} p{1.5cm}}
\caption{Overview of LLM-based Log anomaly detection models} \label{tab:anomaly_table} \\
\hline
Ref. & Name & Approach & Training data & LLM & Parser? & Datasets & F1 \\
\hline
\endfirsthead

\multicolumn{8}{c}{{\bfseries Table \thetable\ (continued)}} \\
\hline
Ref. & Name & Approach & Training data & LLM & Parser? & Datasets & F1 \\
\hline
\endhead

\hline \multicolumn{8}{r}{{Continued on next page}} \\
\endfoot

\hline
\endlastfoot

        \cite{logbert} & LogBERT (2021) & Pre-training from scratch & Normal & BERT & \cmark & HDFS, BGL, and Thunderbird & 0.823, 0.908, and 0.966 \\ \hline
        \cite{bertlog} & BERT-Log (2022) & Pre-training from scratch, and Fine-tuning & Normal and abnormal & BERT & \cmark & HDFS and BGL & 0.996 and 0.994  \\ \hline
        \cite{hanloggpt} & LogGPT (2023) & Pre-training from scratch, and Fine-tuning & Normal & GPT-2 & \cmark & HDFS, BGL, and Thunderbird & 0.901, 0.977, and 0.986  \\ \hline
        \cite{lanobert} & LAnoBERT (2023) & Pre-training from scratch & Normal & BERT & \xmark & HDFS, BGL, and Thunderbird & 0.964, 0.875, and 0.999  \\ \hline
        \cite{qiloggpt} & LogGPT (2023) & Prompt Engineering & - & GPT-3.5 & \cmark & BGL, and Spirit & 0.625 and 0.740  \\ \hline
        \cite{logfit} & LogFiT (2024) & Fine-tuning & Normal & RoBERTa and Longformer & \xmark & HDFS, BGL, and Thunderbird & 0.950, 0.899, and 0.939  \\ \hline
        \cite{sarlog} & SaRLog (2024) & Fine-tuning & Normal and abnormal & Secure-BERT & \xmark & BGL and Thunderbird & 0.988 and 0.999  \\ \hline
        \cite{raglog} & RAGLog (2024) & RAG & Normal & GPT-3.5 & \xmark & BGL and Thunderbird & 0.890  \\ \hline
        \cite{logprompt} & LogPrompt (2024) & Prompt Engineering & - & GPT-3.5 & \cmark & BGL, and Spirit & 0.384 and 0.450  \\ \hline
        \cite{rapid} & Rapid (2024) & RAG & Normal & BERT & \xmark & HDFS, BGL, and Thunderbird & 0.924, 0.999, and 0.997  \\ \hline
        \cite{auditllm} &  Audit-LLM (2024) & LLM-based Agents & - & GPT-3.5 & \xmark & CERT r4.2, CERT r5.2, and PicoDomain & (Acc) 0.961, 0.959, and 0.931  \\ \hline
        \cite{105_loglm} &  LogLM (2025) & Instruction tuning & Normal and abnormal & LLaMA-2 & \cmark & BGL and Spirit & 0.811 and 0.584  \\ \hline
        \cite{100_luk} &  LUK (2025) & Fine-tuning & Normal and abnormal & GPT-4o, LLama3, and BERT & \cmark & BGL & 99.93  \\ \hline
        \cite{guan2025logllm} & LogLLM (2025) & Fine-tuning & Normal and abnormal & BERT and Llama-3 & \xmark & HDFS, BGL, Liberty, and Thunderbird & 0.997, 0.916, 0.958, and 0.966  \\ \hline
        \cite{187-llmlade} & LLM-LADE (2025) & Fine-tuning & Normal and abnormal & LLaMA3 & \cmark & HDFS, BGL, and Thunderbird & 0.990, 0.989, and 0.997  \\ \hline

\end{longtable}
}

\subsection{Network Intrusion}
NIDS is a crucial component for network security, designed to monitor and analyze network traffic to detect malicious activities and security breaches. These systems aim to protect networks against various malicious attacks perpetrated by hackers and other agents, constantly evolving with the changing landscape of novel threats \cite{136_transformers}. In this context, LLM-based models enhance traditional approaches by introducing capabilities such as explainability, semantic understanding, adaptability to new threats, and improved generalizability \cite{89_generative}. Also, While traditional IDSs are often reactive, LLM-based systems can enable a more proactive security posture, identifying subtle signs of network intrusions before they fully manifest \cite{132_beyond, 131_bartpredict}. However, these models also come with some challenges such as limitations in understanding network-specific protocols \cite{94_towards, wang2024lightweight}. 

As illustrated in Table \ref{tab:network_intrusion}, the direct use of prompt engineering method in the field of NID is relatively rare. This limited adoption is primarily attributed to the substantial knowledge gap between the natural language understanding capabilities of LLMs and the low-level, structured, and often protocol-specific nature of network traffic data. One way to bridge this gap is translating raw network data into semantically meaningful representations that LLMs can effectively interpret \cite{73_securitybert, 135_iovbertids, li2025rlfe}. In \cite{73_securitybert}, they propose a novel technique called Privacy-Preserving Fixed-Length Encoding (PPFLE), which transform numerical and categorical data into a textual representation. PPFLE concatenates column names with their respective values and then hashes them, effectively creating a "new language" that BERT can understand, while also preserving data privacy by obscuring sensitive information. Experimental analysis using the Edge-IIoTset cybersecurity dataset demonstrated that SecurityBERT achieved an impressive 98.2\% overall accuracy in identifying fourteen distinct attack types, surpassing the performance of traditional ML and DL models like CNNs and RNNs. Furthermore, it boasts a compact model size of just 16.7MB and an inference time of less than 0.15 seconds on an average CPU, underscoring its practical suitability for real-time traffic analysis and embedded deployment in IoT environments. Moreover, IoV-BERT-IDS \cite{135_iovbertids} is another NID model and Its primary purpose is to detect intrusions in both in-vehicle and extra-vehicle networks within the dynamic Internet of Vehicles environment. This framework includes a Semantic Extractor, which plays a key role by converting raw traffic data into meaningful representations. This is achieved by converting Controller Area Network (CAN) data into CAN byte sentences (CBS) and extra-vehicle network traffic data into traffic byte sentences (TBS). The model is subsequently pre-trained on two specific tasks to capture bidirectional contextual features, and later fine-tuned for intrusion detection in both in-vehicle and extra-vehicle networks using labeled data. Experimental results show that IoV-BERT-IDS delivers superior classification performance and demonstrates strong generalization across various vehicle models in IoV environments. The paper also emphasizes that future research should focus on developing more efficient and lightweight BERT models to address the computational constraints of vehicular networks.

Another study \cite{86_dollm} investigate the application of LLMs on understand non-language network data for detecting unknown malicious flows, specifically focusing on Carpet Bombing DDoS attacks. Carpet Bombing is a recent and growing threat characterized by low-rate, multi-vector attacks targeting numerous victim IPs across subnets, leading to access link congestion and service disruption. In their study, they leverage LLMs' contextual understanding to extract powerful flow representations, moving beyond traditional feature engineering to improve DDoS detection performance. 

There are some research endeavors actively focused on making LLM-based NIDS suitable for practical network operations by mitigating their inherent high computational requirements and inference latency, often through the development of lightweight architectures, prompt optimization, and their integration into hybrid or federated learning frameworks that concurrently enhance explainability and adaptability to evolving threats \cite{li2025rlfe, wang2024lightweight, rezaei2025fedllmguard}.
In a recent study \cite{li2025rlfe}, they propose RLFE-IDS framework to overcome the significant difficulties and computational demands associated with directly deploying and fine-tuning LLMs for intrusion detection, as well as to bridge the structural and semantic gap between network traffic data and text data. 
This framework comprises of RAG and a custom embedding model called FE-Net. FE-Net is specifically designed to transform network traffic data into feature vectors (1024-dimensional) that preserve crucial correlations and are suitable for storage in this knowledge database, enabling the retrieval of relevant information for the LLM. When new network data arrives, RLFE-IDS queries this database using RAG for similar information, which then serves as context for the LLM to classify threats. The paper indicates that while the fine-tuned LLM-IDS achieved a slightly higher optimal accuracy of 99.36\%, the proposed RLFE-IDS framework, which attained 98.02\% accuracy in few-shot tasks, is considered superior. This is attributed to RLFE-IDS's substantial advantages in resource efficiency, reduced deployment complexity, and enhanced adaptability through RAG and the FE-Net embedding model, particularly in resource-constrained environments.

Another common and emerging concept in NIDS, is Federated Learning (FL), which addresses critical limitations of traditional centralized machine learning methods in evolving network environments. FL is an avant-garde machine learning technique that enables distributed agents to collaboratively train a shared model without exposing sensitive local data, thereby mitigating issues like high latency, significant communication overheads, and privacy risks associated with centralized data aggregation. This approach enhances privacy, scalability, and fault tolerance in NIDS deployments, particularly in resource-constrained Internet of Things (IoT) environments \cite{belenguer2025review, beuran2025fedmse}. Recently, researchers have integrated FL with LLMs to further enhance anomaly detection, particularly addressing the limitation of existing FL-based systems that rely heavily on numerical features and thus miss semantic and contextual nuances of evolving threats \cite{adjewa2024efficient, rezaei2025fedllmguard}. An example is FedLLMGuard \cite{rezaei2025fedllmguard}, a novel framework designed for real-time, privacy-preserving, and adaptive anomaly detection in 5G networks by integrating FL with LLMs. This model deploys lightweight LLMs, such as Tiny-BERT, at the client level to analyze text-based traffic logs and generate semantic embeddings, and utilizes soft prompt tuning to reduce computational and communication overhead during the federated optimization loop. FedLLMGuard consistently outperforms state-of-the-art machine learning models even against adversarial attacks, demonstrating superior accuracy, lower false positive rates, and significantly reduced detection latency and mitigation times.

While previous studies effectively detect known attacks and stop ongoing threats, they frequently fall short in predicting multistage attacks and preventing catastrophic damages, highlighting the pressing requirement for innovative approaches that enhance proactive threat prediction capabilities. Proactive and predictive intrusion detection represents a critical evolution in cybersecurity, moving beyond the traditional reactive nature of NIDSs that primarily respond to observed anomalies or specific attack patterns. The core objective of this forward-looking approach is to anticipate and preemptively mitigate malicious activities before they can cause damage, thus significantly reducing the likelihood and impact of successful intrusions \cite{132_beyond, 131_bartpredict}. In this regard, a study \cite{132_beyond} designed a framework to anticipate and mitigate malicious activities before they can cause damage. The main components of this framework are: a fine-tuned GPT-2 model to predict subsequent network packets given current ones, a fine-tuned BERT model to evaluate the validity of the predicted packets, and finally a trained LSTM to classify network packets as either normal or malicious. In the proposed frameworks, BERT is adapted through fine-tuning to perform packet-pair classification. This task focuses on predicting whether a given packet directly follows another, making BERT's comprehensive contextual understanding essential for accurately identifying if a generated packet is indeed the proper "next" packet in the sequence. 
Similarly, another study \cite{131_bartpredict} adopts similar approach but replaces the GPT-2 and LSTM with a BART and BERT, respectively. This modification improves accuracy but also leads to higher computational demands.

LLMs also can offer the ability to generate detailed, human-readable explanations for NID classifications, thereby enhancing security professionals' understanding and trust in the system's predictions. By integrating relevant background knowledge and summarizing complex network events, LLMs facilitate threat response and aid in deciphering intricate attack patterns, ultimately enhancing the interpretability and transparency of NIDS \cite{94_towards}.
A recent study \cite{128_idsagent} have introduced an LLM-powered AI agent for intrusion detection in IoT networks that generates predictions accompanied by detailed explanations. IDS-Agent employs a LLM to generate a sequence of actions and reason based on observed states, acting as a powerful operator and intermediary for intrusion detection. Its comprehensive framework features an abundant toolbox for tasks like data extraction and preprocessing, classification using various machine learning models, and knowledge and memory retrieval, ultimately providing explainable prediction results. Empirical evaluations demonstrate that IDS-Agent achieves F1 scores of 0.975\% on ACI-IoT and 0.750\% on CIC-IoT2023, outperforming state-of-the-art ML-based IDSs. Furthermore, it exhibits a recall of 0.61 in detecting zero-day attacks and enhances interpretability by leveraging inherent and external knowledge to aid decision-making, even in cases of contradictory classifier predictions.
Although LLM-based NID models demonstrate acceptable accuracy, many studies advocate for leveraging them as complementary agents within IDSs to primarily enhance explainability and support human interpretation, a strategic prioritization driven by their considerably higher computational costs and inference times relative to traditional machine learning models. \cite{79_explaining, 99_xgnid, 78_huntgpt}.

{
\scriptsize
\begin{longtable}{p{1.25cm} p{1.75cm} p{1.75cm} p{1cm} p{1.5cm} p{0.75cm} p{2.5cm}}
\caption{Overview of LLM-based Network Intrusion detection models} \label{tab:network_intrusion} \\
\hline
Ref. & Name & Method & LLM & Datasets & F-1 & Key Features \\
\hline
\endfirsthead

\multicolumn{7}{c}{{\bfseries Table \thetable\ (continued)}} \\
\hline
Ref. & Name & Method & LLM & Datasets & F1 & Key Features \\
\hline
\endhead

\hline \multicolumn{7}{r}{{Continued on next page}} \\
\endfoot

\hline
\endlastfoot

         \cite{73_securitybert} & SecurityBERT (2024) & Pre-training from scratch, and Fine-tuning & BERT & Edge-IIoTset & 0.840 & Leveraging a privacy-preserving encoding technique, and Generalizability is not verified  \\ \hline
         \cite{134_bertids} & BERTIDS (2024) & Fine-tuning & BERT & NSL-KDD  & 98.090 & Poor generalization   \\ \hline
         \cite{132_beyond} & - (2024) & Fine-tuning & GPT-2 and BERT & CICIoT2023 & 0.930 & Predictive intrusion detection, and efficiency is not varified  \\ \hline
         \cite{86_dollm} & DoLLM (2024) &  Representations learning & Llama2 & CIC-DDoS2019 & 0.938 & Carpet Bombing DDoS Detection model, Significant computational overhead and inference latency  \\ \hline
         \cite{135_iovbertids} & IOV-BERT-IDS (2024) &  Pre-training from scratch, and Fine-tuning & BERT & CICIDS, BoT-IoT, Car-Hacking,and IVNIDS & 1.000, 1.000, 0.999, and 0.998 & Vehicular network intrusion detection model, and not efficient enough due to limited vehicular resources  \\ \hline
         \cite{128_idsagent} & IDS-Agent (2024) & LLM-based agent & GPT-4o & ACI-IoT and CICIoT2023 & 0.970 and 0.750 & Capable of explicit knowledge integration and explanation, and efficiency is not varified \\ \hline
         \cite{129_llminwireless} & - (2024) & Prompt Engineering & GPT-4 & DTL-IDS [15] & over 0.950 & Evaluates different in-context learning techniques  \\ \hline
         \cite{131_bartpredict} & BARTPredict (2025) & Fine-tuning & BART and BERT & CICIoT2023 & 0.982 & Predictive intrusion detection, and efficiency is not varified  \\ \hline
         \cite{li2025rlfe} & RLFE-IDS (2025) & RAG & Chat-GLM3 & CIC-IDS2017, CSE-CICIDS2018, CIC-DDoS2019 and CICIoT2023 & (Acc) 0.980 & Knowledge Retrieval, and Simplified Deployment  \\ \hline
         \cite{rezaei2025fedllmguard} & FedLLMGuard (2025) & Prompt-tuning & Tiny-BERT & TII-SSRC-23, CICDDoS2019, and NF-UNSW-NB15 & 0.994, 0.997, and 0.989 & FL-LLM Integration  \\ \hline

\end{longtable}
}

\subsection{Phishing Detection}
Phishing attacks represent a significant and persistent threat in the digital world, designed to deceive individuals into revealing sensitive and confidential information. These attacks frequently exploit trust and security by mimicking legitimate sources through social engineering and communication channels like email, websites, SMS, or online social networks \cite{do2022deep}. Phishing causes substantial financial loss and setbacks for individuals, businesses, and economies globally. It is considered the most common type of social engineering and is implicated in approximately 90\% of data breaches for organizations. The financial impact is considerable, as global consumer losses from scams in 2023 were estimated at \$1.026 trillion, with phishing scams representing the largest share \cite{phishlm, knowphish}.
Furthermore, Phishing attacks have shown a marked rise in both frequency and complexity, as evidenced by a 17.3\% increase in phishing emails recorded between September 15, 2024, and February 14, 2025. A significant proportion of these attacks, accounting for 82.6\%, incorporated artificial intelligence, representing a 53.5\% year-over-year increase in the adoption of AI-driven techniques within phishing campaigns. These findings underscore the necessity of adopting more advanced strategies to effectively counter such threats \cite{knowbe42025phishing}.

Phishing detection serves as an active defense against the constantly evolving tactics used by attackers.
The primary LLM-based phishing detection approaches can be categorized into email-based, URL-based, brand-based, and HTML-based methods. The email-based approach involves prediction based on the text of the message and views this task as text classification. Some papers have evaluated the performance of pre-trained knowledge of different language models in this context using the prompt engineering method \cite{2023devising, chataut2024can, patel2024evaluating}, some have fine-tuned a specific language model on these texts \cite{phishlm, chataut2024can}, and some have also presented more innovative models \cite{nahmias2024prompted}. As expected, within this approach, LLMs demonstrate great potential and can reach up to 97.46\% accuracy \cite{chataut2024can}. For instance, in a study \cite{nahmias2024prompted}, they propose a document vectorization technique called prompted contextual document vectors to detect spear-phishing attacks. This technique utilizes an ensemble of LLMs and a set of human-crafted questions designed to probe social engineering tactics and specific spear-phishing indicators. Next, each of them is prompted to reason step by step and quantify its answer to each question as a probability. Lastly, the probabilities generated by all LLMs for each question are subsequently concatenated to create an explainable numerical vector representing each email. These resulting prompted contextual vectors are then used as input for a downstream supervised machine learning model to perform the detection task, aiming to capture the underlying malicious intent. A significant finding is the robustness of these prompted contextual vectors to concept drift, as they capture the underlying malicious intent rather than surface-level stylometric features. The paper also highlights the interpretability of these vectors, which can provide clear explanations for detection decisions.

URL-based approach focuses on analyzing the text of the Uniform Resource Locator (URL) itself to determine if a website is malicious or legitimate. This method treats the URL as the main source of information, often processing it as sequential data and learning relationships among its characters or smaller sub-parts \cite{dephides}. LLM based methods can directly analyze the raw URL text, adopting a fine-tuning \cite{phishbert} or prompt engineering \cite{promptorft, trad2024large, desolda2025apollo} method. LLMs can leverage their contextual understanding to identify suspicious patterns like minor character differences or fake brand names in domain names. A recent study \cite{desolda2025apollo}, introduces APOLLO, a system designed to detect phishing emails and automatically generate user-friendly explanations that warn recipients. It addresses the challenge of manually crafting effective explanations, which is time-consuming and not scalable, by dynamically producing personalized warnings. APOLLO utilizes a URL enricher to gather external threat intelligence, incorporating details like VirusTotal \cite{virustotal} scans and domain geolocation, this enhances the LLM's factual knowledge base. Moreover, it adopts a LLM prompter to employ a two-step prompt chaining approach. The initial prompt is designed to classify the email, identifying persuasion principles and suspicious features present in the content. Following this, the second prompt refines these identified elements into a clear and structured explanation, comprising a feature description, a hazard explanation, and the potential consequences.

HTML-based phishing detection is another approach that examines a webpage’s HTML source code to identify features indicative of phishing activity. This method examines elements such as HTML tags, their structure, layout consistency, textual content, and other features within the code that attackers might use to embed malicious functionalities. These models often requiring preprocessing to simplify lengthy code and fit within model token limits. In this regard,  PhishLang \cite{phishlang} presents a lightweight and efficient anti-phishing framework designed for direct operation on the user's device. PhishLang's methodology involves training on both website source code and URLs to detect phishing attempts. MobileBERT, as its core language model, achieved the highest accuracy (0.96) among tested Small Language Models (SLMs) and boasted the fastest prediction times (median 0.39 seconds per prediction) and lowest memory usage (74MB), making it ideal for real-time deployment on resource-limited devices.

Brand-based phishing detection approach leverages the immutable nature of genuine brand domains to achieve high precision and provide interpretable detection outcomes. It mainly involves two steps: First, identify the brand that the webpage under investigation is trying to impersonate , and second, compare the official URL of that brand to the URL of the webpage. If the webpage presents itself as a particular brand but its domain does not match the legitimate domains of that brand, it is classified as a phishing site. This approach often require analysis of visual elements such as logos on screenshots and textual content within the HTML. Multi-modal language models (MLMs) are particularly well-suited for implementing this approach. Models adopting this approach seek to overcome the limitations of traditional methods, which often depend on static, manually curated knowledge bases or are constrained to a single input modality, such as images or text \cite{lee2024multimodal, knowphish, gepagent, chatphishdetector}. In this context, KnowPhish \cite{knowphish} aims to address two salient limitations inherent in existing brand-based phishing detectors: their reliance on limited, manually curated brand knowledge bases and their exclusive dependence on visual information. To overcome these shortcomings, KnowPhish introduces an automated knowledge collection pipeline to construct a large-scale, multimodal brand knowledge base (BKB). This process proactively identifies potential phishing targets by leveraging brand-industry relationships from Wikidata and enriching brand information with logos, aliases, and domain variants gathered from diverse sources. Furthermore, the study proposes the KnowPhish Detector (KPD), a multimodal phishing detection approach that augments traditional visual detection with an LLM-based component designed to identify textual brand intention from webpage HTML and URLs. This integrated strategy enables KPD to detect phishing webpages regardless of whether they display logos, by first generating an LLM summary of the webpage, then classifying it as a credential-requiring page (CRP), and finally extracting brand intention through either visual or textual analysis to determine if a phishing attempt is present.  Extensive evaluations on both a manually validated dataset and a real-world dataset demonstrated that KnowPhish significantly boosts the F1-score and recall of existing brand-based models and offers substantially faster runtime efficiency due to its offline BKB construction. 

LLMs represent a transformative advancement in phishing detection, moving beyond conventional methods by leveraging sophisticated language understanding and multimodal capabilities to analyze diverse inputs like URLs, HTML, and screenshots. Models such as GPT-4 have demonstrated exceptional accuracy and high confidence, and even in some instances surpassing human detection capabilities in identifying suspicious domains and social engineering tactics \cite{2023devising, lee2024multimodal}. Fine-tuned LLMs, like PhishBERT \cite{phishbert}, have also shown particular strength in phishing URL detection, exhibiting superior performance on imbalanced datasets and robustness against adversarial attacks. However, there are some challenges among presented models, such as token length limitations, which necessitate preprocessing of large raw inputs like email or HTML content, and the high computational load and API costs associated with multimodal LLMs, especially when processing visual data. Researchers are actively attempting to address these issues with ideas such as agentic approaches \cite{wang2024mixture, trad2024large}, optimized prompt engineering \cite{nahmias2024prompted, desolda2025apollo}, and the development of efficient open-source models \cite{chataut2024can}, with expectations that these limitations will be significantly mitigated in the near future.

{
\scriptsize
\begin{longtable}{p{1.5cm} p{1.5cm} p{1.25cm} p{1.5cm} p{1.75cm} p{2cm} p{1cm}}
\caption{Overview of LLM-based Phishing detection models} \label{tab:phishing_table1} \\
\hline
Ref. & Name & Approach & Training methods & LLMs & Datasets & F1 \\
\hline
\endfirsthead

\multicolumn{7}{c}{{\bfseries Table \thetable\ (continued)}} \\
\hline
Ref. & Name & Approach & Training methods & LLMs & Datasets & F1 \\
\hline
\endhead

\hline \multicolumn{7}{r}{{Continued on next page}} \\
\endfoot

\hline
\endlastfoot

         \cite{phishlm} & PHISHLM (2022) & Email-based & Fine-tuning and Priming & GPT-2 & NAZARIO, IWSPA-AP, FRAUD corpus, UNTROUBLED corpus & -  \\ \hline
         \cite{phishbert} & PhishBERT (2023) & URL-based & Pre-training from scratch, and Fine-tuning & BERT & Common Crawl, Open PageRank, and PhishTank & 0.920  \\ \hline
         \cite{2023devising} & - (2023) & Email-based & Prompt engineering & GPT-4, Claude-1, Bard, and ChatLLaMA & Self-built & -  \\ \hline
         \cite{chataut2024can} & CyberGPT (2024) & Email-based & Prompt engineering, and Fine-tuning & GPT-3.5 and GPT-4 & \cite{subhajournal_phishingemails} & 0.967  \\ \hline
         \cite{promptorft} & - (2024) & URL-based & Prompt engineering, and Fine-tuning & GPT-3.5,Claude 2, GPT-2, Bloom, Baby LLaMA, and DistilGPT-2 & \cite{hannousse2021web} & 0.973  \\ \hline
         \cite{patel2024evaluating} & - (2024) & Email-based & Prompt engineering & 15 different models & Fraudulent E-mail Corpus & -  \\ \hline
         \cite{lee2024multimodal} & - (2024) & Brand-based & Prompt engineering & GPT-4, GeminiPro 1.0, and Claude3 & Tranco list and OpenPhish & -  \\ \hline
         \cite{knowphish} & KnowPhish detector (2024) & Brand-based & Prompt engineering, and Fine-tuning & GPT-3.5 and XLM-RoBERTa & Tranco list, OpenPhish and APWG & 0.920 \\ \hline
         \cite{gepagent} & GEPAgent (2024) & Brand-based & Agent based LLM & GPT4 and GPT-4V & Tranco list and OpenPhish & 0.946  \\ \hline
         \cite{chatphishdetector} & ChatPhish-Detector (2024) & Brand-based & Prompt engineering & GPT-4V & Tranco list, OpenPhish, and PhishTank & 0.993  \\ \hline
         \cite{trad2024large} & - (2024) & URL-based + Screenshot-based & Prompt engineering & Gemini 1.5 and GPT-4o & \cite{chiew2018building} & 0.937  \\ \hline
         \cite{nahmias2024prompted} & - (2024) & Email-based & Prompt engineering & GPT-3.5, GPT- 4, and Gemini Pro & Custom & 0.910  \\ \hline
         \cite{phishlang} & PhishLang (2025) & URL-based and HTML-based & Fine-tuning & MobileBERT &  Alexa Rankings list,  PhishPedia, and OpenPhish & 0.940  \\ \hline
         \cite{desolda2025apollo} & APOLLO (2025) & Email based + URL-based & Prompt engineering & GPT-4o &  Nazario, NigerianFraud, and SpamAssassin & 0.999  \\ \hline
         
\end{longtable}
}

\subsection{Vulnerability Detection}
Vulnerability detection constitutes a fundamental process in cybersecurity, focusing on the identification of defects within software code that may be exploited by malicious actors. These unintentional defects can lead to severe consequences such as data breaches, system failures, financial losses, and unauthorized access. Therefore, there is a need for strong proactive defense mechanism to detect these flaws before attackers exploit them. Historically, vulnerability detection relied on traditional methods like manual code inspections, rule-based static analysis, and dynamic analysis (e.g., fuzz testing). However, these approaches often suffer from significant limitations, including low accuracy, high false positive rates, scalability issues with complex modern software, and a demanding need for extensive human expertise and time \cite{sheng2025llms, guo2024outside}. 

To address these limitations, LLMs, with their sophisticated abilities in code comprehension, generation, and summarization, as well as their capacity to capture complex sequential patterns, provide promising solutions that surpass traditional methods \cite{mahyari2024harnessing}. Although, several studies have highlighted the limitations of general-purpose LLMs in this domain. These models frequently process source code as plain text, lacking domain-specific knowledge essential for accurate vulnerability identification. Their performance remains limited, frequently approaching the performance of a random baseline or a naive classifier \cite{steenhoek2024comprehensive, cheshkov2023evaluation, fu2023chatgpt}. As a result, researchers in this field have adopted several methods to enhance their LLM-based models, including the use of domain-specific pre-trained models, fine-tuning on domain-specific datasets, and the application of novel prompt engineering techniques.

A common approach in automated software vulnerability detection involves processing source code as raw, sequential text inputs for LLMs. This methodology typically entails fine-tuning pre-trained LLMs on specialized, task-specific datasets to develop their proficiency in identifying vulnerable code patterns \cite{148_linevul, 18_vuldetect, 154_LLMAO, 151_secureqwen, 145_securefalcon}. Moreover, domain-specific pre-trained models, such as CodeBERT\cite{feng2020codebert}, CodeGen\cite{nijkamp2022codegen}, and CodeT5\cite{wang2021codet5}, are specifically developed by training on extensive corpora of programming languages, often combined with natural language, allowing them to capture both semantic and syntactic relationships within code. By leveraging these specialized models, researchers can develop frameworks that effectively capture complex code patterns, provide more fine-grained vulnerability localization, and handle large codebases, significantly enhancing the accuracy, integrity, and trustworthiness of software systems. LineVul\cite{148_linevul} is one of these models which addresses coarse-grained results limitation of prior models by utilizing the attention mechanism inherent in the CodeBERT architecture, directly pinpointing vulnerable lines. LLMAO\cite{154_LLMAO} also performs vulnerability detection at the line level. It uses the final hidden states for each newline token in code sample from CodeGen to train them in bidirectional adapter layers and determine vulnerable lines. Furthermore, A recent study introduced SecureFalcon \cite{145_securefalcon}, demonstrating promising outcomes in this task. SecureFalcon seeks to address the significant computational overhead and time inefficiency typically linked to formal verification techniques, as well as problems arising from training data quality and inherent biases. SecureFalcon mitigates the first challenge by employing a lightweight LLM architecture with only 121 million parameters, derived from the Falcon-40B model. This architecture enables near-instant inference times, specifically a Time to First Token (TTFT) of 0.3 seconds and a total classification time of 0.6 seconds for typical inputs. This efficiency dramatically reduces vulnerability detection times, making the approach practical for real-time applications like instant code completion frameworks. In addition, SecureFalcon addresses the second challenge by training on two comprehensive resources: the FormAI dataset, which consists of synthetically generated and formally verified code samples, and FalconVulnDB, an aggregated dataset compiled from diverse, publicly available, human-curated sources.

Another common approach in this domain involves extending beyond raw source code text by explicitly incorporating and leveraging the semantic, syntactic, and structural relationships inherent in the code. This approach designed to overcome the limitations of existing methods that struggle with the complexity and diversity of modern codebases \cite{141_safe, 137_enstack}. In this regard, The core innovation of SAFE \cite{141_safe} lies in its ability to enable LLMs to not only acquire fundamental knowledge from source code but also to adeptly utilize semantic and syntactic relationships present within the code. To achieve this, SAFE proposes a two-phase architecture that incorporates a distillation mechanism. In the first phase, two lightweight deep-learning models act as "teachers" to delicately learn these relationships: Teacher-A focuses on semantic relationships, while Teacher-B concentrates on syntactic relationships, specifically utilizing Abstract Syntax Trees (ASTs). In the second phase, a "Student-S" model, which is an LLM backbone, is trained to emulate the behaviors and distilled knowledge from these two specialized teacher models. This integrated approach allows the student model to not only leverage the LLM's inherent capacity for fundamental knowledge extraction but also benefit from the enriched semantic and syntactic understanding provided by the teacher models. The methodology of EnStack \cite{137_enstack} also involves leveraging the distinct strengths of multiple pre-trained LLMs: CodeBERT for semantic understanding, GraphCodeBERT for handling the graph-based structure of code, and UniXcoder for understanding of the syntactic and semantic structures of code snippets. These models are fine-tuned separately on the Draper VDISC dataset, and their predictions are then combined using meta-classifiers to effectively capture complex code patterns. Empirical evaluations demonstrated significant performance gains, with the ensemble achieving higher evaluation metrics over individual models. For instance, combining GraphCodeBERT and UniXcoder with SVM yielded an accuracy of 82.36\% and an F1-score of 82.28\%, highlighting the framework's enhanced robustness and reliability in detecting vulnerabilities.

Employing prompt engineering for vulnerability detection offers a cost-effective alternative to fine-tuning by harnessing the generalization capabilities of LLMs, enhancing interpretability, and mitigating the risk of catastrophic forgetting of pre-trained knowledge \cite{150_grace, 144_dlap, 152_smartguard, 143_anvil}. In this regard, Grace \cite{150_grace} integrates graph structural information and in-context learning to enhance LLMs' capability in software vulnerability detection. It improves LLM prompting by incorporating domain-specific and identity information, along with the generated graph structures and in-context learning examples. Experiments showed that incorporating graph structural information had a significant impact, leading to improvements in F1 scores by 14.82\%, 24.64\%, and 73.8\% across the FFmpeg+Qemu, Reveal, and Big-Vul datasets, respectively. Similarly, SmartGuard \cite{152_smartguard} methodology involves leveraging LLMs to detect smart contract vulnerabilities by first retrieving semantically similar code snippets and then generating corresponding CoT for these snippets to facilitate in-context learning. Smart contracts are small programs deployed on the blockchain that automatically enforce terms and conditions between two untrusted parties. These contracts manage funds or data on the blockchain using predefined business logic. Detecting vulnerabilities in smart contracts is essential, as flaws in their code logic can severely compromise blockchain security. Unlike conventional software, smart contracts are immutable once deployed, making it difficult to fix vulnerabilities after deployment. In their study, they also highlight the broad applicability of SmartGuard in domains such as finance and healthcare, and report the superiority of it against existing static analysis tools. 

In a recent study \cite{143_anvil}, several researchers propose a novel approach to detect vulnerabilities by reframing it as reconstruction-based anomaly detection problem. CodeLlama is pre-trained on large-scale, unlabeled code corpora with the objective of generating code consistent with its training distribution. The underlying assumption is that deviations from this learned "normal" distribution indicate potential anomalies. The ANVIL framework involves using CodeLlama to perform a masked code reconstruction task. For each line of code, ANVIL masks the original line and instructs the LLM to reconstruct it based on the surrounding context. The divergence between the LLM-generated code and the original ground truth is quantified by a hybrid anomaly score. Moreover, as ANVIL does not require training, its reconstruction-based detection strategy demonstrates strong generalization across different projects. 
Another study \cite{147_vulrag} conduct a empirical study and highlights the limitations of LLMs in distinguishing between vulnerable code and its semantically similar but benign patched counterparts. Subsequently, the study introduces a RAG-based framework designed to address this limitation.
The core concept of VulRAG lies in its capability to distill high-level, generalizable vulnerability knowledge from historical vulnerabilities and corresponding fixes, thereby enabling LLMs to more accurately interpret code behavior. This is accomplished through a multi-dimensional representation of vulnerability knowledge that encompasses functional semantics, root causes of vulnerabilities, and fixing solutions. This advanced knowledge representation allows VulRAG to retrieve semantically relevant vulnerabilities more effectively during the retrieval phase and to prompt LLMs more efficiently during inference, surpassing the performance of approaches that rely solely on plain code pair representations in code-level RAG frameworks. An ablation study in their paper confirmed the superiority of knowledge-level RAG over code-level RAG and fine-tuning baselines, with 16\%-27\% and 22\%-26\% increases in pair accuracy, respectively. Pair accuracy measures the proportion of vulnerable–patched code pairs in which both the vulnerable and patched code are correctly identified.

Several studies have also explored solutions against the lack of explainability in LLM-based models. A study \cite{149_vulnerability} uniquely employs a combination of SHAP (SHapley Additive exPlanations), LIME (Local Interpretable Model-agnostic Explanations), and heat map of attention weights to provide insights into the model's decision-making process. Another work \cite{146_llmvulexp} proposes LLMVulExp, an automated framework designed to specialize LLMs for the dual tasks of software vulnerability detection and explanation. LLMVulExp employs prompt-based methods to automatically generate vulnerability explanations and then, fine-tunes a LLM through instruction tuning combined with LoRA. The training objectives are divided into three distinct sub-tasks: classifying the presence of vulnerabilities, localizing the vulnerable code segments, and generating specific explanations for the identified vulnerabilities. During the evaluation phase, the proposed model’s performance is assessed on binary and multi-class vulnerability detection tasks derived from the SeVC dataset, as well as on a CWE type-based multi-label classification task obtained from the DiverseVul dataset. The study demonstrates that integrating vulnerability explanations into the fine-tuning process preserved detection capabilities, while the extraction of key code segments significantly improved detection accuracy.

The burgeoning volume of research on software vulnerability detection is primarily motivated by the escalating prevalence and inherent complexity of software vulnerabilities within contemporary, rapidly evolving systems \cite{guo2024outside}. While general-purpose LLMs often exhibit limitations in direct vulnerability detection \cite{fu2023chatgpt, steenhoek2024comprehensive}, their effectiveness is considerably enhanced through the strategic integration of domain-specific knowledge by employing advanced contextual understanding techniques \cite{150_grace, 137_enstack}, explainability mechanisms \cite{139_lprotector, 147_vulrag}, as well as specialized fine-tuning \cite{18_vuldetect, 145_securefalcon} and multi-agent systems \cite{mao2024multi, wei2025advanced}.
Moreover, this surge in LLM based software vulnerability detection research is driven by their impressive capabilities in code comprehension and generation, which effectively address the limitations of traditional detection methods. LLMs offer significant opportunities to automate analysis and reduce manual feature engineering by learning complex patterns from code, overcoming the low accuracy, high false positive rates, and poor generalization often exhibited by earlier static analysis tools \cite{sheng2025llms}. However, this domain is not without its challenges. A prominent issue is that most datasets are structured at the function level, hindering the precise detection of vulnerable codes down to the individual line, despite some research, such as LineVul \cite{148_linevul} and LLMAO \cite{154_LLMAO}, explicitly targeting line-level prediction. Furthermore, a common simplification is that many studies assume there is only one vulnerability in a code snippet, overlooking real-world complexities where multiple vulnerabilities may coexist within a single file or across an entire repository \cite{152_smartguard}. Addressing these fundamental challenges, particularly through the curation of high-quality, fine-grained, and diverse datasets, is crucial. Overcoming these limitations is essential for LLMs to transition from academic exploration to reliable, practical deployment in automated software vulnerability detection.

{
\scriptsize
\begin{longtable}{p{1.5cm} p{1.5cm} p{1.5cm} p{1.5cm} p{1.5cm} p{1.5cm} p{1.5cm}}
\caption{Overview of LLM-based Vulnerability detection models} \label{tab:vulnerability_det} \\
\hline
Ref. & Name & Method & LLM & Input Types & Datasets & F-1 \\
\hline
\endfirsthead

\multicolumn{7}{c}{{\bfseries Table \thetable\ (continued)}} \\
\hline
Ref. & Name & Method & LLM & Input Types & Datasets & F-1 \\
\hline
\endhead

\hline \multicolumn{7}{r}{{Continued on next page}} \\
\endfoot

\hline
\endlastfoot

    \cite{148_linevul} & Linevul (2022) & Fine-tuning & CodeBERT & Raw code & Big-Vul & 0.920 \\ \hline
    \cite{18_vuldetect} & VulDetect (2023) & Fine-tuning & GPT-2 & Raw code & SARD and SeVC & 0.924 and 0.829  \\ \hline
    \cite{154_LLMAO} & LLMAO (2024) & Fine-tuning & CodeGen & Raw code & Devign & -  \\ \hline
    \cite{150_grace} & GRACE (2024) & Prompt Engineering & CodeT5, and GPT-4 & Raw code, AST, PDG, CFG & Devign, Reveal, and Big-Vul & 0.651, 0.431, and 0.355  \\ \hline
    \cite{141_safe} & SAFE (2024) & Distillation &  RoBERTa and CodeT5 & Raw code, AST, DFG & ReVeal, D2A, and Devign & 0.667, 0.490, and 0.707 \\ \hline
    \cite{139_lprotector} & LProtector (2024) & RAG & GPT-4o & Raw code & Big-Vul & 0.335 \\ \hline
    \cite{149_vulnerability} & - (2024) & Fine-tuning & BERT & Raw code & DiverseVul & 0.879 \\ \hline
    \cite{137_enstack} & EnStack (2024) & Fine-tuning & CodeBERT, GraphCodeBERT, and UniXcoder & Raw code & Draper VDISC & 0.823 \\ \hline
    \cite{144_dlap} & DLAP (2025) & Prompt Engineering & LineVul, and GPT-3.5 & Raw code, AST, DFG/CFG & custom & 0.584 \\ \hline
    \cite{151_secureqwen} & SequreQwen (2025) & Fine-tuning & CodeQwen1.5 & Raw code, AST &  PythonVulnDB & 0.950 \\ \hline
    \cite{146_llmvulexp} & LLMVULEXP (2025) & Prompt Engineering and Fine-tuning & CodeLlama, and GPT-3.5 & Raw code & SeVC, and DiverseVul & 0.924  \\ \hline
    \cite{145_securefalcon} & SecureFalcon (2025) & Fine-tuning & Falcon & Raw code & FormAI and FalconVulnDB & 0.940, and 0.920 \\ \hline
    \cite{152_smartguard} & SmartGuard (2025) & Prompt Engineering & CodeBERT, and GPT-3.5  & Raw code & SolidiFI & 0.949 \\ \hline
    \cite{143_anvil} & ANVIL (2025) & Prompt Engineering & CodeLlama & Raw code & Magma, and PRIMEVUL & (ROC) 61.8\%   \\ \hline
    \cite{147_vulrag} & VulRAG (2025) & Prompt Engineering, and RAG & Various models & - & PairVul & (Pair Acc) 0.320   \\ \hline 

\end{longtable}
}

\subsection{Others}
There are also several studies that explore the integration of LLMs into tasks beyond those discussed earlier. Nevertheless, their number remains limited, and further research is required in these areas. MalBERT \cite{rahali2021malbert} and SeMalBERT \cite{liu2024semalbert} both leverage BERT for static analysis-based malware detection, aiming to overcome the limitations of traditional methods that struggle with imprecise representations and a lack of contextual understanding in identifying malicious software. 
Effective spam detection is also critical for cybersecurity and security teams as unsolicited messages frequently serve as entry points for cyberattacks. Beyond this, the widespread delivery of spam emails and SMS disrupts user productivity, wastes network resources, and introduces a myriad of security risks to individuals and organizations \cite{jamal2024improved, si2024evaluating}. ExplainableDetector \cite{uddin2025explainabledetector} is a fine-tuned transformer-based Language Model which focuses on accurately detecting SMS spam messages and provides transparency into the model's decision-making process. Moreover, Mad-LLM \cite{du2024mad} addresses the challenge of detecting multi-stage attacks. These complex attacks involve multiple steps to achieve an attacker's ultimate goal, but their detection is hindered by the sheer volume of alerts from diverse sources, their heterogeneous formats, and often weak correlations. MAD-LLM's methodology for alert-based multi-stage attack detection is systematically organized into Prompt Construction, Alert Aggregation, and Alert Correlation. The process begins with the careful design of prompts incorporating elements like task objectives, standards, and MITRE ATT\&CK knowledge to effectively guide the LLM, followed by the aggregation of similar alerts to reduce redundancy and identify the attack type and stage. Finally, these similar alerts are logically correlated using domain-specific knowledge to uncover causal relationships and construct multi-stage attack chains. Experimental evaluations using the AIT Alert dataset \cite{landauer2024introducing} demonstrated MAD-LLM's superior performance, achieving an 87.49\% F1 score, and a 100\% attack stage detection rate, even for attack stages with minimal and infrequent alerts. Several researchers have also investigate the application of LLMs in Web content classification \cite{voros2023web}. This task involves categorizing web resources, such as URLs, into predefined content types to manage access, enforce policies, and ensure network security and regulatory compliance within organizations. It ultimately fosters a secure and professional work environment, enhancing operational efficiency and protecting the organization's reputation.

{
\scriptsize
\begin{longtable}{p{1.5cm} p{2.5cm} p{2cm} p{1.5cm} p{2cm}}
\caption{Overview of LLM-based models within Different Detection tasks} \label{tab:others_detect} \\
\hline
Ref. & Name & Method & LLM & Downstream Tasks \\
\hline
\endfirsthead

\multicolumn{5}{c}{{\bfseries Table \thetable\ (continued)}} \\
\hline
Ref. & Name & Method & LLM & Downstream Tasks \\
\hline
\endhead

\hline \multicolumn{5}{r}{{Continued on next page}} \\
\endfoot

\hline
\endlastfoot

         \cite{rahali2021malbert} & MalBERT (2021) & Fine-tuning & BERT & Malware Detection \\ \hline
         \cite{seyyar2022attack} & - (2022) & Representation learning & BERT & Anomalous HTTP requests Detection \\ \hline
         \cite{voros2023web} & - (2023) & Knowledge Distillation & T5, GPT-3, BERTiny, and eXpose & Web content classification \\ \hline
         \cite{liu2024semalbert} & SeMalBERT (2024) & Representation learning & BERT & Malware Detection \\ \hline
         \cite{jamal2024improved} & IPSDM (2024) & Fine-tuning & DistilBERT and RoBERTa & Spam Detection \\ \hline
         \cite{du2024mad} & MAD-LLM (2024) & Prompt Engineering & GPT-4 & Multi-stage Attack Detection \\ \hline
         \cite{si2024evaluating} & - (2025) & Prompt Engineering & GPT-3.5 & Spam Detection \\ \hline
         \cite{uddin2025explainabledetector} & Explainable-Detector (2025) & Fine-tuning & DistilBERT and RoBERTa & Spam Detection \\ \hline

\end{longtable}
}

\subsection{Answer of RQ3: }In the detection phase, among the LLM-based models proposed for various tasks, log parsing appears to be the only task sufficiently mature for integration into operational environments. Following this, log anomaly detection and vulnerability detection have been extensively studied and achieved satisfactory performance; however, many of these models require substantial computational resources, which may not be feasible for all organizations. In the areas of NID and Phishing Detection, models with acceptable performance on specific datasets have been developed, but additional research and evaluation on larger datasets are necessary. Finally, malware detection and spam detection represent tasks with limited existing work, although their development is expected to grow in the near future.

\section{RQ4: How are LLMs adapted for the analysis phase of SOCs?}\label{sec:rq4}

The analysis phase of SOCs faces growing challenges due to the increasing volume and complexity of cyber threats, which often overwhelm human analysts and contribute to alert fatigue. LLMs are increasingly applied in SOCs to improve tasks such as vulnerability assessment \cite{aghaei2023automated}, root cause analysis \cite{159_automated}, and rule generation \cite{39_llmcloudhunter}. Moreover, their capabilities help reduce analyst workload, accelerate decision-making, and improve the overall accuracy and efficiency of SOC operations. In the following sections, we first introduce LLMs specialized for cybersecurity and then examine their applications in CTI extraction, knowledge base mapping to support threat understanding, answering analysts’ queries, log analysis, and risk assessment.

\subsection{Domain Specific Language Models}
Domain-specific language models (LMs) are specialized language models designed to effectively understand and process text within a particular domain, such as cybersecurity. They are pivotal because general-purpose LLMs like BERT, though powerful for broad applications, are often ineffective for domain-specific text due to significant differences in terminology, writing styles, and word distributions \cite{2_cybert, 13_ctibert}. For example, the meaning of the word virus in general models might be a microscopic organisms that can infect hosts, like humans, while in fact it is a type of malware in the cybersecurity \cite{16_cysecbert}. Moreover, incorporating domain-specific language models into the architecture of analytical models in this field often leads to improved performance compared to their general-purpose counterparts \cite{59_ttphunter, 6_ttppredictor}.

Table \ref{tab:languagemodels} lists domain specific language models which proposed in the field of cybersecurity. These models can be categorized based on their training methods: Continual Pre-training \cite{2_cybert, 5_securebert, 16_cysecbert} and Pre-training from scratch \cite{13_ctibert}. Continual pre-training method typically requires significantly less domain-specific data and time compared to pre-training from scratch. However, because the volume of domain-specific data is relatively small in comparison to the data used in general pre-training, the semantic representations of many common words remain largely unchanged from those in general-purpose models. 
In SecureBERT \cite{5_securebert}, they attempts to address this challenge by introducing a small noise  to the initial pre-trained weights of mutual tokens from a general language model and it is capable of outperforming RoBERTa-large, despite the latter having nearly three times as many parameters. 

{
\scriptsize
\begin{longtable}{p{1.5cm} p{1.5cm} p{1.5cm} p{1.5cm} p{2.5cm} p{2.5cm}}
\caption{Overview of Cybersecurity Language Models} \label{tab:languagemodels} \\
\hline
Ref. & Name & Method & LLM & Downstream Tasks & Sources \\
\hline
\endfirsthead

\multicolumn{6}{c}{{\bfseries Table \thetable\ (continued)}} \\
\hline
Ref. & Name & Method & LLM & Downstream Tasks & Sources \\
\hline
\endhead

\hline \multicolumn{6}{r}{{Continued on next page}} \\
\endfoot

\hline
\endlastfoot

         \cite{2_cybert} & CyBERT (2021) & Continual Pre-training & BERT & Named Entity Recognition, Threat Classification, and Cybersecurity Knowledge Graph & Krebs on Security, CVEs, and APTNotes \\ \hline
         \cite{5_securebert} & SecureBERT (2022) & Continual Pre-training & RoBERTa & Sentiment Analysis and Named Entity Recognition & Websites, Security Reports, Whitepapers, Books, Articles, Surveys, and Videos (YouTube)  \\ \hline
         \cite{13_ctibert} & CTI-BERT (2023) & Pr-training from scratch & BERT & ATT\&CK Technique Classification, IoT App Description Classification, Malware Sentence Detection, Malware Attribute Classification, Coarse-grained and Fine-grained Security Entities, and Token Type Classification & Attack Description (MITRE ATT\&CK), Security Textbook, Academic Paper, Security Wiki, Threat Report, Vulnerability Description (CVEs and CWEs) \\ \hline
         \cite{16_cysecbert} & CySecBERT (2024) & Continual Pre-training & BERT & Clustering, Word Similarity, Named Entity Recognition, Relevance Classification, and Specialised CTI Classification  & Blogs, arXiv, NVD, and Twitter  \\ \hline

\end{longtable}
}

There are more pre-trained language models in other domains which could be applicable in SOC, either. For log data, models like Biglog \cite{tao2023biglog} provide a unified vectorized representation for log analyzing tasks, offering significant benefits in few-shot learning and domain adaptability. They have been successfully deployed in industrial network management systems for customized log analysis, including classifying anomalies and recognizing key attributes. KnowLog \cite{ma2024knowlog} further enhances log understanding by integrating domain knowledge (like abbreviations and natural language descriptions) and employing contrastive learning to achieve universal representations, proving effective in various log understanding tasks, low-resource scenarios, and cross-vendor transfer learning. BERTOps \cite{gupta2023learning} is an LLM for AIOps, and pre-trained on extensive public and proprietary log data, excelling in log format detection, golden signal classification, and fault category prediction, and is designed for plug-and-play integration into AIOps platforms. Additionally, PreLog \cite{le2024prelog}, a pre-trained sequence-to-sequence model for log analytics, uses log-specific pre-training objectives to capture semantic meaning and handle unstable log data. It also can be uniformly applied to various downstream tasks like log parsing, anomaly detection, fault localization, and failure identification through a prompt tuning paradigm.

Regarding code, CodeBERT \cite{feng2020codebert} is a bimodal pre-trained model for programming and natural languages, supporting applications like natural language code search and code documentation generation, essential for comprehending vulnerabilities in software. CODEGEN \cite{nijkamp2022codegen} assists in program synthesis and multi-turn code generation, which could be leveraged for automating exploit implementation and penetration testing, though it also carries the risk of generating vulnerable code. Lastly, CodeT5 \cite{wang2021codet5}, an encoder-decoder model, enhances code understanding and generation by leveraging developer-assigned identifiers, contributing to tasks like defect detection and clone detection, which are crucial for identifying software vulnerabilities.

For network traffic, ET-BERT \cite{lin2022bert} pre-trains deep contextualized datagram-level representations, achieving state-of-the-art performance in encrypted traffic classification tasks such as identifying encrypted malware and managing traffic on VPN, Tor, and TLS 1.3, which is vital for network security and management even with invisible content. These specialized LLMs offer robust capabilities for enhancing automation, threat intelligence extraction, and overall operational efficiency in SOC environments.

\subsection{CTI Extraction}
\label{sec:extraction}

Cyber Threat Intelligence refers to evidence-based knowledge encompassing context, mechanisms, indicators, potential impacts, and actionable recommendations related to existing or emerging threats to assets. This information is intended to support informed decision-making in response to such threats \cite{gartnerThreatIntelligence}. It involves the systematic collection, analysis, and dissemination of data from Various open-source platforms, such as security company content platforms \cite{talos2025, unit42_2025}, security news \cite{hackernews, bleepingcomputer}, and influential personal security blogs \cite{krebsonsecurity, schneier}. CTI is crucial for providing actionable insights to help organizations proactively defend against attacks, enhance incident response, and identify and mitigate vulnerabilities. Key components of CTI include Indicators of Compromise (IoCs) and Tactics, Techniques, and Procedures (TTPs). IOCs are forensic artifacts like IP addresses, domain names, file hashes, and CVEs that indicate a potential compromise. Moreover, TTPs represent the strategies, methodologies, and attack patterns employed by threat actors \cite{43_shah2024ai}. 

As the volume of CTI on the web continues to grow, cybersecurity text mining is emerging as an indispensable task within SOCs. These security teams must effectively leverage the extensive body of CTI to gain insights into adversary behaviors, motives, and targets. A major challenge, however, is that much of this intelligence is contained within large collections of unstructured natural language reports. Manually analyzing these reports, is time-consuming, labor-intensive, and can be inefficient \cite{50_ttpdrill}. Traditionally, CTI has relied on extracting static IOCs from reports; nevertheless, their short lifespan, ease of evasion, and limited contextual information made them ineffective as a standalone approach \cite{55_vulcan, li2019reading}. Purpose of CTI extraction is to automates the process of converting this unstructured data into a structured format, allowing teams to move beyond relying solely on limited and short-lived IoCs. By extracting entities, relationships, and TTPs, security teams can construct structured representations like knowledge graphs or STIX bundles. This structured, actionable intelligence can then be readily integrated into automated tools and workflows for intrusion detection, threat hunting, informing security measures, guiding incident responses, prioritizing threats, and enabling a more proactive and effective defense against cyber threats \cite{51_extractor, 54_tricti}.

As shown in Table \ref{tab:cti_extraction}, TTP classification and Threat entity/relation extraction are the two main objectives of the proposed models in this field. Models with the aim of only extracting relevant TTPs, usually involves LLMs to understand context of sentences and then map them to relevant TTPs using deep learning models \cite{53_tim, 59_ttphunter, 58_ttpxhunter}. For instance, TTPHunter \cite{59_ttphunter} utilizes combination of RoBERTa and linear classifier to predict the corresponding TTP class for each sentence. It also employs thresholding technique to the classifier's confidence score to filter out irrelevant sentences that do not strongly describe any TTPs. TTPHunter limited to only identifying 50 most frequent TTPs due to data scarcity for others. This made the authors of it propose refined version, called TTPXHunter \cite{58_ttpxhunter} to address this limitation and improve its performance. TTPXhunter utilizes domain-specific language model, SecureBERT, for generating more accurate sentence embedding and augmenting the training dataset by incorporating data augmentation method. The proposed model also convert the extracted TTPs into the STIX machine-readable format, facilitating threat intelligence sharing and integration.

For models designed to extract threat entities and relationships in conjunction with TTPs, two primary approaches can be identified: Fine-tuning and Prompt engineering. Vulcan \cite{55_vulcan}, LADDER \cite{56_ladder}, and LLM-TIKG \cite{33_llmtikg} are among fine-tuning based models. Vulcan \cite{55_vulcan} utilizes combination of BERT and DL models to identify threat entities in a sentence and then, employ BERT to identify the relations among each possible pairs in a sentence. Vulcan stored all extracted knowledge in a graph database and provide a set of search APIs that security practitioners can use to query the database. These APIs serve as the foundation for building various applications such as evolution identification and threat profiling aimed at supporting threat analysis. 
Ladder \cite{56_ladder} fine-tunes two separate language models on each sentence to identify threat entities and the relationship between them. Since, identifying attack patterns may comprise larger text blocks that a single sentence, they also propose TTPClassifier. This classifier identifies relevant sentences describing attack actions, extracts these attack phrases using a sequence tagging model, and then maps them to standardized MITRE ATT\&CK technique IDs using a semantic similarity-based approach. In another study \cite{33_llmtikg}, they utilize the few-shot learning capability of a LLM with designed prompts for data annotation and augmentation and then a smaller language model fine-tuned using a LoRA-based instruction tuning approach on this generated dataset. The proposed model eliminates the need for labeled data and the significant manual effort required for annotation.
Furthermore, experimental results show that the model attained F1-scores of 0.859 for named entity recognition and 0.982 for TTP classification, respectively.
Ignoring relationships outside of a sentence and the high amount of processing required to examine each pair of threat entities in a sentence are some existing challenges among the models with Fine-tuning approach.

More recent models like aCTIon \cite{19_action}, AttacKG+ \cite{25_attackg+}, LLMCloudHunter \cite{39_llmcloudhunter}, and CTINEXUS \cite{42_ctinexus} leverage Prompt engineering method Because of its applicability on diverse datasets. The core methodology of aCTIon \cite{19_action} is structured around a two-step LLM querying procedure. The first step involves a preprocessing phase that employs iterative summarization to condense the input report, ensuring it fits within LLM context limits and filters out irrelevant information. The second step is an extraction and self-verification phase where the LLM identifies and classifies target entities and attack patterns. The framework incorporates distinct pipelines for general entity and relation extraction, as well as for specific attack pattern extraction, utilizing various prompt strategies.
LLMCloudHunter\cite{39_llmcloudhunter} presents a framework for automated extraction of detection rule candidates from cloud-based open-source cyber threat intelligence (OSCTI).
It distinguishes itself from prior models through several notable advancements. Foremost, its unique design incorporates the ability to process visual components, such as images, present in OSCTI data, a capability largely overlooked by previous research. This framework broadens the scope of OSCTI analysis beyond traditional text-centric methodologies. Furthermore, LLMCloudHunter places a specific emphasis on cloud security environments, contrasting with many existing methodologies that primarily focus on on-premise settings. This addresses a gap given the increasing importance of cloud-centric operations and their unique security challenges. Finally, the framework produces its outputs as Sigma rule candidates, a generic and open signature format widely recognized for its seamless integration into various SIEM systems. This ensures that the extracted threat intelligence translates into actionable detection rules for practical defense mechanisms. 
Morover, CTINEXUS \cite{42_ctinexus} introduces a novel framework for automated CTI knowledge extraction and high-quality cybersecurity knowledge graph (CSKG) construction. CTINEXUS features a sophisticated pipeline that includes automatic prompt construction, a hierarchical entity alignment technique to canonicalize and remove redundancy, and a long-distance relation prediction technique to complete CSKGs with missing links. This approach allows for effective learning from minimal examples without requiring model weight updates, significantly simplifying generalization to various ontologies. Extensive evaluations using 150 real-world CTI reports from multiple recognized platforms demonstrate the framework's effectiveness. CTINEXUS achieved impressive F1-scores of 87.65\% in cybersecurity triplet extraction, 89.94\% in coarse-grained entity grouping, 99.80\% in fine-grained entity merging, and 90.99\% in relation prediction. Furhtemore, their results demonstrate that CTINEXUS surpasses methods such as EXTRACTOR and LADDER in both quantitative and qualitative evaluations.

This field currently faces several challenges that need to be addressed prior to model development.
One of them is the dynamic nature of cyber threat intelligence frameworks, such as MITRE ATT\&CK which presents a challenge to current LLM-based analysis due to their continuous updates. This constant evolution renders traditional fine-tuning approaches less suitable, as they necessitate large, labor-intensive labeled datasets for retraining, leading to performance degradation and costly, perpetual retraining to combat concept drift. Compounding this, a critical limitation observed in existing LLM-based systems is their adherence to a one-to-one classifier model. This design restricts the mapping of a given sentence to only a single TTP, thereby struggling with the inherent ambiguity and complexity of real-world cyberattack descriptions, where a single event can correlate to multiple TTPs or distinct campaign stages. Consequently, these models may overlook critical insights or misclassify information, preventing a comprehensive understanding of multi-faceted threat behaviors. Addressing these limitations is crucial for the advancement of truly intelligent and adaptive CTI automation \cite{59_ttphunter, 58_ttpxhunter}.

{
\scriptsize
\begin{longtable}{p{1.5cm} p{1.75cm} p{2.5cm} p{1.5cm} p{1.75cm} p{1.5cm}}
\caption{Overview of LLM-based CTI extraction models} \label{tab:cti_extraction} \\
\hline
Ref. & Name & Method & LLM & Extracted CTI Types & Output \\
\hline
\endfirsthead

\multicolumn{6}{c}{{\bfseries Table \thetable\ (continued)}} \\
\hline
Ref. & Name & Method & LLM & Extracted CTI Types & Output \\
\hline
\endhead

\hline \multicolumn{6}{r}{{Continued on next page}} \\
\endfoot

\hline
\endlastfoot

    \cite{53_tim} & TIM (2022) & Representation Learning, and ML-based classification & Sentence-BERT & TTPs & STIX and Sigma Rules \\ \hline
    \cite{54_tricti} & TriCTI (2022) & Fine-tuning & CBERT and BERT & IOCs and their associated campaign stages & Labeled IOCs \\ \hline
    \cite{55_vulcan} & Vulcan (2022) & Fine-tuning & BERT & Threat entities and relationships + TTPs & Graph Database \\ \hline
    \cite{59_ttphunter} & TTPHunter (2023) & Fine-tuning & BERT and RoBERTa & TTPs (50 most frequently used) & Knowledge Graph \\ \hline
    \cite{56_ladder} & LADDER (2023) & Fine-tuning & BERT, RoBERTa and XLM-RoBERTa & Threat entities and relationships + TTPs & Knowledge Graph \\ \hline
    \cite{19_action} & aCTion (2023) & Prompt Engineering & GPT-3.5 & Threat entities and relationships + TTPs & Knowledge Graph \\ \hline
    \cite{25_attackg+} & AttackKG+ (2024) & Prompt Engineering & Several commercial LLMs & Threat entities and relationships + TTPs & Knowledge Graph  \\ \hline
    \cite{33_llmtikg} & LLM-TIKG (2024) & Prompt Engineering, and Instruction tuning & GPT-3.5 and Llama 2 & Topic + Threat entities and relationships + TTPs & Knowledge Graph  \\ \hline
    \cite{58_ttpxhunter} & TTPXHunter (2024) & Fine-Tuning & SecureBERT & TTPs & STIX  \\ \hline
    \cite{39_llmcloudhunter} & LLM CloudHunter (2025) & Prompt Engineering & GPT-4o & Threat entities and relationships + TTPs & Sigma Rules  \\ \hline
    \cite{42_ctinexus} & CTINEXUS (2025) & Prompt Engineering & GPT-4 & Threat entities and relationships + TTPs & Knowledge Graph  \\ \hline
         
\end{longtable}
}

Another task related to this domain is CTI generation, which involves producing comprehensive cybersecurity reports in natural language by utilizing formal or structured representations of threat intelligence as input. This process offers the advantages of being easily comprehensible to personnel without specialized expertise and facilitating straightforward dissemination across the organization. In this regard, AGIR \cite{21_agir} utilize template based module that parses the input STIX graph and fills in predefined report templates according to specific rules, generating a preliminary report and then, rewrite the text for improved fluency and overall quality using ChatGPT. Automating CTI generation can bring benefits such as Reduction of Manual Effort, time, and expert demands \cite{26_2024chatgpt}.

\subsection{Mapping}
\label{sec:mapping}

Automated mapping in cybersecurity involves linking diverse cybersecurity knowledge bases such as CVEs, CWEs, CAPEC, and the MITRE ATT\&CK framework \cite{14_bron}. This process is essential given the increasing complexity of modern computing frameworks and the rising number of reported cybersecurity vulnerabilities. Although each of the mentioned knowledge sources offers valuable information for SOC teams, they exhibit inherent limitations when considered independently \cite{31_hwrex}. For example, CVEs often provide little or no information on how to combat the vulnerability, necessitating linkage to frameworks like ATT\&CK which offer mitigation strategies \cite{17_cvet}. Mapping between this sources could provide a holistic view of the entire cyber landscape, encompassing threats, vulnerabilities, weaknesses, attack patterns, and affected systems and help cybersecurity defense teams to better prioritize alerts. Nevertheless, performing this task manually is inefficient since it requires extensive effort, considerable time, and substantial domain expertise. To address these challenges, several studies have explored automating this process using LLMs.

As demonstrated in table \ref{tab:mappings}, most mapping frameworks focus on establishing connections between CVEs and MITRE ATT\&CK techniques. Combining these two complementary resources is vital for enhancing strategic analysis, contextualizing security alerts, supporting product security evaluations, and enabling proactive threat hunting and risk prioritization. Moreover, the frequent reference to the MITRE ATT\&CK framework can also be explained by its broad integration into industry-standard SIEM tools. In this regard, SMET \cite{125_smet} and TTPpredictor \cite{6_ttppredictor} utilize SRL techniques to extract attack vectors and as a result, reducing noise by focusing on them and circumventing the reliance on pre-annotated datasets. Then, both models fine-tune a language model to generates semantically meaningful vector representations of these extracted attack vectors. Crimson \cite{12_crimson} is another framework which it's main idea revolve around Retrieval-Aware Training (RAT). RAT integrates real-time data retrieval directly into the LLM training process, enriching both the training and inference phases with current and relevant cybersecurity data. This mechanism ensures that the model is up-to-date with the latest threats and information. The evaluation results show that Crimson achieves accuracy comparable to GPT-4 while producing fewer hallucinations, which is noteworthy given that it contains nearly 250 times fewer parameters than GPT-4.
Furthermore, several researchers \cite{4_vwc-map} propose a two tier stage model for mapping between vulnerabilities and attack patterns. In first stage, It maps CVEs to CWEs through V2W-BERT \cite{1_v2wbert} model which consist of RoBERTa in a Siamese network architecture. In second stage, T5 Generates the CAPEC text description using the CWE text description as input, and ultimately only a similarity measure is required. Requirement for intensive computational resources, the limited size of the evaluation dataset, and reliance on manual qualitative assessment are some potential limitations of this model.

Besides CVEs, there are some other knowledge sources that mapping from could be beneficial. For instance, some researchers conducted a comparative analysis to address the significant challenge of cybersecurity practitioners face in understanding the complex and often ambiguous descriptions within the ATT\&CK framework \cite{20_advancingttpanalysis}. They have done this through interpreting cyberattack TTPs, particularly by mapping procedure descriptions to MITRE ATT\&CK tactics. Their results indicate that large-scale decoder-only LLMs with RAG perform better than small-scale encoder-only LLMs with fine-tuning. Moreover, linking NIDS Rules with MITRE ATT\&CK techniques is another important mapping task mentioned in a recent study \cite{40_nidswithattack}. Automation of this task aims to improve the efficiency of SOC alert analysis while reducing the operational burden on analysts. The propose methodology involves a comparative analysis between two approaches: one is the LLM-based approach that focuses on prompt engineering, and the other one is the ML-based approach, in which all steps, including selecting suitable models, writing code, and executing the tasks, are performed by LLM. Ultimately, the results highlight the strengths of each approach: the LLM-based method excels in flexibility and explainability, while the ML-based approach demonstrates higher accuracy. This outcome suggesting the potential of a hybrid LLM-ML approach to combine respective strengths.

{
\scriptsize
\begin{longtable}{p{1.5cm} p{1.5cm} p{2.5cm} p{1.5cm} p{1cm} p{1.5cm} p{1cm}}
\caption{Overview of LLM-based Mapping models} \label{tab:mappings} \\
\hline
Ref. & Name & Method & LLM & From & To & F1 \\
\hline
\endfirsthead

\multicolumn{7}{c}{{\bfseries Table \thetable\ (continued)}} \\
\hline
Ref. & Name & Method & LLM & From & To & F1 \\
\hline
\endhead

\hline \multicolumn{7}{r}{{Continued on next page}} \\
\endfoot

\hline
\endlastfoot

         \cite{17_cvet} & CVET (2021) & Self-Knowledge Distillation Fine-Tuning & RoBERTa & CVE & MITRE ATT\&CK Tactics & 0.762 \\ \hline
         \cite{1_v2wbert} & V2W-BERT (2021) & continual pre-training, and Fine-tuning & BERT & CVE & CWE & 0.972 \\ \hline
         \cite{4_vwc-map} & VWC-MAP (2022) & Fine-tuning & RoBERTa and T5 & CVE & CAPEC & - \\ \hline
         \cite{3_cve2attack} & CVE2 ATT\&ACK (2022) & Fine-tuning & SecBERT & CVE & MITRE ATT\&CK Techniques & 0.478 \\ \hline
         \cite{125_smet} & SMET (2023) & Adopting SRL technique, and Fine-tuning & Sentence BERT & CVE & MITRE ATT\&CK Techniques & - \\ \hline
         \cite{6_ttppredictor} & TTPpredictor (2023) & Adopting SRL technique, and Fine-tuning & SecureBERT & CVE & MITRE ATT\&CK Techniques & 0.980 \\ \hline
         \cite{12_crimson} & Crimson (2024) & Retrieval-Aware Training & Claude2, GPT-4, and LLaMA-2 & CVE & MITRE ATT\&CK Techniques & 0.729 \\ \hline
         \cite{44_cyberknowledge} & - (2024) & RAG & Llama 3 & CAPEC & MITRE ATT\&CK Techniques & - \\ \hline
         \cite{20_advancingttpanalysis} & - (2024) & RAG & GPT-3.5 & TTPs  & MITRE ATT\&CK Tactics & 0.950 \\ \hline
         \cite{40_nidswithattack} & - (2025) & Prompt Engineering  & GPT-4, Claude-3, and Gemini & NIDS Rules  & MITRE ATT\&CK Tactics and Techniques & - \\ \hline
         
\end{longtable}
}

\subsection{Question-Answering models}
SOC analysts face significant challenges in their workload due to the large volume and complexity of cyber threat information. This complexity increases significantly during cyber attacks, as they need to respond to threats in the shortest possible time and make the best decision to stay safe from the risks. Mistakes in these situations can cause significant damage to the organization and individuals. This has resulted in the individuals assigned to this role possessing a high level of expertise, while simultaneously having to devote a substantial portion of their time to labor-intensive tasks, such as sifting through large volumes of textual data to remain up to date with evolving threats. To address this challenges, Question-Answering (QA) models have been proposed.
QA models are computational systems that allow users to obtain answers to their questions expressed in natural language, serving as an advanced approach to information retrieval. For Small and Medium-sized Enterprises, QA models are crucial for providing intuitive and user-friendly access to complex cybersecurity information, compensating the limited technical expertise and personnel, and enabling rapid, informed decision-making around the clock \cite{franco2020secbot}.

Traditional QA models faced significant limitations primarily due to their struggle to fully comprehend the user's query context and content meaning. These systems often relied on predefined rules and patterns, which made them inflexible and unable to adapt to diverse user interactions or handle complicated and unclear questions effectively. With the advent of LLMs, the first idea that came to the minds of many researchers was to use these general purpose models as chatbots in the field of cybersecurity. However, this idea also failed to meet expectations due to problems such as tendency to hallucinate, outdated knowledge, and high sensitivity to engineering prompts \cite{23_failureanalysis, shafee2025evaluation}. 

As can be seen in Table \ref{tab:questionanswering}, RAG-based methods are the most popular approach among researchers in this field.
RAG models work by retrieving and integrating relevant, up-to-date context from external knowledge bases before generating a response, thereby combining the generative power of LLMs with robust information retrieval. This approach directly mitigates hallucination and provides access to current information, crucial for a rapidly evolving field like cybersecurity where factual accuracy and timeliness are paramount. Furthermore, RAG models can provide the source of their answers, enhancing transparency and allowing cybersecurity analysts to verify the information, which is vital for informed decision-making \cite{22_ragquestion, 38_chatnvd}. 

In this regard, a study \cite{48_attackqa} released a specialized Q\&A dataset, AttackQA, derived from the MITRE ATT\&CK knowledge base. AttackQA constructed using a hybrid methodology: 20\% manually via heuristics and 80\% automatically by a lightweight open-source LLM. They also fine-tuned a LLM as a quality controller to detect and reject low-quality Q\&A pairs in the dataset. This dataset is then used to fine-tune both embedding and generation models for RAG applications, aiming to assist SOC analysts. The evaluation results indicate the superiority of fine-tuned open-source models over proprietary models like GPT-4o in accuracy and speed. A prevalent architectural pattern among RAG models involves the simultaneous or sequential use of multiple knowledge bases to retrieve pertinent information for distinct use cases. This approach is exemplified by systems such as CyberGuardian \cite{7_cyberguardian}, which employs separate vector databases for internal organizational knowledge and a broader base dataset. In another study, LOCALINTEL \cite{34_localintel} leverages both publicly available global cyber threat intelligence repositories and private local organizational knowledge databases. In more detail, Upon receiving an input Prompt from a SOC analyst, an LLM agent generates search queries to retrieve relevant global cyber threat intelligence from publicly available repositories like CVE and NVD. Subsequently, the system generates local search queries by combining the initial prompt with the newly retrieved global knowledge. This query is then executed against a vectorized local knowledge database, which contains private organizational information and trusted threat reports. Finally, after retrieving pertinent local knowledge, LLM integrates both the retrieved global and local knowledge to produce a comprehensive contextualized Response Completion that is specifically tailored to the organization's unique operational context. Depending on the implementation, a two-database approach can offer several advantages, including enhanced efficiency and faster retrieval, improved accuracy through organization-specific contextualized responses, and better support for data segregation and privacy.

Fine-tuning on large-scale corpora of question–answer pairs represents another approach within QA models. In a work \cite{36_sevenllm}, they collect over ten thousand bilingual (English and Chinese) cybersecurity incident websites from leading vendors and news outlets, curating 6,706 English and 1,779 Chinese high-quality reports as foundational data. They also employ Select-Instruct to create the instruction corpus, wherein a raw corpus is combined with prompts to guide GPT-4 in selecting a task from the pool and subsequently generating structured query-response pairs for the selected task, accompanied by a reasoning process. SecKnowledge \cite{8_cyberpal} is another cybersecurity instruction dataset that involves a multi-phase generation process. The first generation step focuses on building high-quality instructions based on predefined schemas established through extensive domain expertise and in-depth analysis of diverse security datasets and then on second step, expands these instructions through hybrid synthetic content-based data generation process. CyberPal.AI achieved an average improvement of 9-10\% over baseline counterparts on public cybersecurity benchmarks, indicating its ability to generalize effectively to new tasks and its deep understanding of the cybersecurity domain.

Ultimately, a noteworthy observation emerging from the review of the examined models is the variation in their evaluation methodologies. Some studies employ LLMs as evaluators or "judges" \cite{22_ragquestion, 7_cyberguardian, 48_attackqa}, while some others adopt more novel evaluation strategies. For example, IntellBot \cite{41_intellbot} employs a two-stage evaluation approach to thoroughly assess its performance. In the first stage, it conduct a indirect evaluation which it generates five question based on the chatbot's answer and then, computes the BERT score between the original user query and each of these five generated questions to quantify their semantic alignment. The second stage involves calculating the cosine similarity between the chatbot-generated responses and the corresponding manual answers. Furthermore, as LLMs are increasingly applied to complex, domain-specific tasks like cybersecurity, the development of robust and diverse evaluation benchmarks has become crucial to accurately assess their capabilities and performance. CyberMetric \cite{cybermetric}, SecKnowledge-Eval \cite{8_cyberpal}, and SEVENLLM-Bench \cite{36_sevenllm} are some proposed cybersecurity benchmarks in this domain. These benchmarks play a vital role in advancing the field by providing standardized tools to measure LLMs' understanding, reasoning, and practical applicability in real-world cybersecurity scenarios, especially regarding cyber threat intelligence.

{
\scriptsize
\begin{longtable}{p{1.5cm} p{2cm} p{1.75cm} p{1.25cm} p{1.5cm} p{3cm}}
\caption{Overview of Question-Answering models. Generation method column describes the method of constructing questions that form the dataset for evaluating or training models.} \label{tab:questionanswering} \\
\hline
Ref. & Name & Method & LLM & Generation method & Key Features \\
\hline
\endfirsthead

\multicolumn{6}{c}{{\bfseries Table \thetable\ (continued)}} \\
\hline
Ref. & Name & Method & LLM & QA Generation method & Key Features \\
\hline
\endhead

\hline \multicolumn{6}{r}{{Continued on next page}} \\
\endfoot

\hline
\endlastfoot

        \cite{28_cybersentinel} & Cyber Sentinel (2023) & Prompt Engineering & GPT-4 & Human-curated & Designed to handle a number of security actions  \\ \hline
         \cite{22_ragquestion} & - (2024) & RAG & Mistral & Human-curated & Focused on cyber-Attack investigation and attribution \\ \hline
         \cite{36_sevenllm} & SEVENLLM (2024) & Select-Instruct, and Fine-Tuning & GPT-4, Llama 2, and Qwen-1.5 & LLM-generated & Creation of bilingual instruction corpus \\ \hline
         \cite{7_cyberguardian} & CyberGuardian (2024) & Fine-Tuning, and RAG & Llama 2 & Human-curated & Has a plugin architecture \\ \hline
         \cite{48_attackqa} & - (2024) & Fine-Tuning, and RAG & Llama 3 & Hybrid & Creation of a Q\&A dataset based off the MITRE ATT\&CK database \\ \hline
         \cite{41_intellbot} & IntelBot (2024) & Prompt Engineering, and RAG & BERT, and GPT-3.5 & Human-curated & Ensemble retriever \\ \hline
         \cite{34_localintel} & LOCALINTEL (2025) & LLM-based Agents, and Prompt Engineering + RAG & GPT-3.5 & Human-curated & Leveraging global and local knowledge  \\ \hline
         \cite{8_cyberpal} & CYBERPAL.AI (2025) & Fine-Tuning & Mistral, Llama-3, and Phi-3 & Hybrid & Comprehensive instruction dataset \\ \hline
         \cite{38_chatnvd} & ChatNVD (2025) & Prompt Engineering, and RAG & GPT-4o Mini, Gemini 1.5 Pro, and LLaMA 3 & Human-curated & Streamline vulnerability analysis for diverse users \\ \hline

\end{longtable}
}

\subsection{Log analyzing}
Log analysis plays a integral role in diagnosing incidents, allowing engineers to identify the underlying root causes of disruptions and help them to make the best decisions in shortest possible time \cite{ahmed2023recommending}. Given the sheer volume of these logs, manual inspection is impractical, even with basic search tools, leading to a fast-growing demand for automatic log analysis frameworks \cite{zhu2021unilog}. 

As shown in Table \ref{tab:loganalysing}, the Root Cause Analysis (RCA) task is the most popular topic among researchers in this field. This task aims to identify the underlying cause of a abnormal behavior. RCA is a critical but often laborious, error-prone, and challenging task for on-call engineers (OCEs), especially given the escalating scale and complexity of modern cloud systems and the huge volume of logs they generate \cite{161_automatic}. A shared characteristic among all models proposed for this task is their focus on addressing or mitigating the cumbersome effects of these challenges. Within the studies investigating the application of LLMs to this specific task, fine-tuning has not emerged as a widely adopted approach. This can be attributed to the particular prerequisites that the model must satisfy in order to attain optimal performance for this task. Firstly, it should be able to interact with the external environment. This is crucial because incident reports often lack the comprehensive diagnostic data needed for a conclusive RCA.
Secondly, the model should be capable of distill relevant insights from high-volume, multi-source data (logs, metrics, traces) into succinct, actionable outputs, thereby alleviating the significant manual effort traditionally required from on-call engineers. Thirdly, the model should effectively adapt to specific domains and evolving incident types without constant, costly training. Lastly, it should provide explanations and detailed diagnostic processes, rather than just a predicted root cause. Unlike fine-tuning, all the mentioned requirements are well-aligned with LLM-based agents which make it ideal choice for RCA models \cite{157_exploring, 159_automated, 164_rcagent, 166_flow}.

RCACopilot \cite{161_automatic} is an on-call system powered by LLMs for automating RCA of cloud incidents. RCACopilot is designed to match incoming incidents to corresponding incident handlers based on their alert types, aggregate critical runtime diagnostic information from multiple sources, predict the incident’s root cause category, and then provide an explanatory narrative. This model utilizes GPT-4 to summarize diagnostic information and retrieves top-k similar historical incidents from different categories as demonstrations. Then, use summarized diagnostic information in addition to demonstrations and COT prompting to choose the most likely incident with the same root cause and generate an explanation for its reasoning. Although this model has been deployed at Microsoft, it exhibits certain limitations, including its dependence on manually crafted predefined handlers and its focus on predicting general root cause categories rather than precise diagnoses. Several researchers developed another model \cite{157_exploring} to overcome these limitations. The proposed model also address the inability to dynamically collect additional diagnostic information by utilizing LLM-based agents, which can reason, plan, and interact with the external environments. Furthermore, the study presents a practical case study with an Azure Fundamental Team at Microsoft, showing how the ReAct agent, when given access to team-specific diagnostic tools and knowledge base articles, can autonomously assist engineers in finding incident root causes in a real-world scenario. 

RCAgent \cite{164_rcagent} is a framework developed to operate on internally deployed models, with the objective of enabling privacy-preserving applications in industrial settings. This framework utilizes two primary types of agents: Controller Agent and Expert Agents. Controller Agent is the main LLM agent that orchestrates the overall RCA process and operates based on a thought-action-observation loop. Expert Agents are LLMs employed as specialized tools by the controller agent to provide domain-specific knowledge and analytical capabilities. They extend the domain knowledge and reasoning abilities of the controller agent. Furthermore, Observation Snapshot Key (OBSK) is designed to address the limited context length of LLMs. OBSK Instead of feeding the entire, lengthy observation directly to the main "controller agent", it only presents a short summary of the observation. Along with this summary, it provides a unique hash ID, called a snapshot key that acts like a reference number for the full, detailed observation. All these full observations are stored in a separate key-value store. If the controller agent, while making decisions, decides it needs to see the complete details associated with a particular summary, it can use the snapshot key to query this store and retrieve the full observation. Evaluated on the Real-time Compute Platform for Apache Flink at Alibaba Cloud, RCAgent demonstrated evident and consistent superiority over the ReAct framework in predicting root causes, solutions, evidence, and responsibilities, as verified by both automated metrics and human evaluations. 

In a recent work, Flow-of-Action framework \cite{166_flow} was proposed. Flow-of-Action integrates a Standard Operating Procedure (SOP) with an LLM-based multi-agent system, designed to guide the RCA process, mitigate hallucinations, and enhance operational efficiency. It operates with three key design components: an SOP flow, an action set, and a multi-agent system (MAS). The SOP flow integrates expert knowledge and predefined diagnosis steps into the LLM's knowledge base, imposing constraints at crucial junctures to guide the RCA towards the correct trajectory and significantly reduce hallucinations. The action set mechanism addresses the challenge of irrelevant actions by first generating a reasoned set of actions before making a final decision. The MAS comprises a MainAgent that orchestrates the process, supported by auxiliary agents such as JudgeAgent and ObAgent. These two assist the MainAgent in determining whether the root cause has been identified and in extracting fault types and key information from large volumes of data, respectively. Flow-of-Action was benchmarked against open-source Kubernetes RCA tools like K8SGPT \cite{k8sgpt} and HolmesGPT \cite{holmesgpt}, as well as general-purpose LLM frameworks such as CoT, ReAct, and Reflexion. The results indicated that Flow-of-Action attained the highest performance, achieving a localization accuracy of 64.01\%, which represents a substantial improvement over ReAct’s 35.50\%. This performance meets real-world RCA accuracy requirements and validates the framework’s effectiveness.

Failure diagnosis constitutes another crucial task within the analysis stage, involving the classification of specific types of anomalies. While anomaly detection focuses on the initial identification of deviations in system behavior by flagging anomalous log sessions, fault diagnosis is the subsequent, more precise process of understanding the nature and location of the problem. In this regard, several researchers conducted a preliminary study at a major cloud vendor, revealing that engineers primarily focus on two categories of information for diagnosis: Fault-Indicating Descriptions (FID) and Fault-Indicating Parameters (FIP). FID describe the symptoms of a fault such as error messages, missing components, abnormal behavior, and wrong status. In contrast, FIP pinpoints the exact location of the fault that demands investigation, such as addresses, component IDs, or parameter names. Subsequently, they propose LoFI \cite{163_lofi}, a two-stage approach for efficient and accurate extraction of FID and FIP from anomalous log sessions. The first stage, log selection, performs coarse-grained filtering to reduce the overwhelming volume of logs by selecting severe logs and semantically similar logs. The second stage, prompt-based extraction, leverages a PLM, specifically UniXcoder, which has been fine-tuned to extract fine-grained fault-indicating information including FID and FIP. Results showed that LoFI significantly outperformed all baseline methods including advanced LLM approaches like ChatGPT utilizing in-context learning. A similar study \cite{155_scalalog} proposed a scalable log-based failure diagnosis method to significantly enhance diagnosis accuracy without requiring explicit training. ScalaLog utilize a RAG-based method where historical log data is first summarized using an LLM-based summarization method and stored in a database. Then, in prediction phase, retrieved similar historical incidents along with the current log summary and all potential failure types form a CoT prompt. This prompt guides the LLM in accurately identifying the specific failure type and articulating the rationale behind its decision.

Several studies have also employed fine-tuning methods to enhance performance on their respective downstream tasks.
CrashEventLLM \cite{162_crasheventllm} is a framework leveraging LLMs for predicting system crashes and their potential causes. Failure prediction adopts a proactive approach by aiming to generate early warnings concerning potential system failures. Another study \cite{158_adaptivelog} highlights the dual challenges in automated log analysis: the limited capabilities of SLMs like BERT on complex logs, and the computational inefficiency and high cost of LLMs such as ChatGPT. To balance performance and inference costs, they propose a framework that integrates both models by strategically assigning LLMs to handle complex log instances, while simpler logs are processed by fine-tuned SLMs. A significant finding of this study is the substantial cost efficiency of the proposed method, which decreases LLM inference expenses by an average of 73\% and achieves runtime savings of up to 66.27\% compared to using LLMs for all samples.

{
\scriptsize
\begin{longtable}{p{1.75cm} p{2.5cm} p{2cm} p{1.5cm} p{3.25cm}}
\caption{Overview of LLM-based log analyzing Models} \label{tab:loganalysing} \\
\hline
Ref. & Name & Method & LLM & Downstream Tasks \\
\hline
\endfirsthead

\multicolumn{5}{c}{{\bfseries Table \thetable\ (continued)}} \\
\hline
Ref. & Name & Method & LLM & Downstream Tasks \\
\hline
\endhead

\hline \multicolumn{5}{r}{{Continued on next page}} \\
\endfoot

\hline
\endlastfoot

         \cite{165_failuremode} & - (2023) & Prompt Engineering, and Fine-tuning & GPT-3.5 & Failure Mode Classification \\ \hline
         \cite{161_automatic} & RCACopilot (2024) & Prompt Engineering & GPT-4 & Root Cause Analysis \\ \hline
         \cite{10_triageprocess} & - (2024) & LLM-Based Agents & Various models & Alert Labeling \\ \hline
         \cite{157_exploring} & - (2024) & LLM-Based Agents & GPT-4 & Root Cause Analysis \\ \hline
         \cite{159_automated} & - (2024) & Prompt Engineering & GPT-4 & Root Cause Analysis \\ \hline
         \cite{162_crasheventllm} & CrashEventLLM (2024) & Fine-tuning & Llama2 & Failure Prediction \\ \hline
         \cite{156_logconfigcocalizer} & LogConfigLocalizer (2024) & Prompt Engineering & GPT-4 & Localize Configuration Errors \\ \hline
         \cite{164_rcagent} & RCAgent (2024) & LLM-Based Agents & Vicuna & Root Cause Analysis \\ \hline
         \cite{163_lofi} & LoFI (2024) & Prompt Tuning & UniXcoder & Failure Diagnosis \\ \hline
         \cite{wang2024large} & COMET (2024) & Prompt Engineering & GPT-4 & Incident triage \\ \hline
         \cite{105_loglm} & LogLM (2025) & Instruction tuning & GPT-4, and LLaMA-2 & Log Interpretation, and Root Cause Analysis \\ \hline
         \cite{158_adaptivelog} & AdaptiveLog (2025) & Fine-tuning, and Prompt Engineering & BERT, and GPT-3.5 & Failure Identification, Module Classification, Level Prediction, Log and Description Semantic Matching, and Log and Possible Cause Ranking \\ \hline
         \cite{100_luk} & LUK (2025) & Multi-expert collaboration, Continual pre-training, and Fine-tuning & BERT, GPT-3.5 and GPT-4o , and Llama-3.3 & Failure Identification, Module Classification, Fault Phenomenon Identification, Log and Description Semantic Matching, and Log and Possible Cause Ranking \\ \hline
         \cite{155_scalalog} & ScalaLog (2025) & RAG & GPT-3.5 & Failure Diagnosis \\ \hline
         \cite{166_flow} & Flow-of-Action (2025) & LLM-Based Agents & GPT-3.5, and GPT-4 & Root Cause Analysis \\ \hline

\end{longtable}
}

\subsection{Risk Assessment}
 
Cybersecurity risk assessment involves evaluating factors such as impact, likelihood, vulnerability, and exposures to determine the overall risk. This process is crucial for organizations to comply with regulations and to prevent operational disruptions in IT/OT environments. Traditionally, these assessments are often manual, expert-driven, and inherently subjective, highlighting the importance of more streamlined and automated approaches for scalability and efficiency. A recent study \cite{iyenghar2025feasibility} investigates the use of LLMs with CoT prompting to automate cybersecurity risk assessments for embedded systems and compares it with expert's risk assessments. The findings demonstrate the potential of LLMs with CoT prompting to not only automate but also enhance the consistency and reliability of cybersecurity risk assessments in operational technology environments, offering a foundation for AI-assisted risk evaluations. Moreover, ExBERT \cite{yin2020apply} and CVEDrill \cite{aghaei2023automated} are two proposed framework to automate cybersecurity vulnerability analysis. ExBERT is an enhanced BERT model specifically designed for predicting whether a vulnerability will be exploited or not. It's methodology involves a BERT model to effectively capture the semantic meaning within vulnerability descriptions and an LSTM model to predict exploitability. CVEDrill also automates the prediction of CVSS vectors and classify CVEs into the appropriate CWE hierarchy directly from unstructured textual descriptions of vulnerabilities. This enables organizations to implement cybersecurity countermeasure mitigation with enhanced accuracy and timeliness, surpassing the capabilities of general-purpose tools like ChatGPT in this specialized domain.

\subsection{Answer of RQ4: }In the analysis phase, nearly all tasks still require further research before they can be fully relied upon with confidence. However, Question-Answering models may serve as a supplementary knowledge source for minor or non-critical tasks.

\section{RQ5: How are LLMs adapted for the response phase of SOCs?} \label{sec:rq5}
In recent years, the complexity and frequency of cyber incidents have escalated, necessitating more advanced and automated solutions for all attackers and defenders in cybersecurity. Businesses are increasingly exposed to cyberattacks as their reliance on Information and Communications Technologies (ICT) grows, leading to significant investments in cybersecurity to avoid negative impacts \cite{franco2020secbot}. Despite continuous reliability efforts, cloud services still experience inevitable severe incidents, which can be costly in terms of financial penalties and engineering efforts \cite{ahmed2023recommending}. Traditional SOCs and incident response (IR) practices, which are vital components of modern cybersecurity frameworks, often struggle to keep pace with these evolving threats \cite{185_ircopilot}. They face challenges such as an overwhelming volume of security alerts, high false-positive rates, slow incident response times, and a shortage of skilled cybersecurity professionals, making manual threat analysis and response insufficient \cite{rahmani2024integrating}. Moreover, Publicly available data from the NVD indicates a significant and accelerating rise in the number of published CVEs. This increase has been dramatic, escalating from approximately 7,928 CVEs in 2014 to 40,287 in 2024, demonstrating an exponential upward trend. Furthermore, over the same decade, the cumulative total of CVEs surged from less than 70,000 to over 262,000. This proliferation of vulnerability intelligence data presents a massive challenge for cybersecurity teams to manage effectively \cite{185_ircopilot}. 
By leveraging capabilities like NLP and ML techniques, AI-driven SOCs can improve situational awareness and accelerate incident response tasks. This includes intelligent incident prioritization by correlating alerts and assigning risk scores, and automated threat mitigation actions like blocking suspicious IPs or isolating compromised endpoints. Such automation significantly reduces mean time to respond (MTTR), thereby decreasing the manual effort required from human analysts.

\subsection{Vulnerability Repair}

Automatic Vulnerability Repair (AVR) is an area of growing importance in software engineering, focused on automatically fixing security flaws in source code without human debugging effort. The sheer volume and complexity of modern software systems means that manually identifying and repairing these vulnerabilities is an extensively effort-demanding and time-intensive task \cite{zhang2023pre}. For instance, the GitHub 2020 security report found that it takes approximately 4.4 weeks to release a fix after a vulnerability is identified \cite{chen2022neural}. This time provides a great opportunity for attackers to exploit these weaknesses, posing enormous risks and contributing to billions of dollars in cybercrime costs \cite{fakih2025llm4cve}. Moreover, although AI-driven automation tools like GitHub Copilot have demonstrated substantial improvements in developer productivity for functional code generation, their effectiveness in vulnerability repair remains limited. In some cases, these tools may even introduce security flaws. This limitation is largely attributed to their pre-training on publicly available code repositories, which frequently contain insecure or vulnerable code patterns \cite{170_secrepair}. Therefore, by accelerating the fixing process, AVR minimizes the exposure of software systems to potential attacks, enhances overall security, and offers a promising solution to address vulnerabilities, thereby significantly reducing the burden on security experts.

Fine-tuning is the most common and often preferred method for automated vulnerability repair models. Despite the massive scale and pre-training of LLMs, they often struggle with the specialized domain of vulnerability repair without further adaptation, as their generic knowledge is not sufficient for this complex task \cite{zhang2023pre}. Vulnerability repair datasets are typically small and scarce, making traditional deep learning approaches difficult to train effectively from scratch \cite{chen2022neural}. Fine-tuning, especially when coupled with transfer learning from larger, related bug-fixing corpora, allows models to acquire generic knowledge from extensive codebases and then specialize it for the unique patterns and intricacies of security vulnerabilities, overcoming the data scarcity problem and yielding more reliable performance \cite{wu2023effective}. This process enables models to learn task-specific knowledge and adapt to the precise output formats required for generating accurate and compilable patches. Conversely, prompt engineering alone is generally ineffective for models like GPT-4 in the context of vulnerability repair \cite{fu2023chatgpt}. Studies demonstrate these general purpose models fail to generate any correct repair patches for vulnerable functions or achieve very low accuracy, significantly trailing fine-tuned models \cite{fu2023chatgpt, wu2023effective, zibaeirad2024vulnllmeval}. Therefore, while prompting can offer some modest improvements by providing security context or vulnerability details, it does not compensate for the lack of specialized training needed for robust vulnerability repair \cite{khan2025code}. 

VulRepair\cite{167_vulrepair} represents one of the earliest deployments of LLMs in this domain. Its main ideas lies in the adoption of Byte-Pair Encoding tokenization to address out-of-vocabulary effectively, along with fine-tuning the CodeT5 model. Through extensive evaluations on datasets like CVEFixes and Big-Vul, VulRepair achieved a Perfect Prediction accuracy of 44\%, which was 13\%-21\% more accurate than competitive baseline approaches including VRepair\cite{chen2022neural} and CodeBERT\cite{feng2020codebert}. A notable limitation of VulRepair, like other Transformer-based models, is its struggle to handle lengthy vulnerable code. Specifically, VulRepair imposes a maximum limit of 512 code tokens, yet a substantial portion of real-world vulnerable code surpasses this limit. Subsequently, extra tokens beyond the 512-token window are not processed by the model, which negatively impacts the accuracy of repairs. 
To address this challenges,  VulMaster\cite{171_vulmaster} adopts the Fusion-in-Decoder (FiD) framework, originally used in NLP for open-domain question-answering. It employs a divide-and-conquer approach and instead of truncating, FiD allows VulMaster to process the entire vulnerable code, regardless of its length. It works by individually encoding multiple input components such as segments of the vulnerable code's token sequence, its AST node sequence, and various forms of CWE expert knowledge (CWE name, vulnerable examples, and their fixes). The contextual embeddings from these individually encoded segments are then concatenated into a composite representation, which is subsequently fed into a single decoder to generate the fixed code. 
Moreover, another study \cite{169_enhanced} also includes an innovative and efficient format for representing code repair modifications. This new representation specifically aims to mitigate the ambiguity and localization errors inherent in older token-based approaches. Their study present a meticulously designed data format to optimize how LLMs process and generate fixes for C/C++ vulnerabilities. For the input, the framework structures the prompt by clearly identifying the vulnerable line numbers, providing contextual information about the vulnerability type (CWE), and including the complete source code of the function with its line numbering. For the output, the format precisely indicates the lines preceding and following the intended code insertion point, with the new code segment clearly marked for straightforward application. This innovative approach offers several advantages: it provides versatility, enabling the model to suggest any type of code modification such as insertions, replacements, or deletions with minimal redundant lines; it ensures clarity in the model's output, making proposed changes easy to identify, understand, and apply to the original code; it significantly decreases redundancy by only generating the new code segments and avoiding the repetition of unchanged code, which in turn improves inference speed due to fewer tokens and reduced computational load; and lastly, it mitigates the risk of misplaced modifications that were common in prior approaches, by clearly delimiting the exact insertion points, thereby enhancing the accuracy and efficiency of the code correction process. The experimental results show that, even with a beam size of only 5, the proposed model outperformed VulRepair by a substantial margin of 13\%, despite VulRepair employing a beam size of 50.
Furthermore, SecRepair\cite{170_secrepair} is a multi-purpose code vulnerability analysis system powered by the CodeGen2. SecRepair aims to assist developers not only in identifying and generating fixed code but also in providing a complete description of the vulnerability along with a concise code comment.

The "Granularity" column in the Table \ref{tab:vulrepair} indicates whether the vulnerability within the input code snippet provided for repair is identified at the function level or at the line level. Models with Line-level granularity assume they know the vulnerable line because they heavily rely on vulnerability repair datasets where precise vulnerability locations are specified at the line level. However, this assumption is largely not applicable in real-world scenarios due to significant challenges in achieving accurate line-level vulnerability localization and without them, the repair process either lacks the precise target required for an effective fix or becomes heavily reliant on time-consuming and error-prone human efforts to bridge the gap between general detection and specific repair. Therefore, accurate detection and localization are essential prerequisites in vulnerability repair. In this regard, several studies have explored the potentials of multi-task learning frameworks, allowing them to be simultaneously trained for both vulnerability detection and repair tasks in this area \cite{173_dcodebert, 174_linejlocrepair}. LineJLocRepair\cite{174_linejlocrepair} is an end-to-end LLM-based approach for automated software vulnerability localization and repair. It was developed to address the limitations of previous AVR methods, such as their heavy reliance on datasets with already specified precise vulnerability locations and their suboptimal performance in generating long patch sequences. LineJLocRepair tackles these issues through a joint training framework that seamlessly integrates both vulnerability localization and repair tasks, enabling their simultaneous training. This architecture allows the model to accept a vulnerable function as input and directly generate patched code as output. In their study, LineJLocRepair demonstrates superior performance compared to state-of-the-art baselines. It achieves an impressive 99\% accuracy in pinpointing vulnerable code lines, representing a 32\% improvement over leading line-level vulnerability localization methods like LineVul, and in terms of vulnerability repair, it shows a 12\% improvement over VulRepair.

Furthermore, some studies have focused on providing actionable, human-understandable guidance for addressing software vulnerabilities. In a study \cite{172_vuladvisor}, several researchers have proposed VulAdvisor, a developer-centric approach for automated software vulnerability repair. It generates natural language suggestions, rather than just code patches. Their proposed method was developed to overcome shortcomings of existing AVR tools, which often produce patches that may not align with developers' practical needs or expectations. Beyond providing direct guidance to developers, VulAdvisor's generated suggestions can also enhance the patch generation capabilities of existing AVR tools. 
In another study \cite{177_proverag}, a provenance-driven system was developed to improve real-time software vulnerability analysis by generating actionable mitigation strategies. By emulating expert analytical workflows, it addresses key limitations of traditional LLMs such as such as hallucinations, omissions, and outdated knowledge. It also helps manage the growing volume of emerging vulnerabilities faced by security analysts. The core idea of this model centers on its automated RAG capabilities combined with a self-critique mechanism which enables the LLM to self-evaluate its own generated responses. ProveRAG systematically retrieves and summarizes up-to-date information from authoritative web sources such as NVD, CWE, and associated hyperlinks. This summarizing retrieval technique is specifically designed to manage information overload and overcome context window limitations inherent in LLMs, proving more efficient than simple chunking retrieval. The system also incorporates provenance, explicitly documenting the source of every piece of information and using a Chain-of-Thought technique within its prompts to compel the LLM to provide verifiable evidence and rationales for its responses.

Before concluding this section, we highlight two common characteristics observed across studies in this domain. The first one is Perfect Prediction, a widely adopted evaluation metric in AVR models. This metric defines a repair as correct only if the model-generated patch precisely matches the human-written ground-truth in an exact, token-for-token sequence. While it sets a rigorous standard for accuracy, this approach has significant limitations in reflecting real-world utility.  It fails to account for functionally correct yet syntactically different solutions, potentially underestimating model performance. The metric’s reliance on exact textual matches disregards semantic correctness and the presence of multiple valid fixes, making comprehensive dataset construction difficult and potentially penalizing correct but non-identical repairs \cite{169_enhanced}. 
The second one is Beam search algorithm, a commonly adopted decoding strategy in sequence-to-sequence models, including those was reviewed in this section. Unlike greedy decoding, which selects only the most probable token at each step, beam search maintains multiple candidate sequences (determined by beam width) throughout the generation process, thereby increasing the likelihood of generating high-quality repairs. This is particularly valuable in code generation, where multiple functionally equivalent patches may exist. In AVR systems, beam search facilitates the generation of multiple candidate patches, which can subsequently be evaluated through automated test suites or manual inspection. Additionally, it enhances performance metrics such as Perfect Prediction by increasing the probability that at least one candidate matches the ground-truth fix. However, its improved accuracy comes at the cost of higher computational overhead, posing limitations for real-time repair scenarios.

{
\scriptsize
\begin{longtable}{p{1cm} p{2cm} p{1.5cm} p{1.25cm} p{1.5cm} p{1.5cm} p{1.5cm} p{0.75cm}}
\caption{Overview of Vulnerability Repair models} \label{tab:vulrepair} \\
\hline
Ref. & Name & Method & LLM & Code & Granularity & Datasets & PP \\
\hline
\endfirsthead

\multicolumn{8}{c}{{\bfseries Table \thetable\ (continued)}} \\
\hline
Ref. & Name & Method & LLM & Code & Vulnerability Granularity & Datasets & Prefect Prediction \\
\hline
\endhead

\hline \multicolumn{8}{r}{{Continued on next page}} \\
\endfoot

\hline
\endlastfoot
         
         \cite{167_vulrepair} & VulRepair (2022) & Fine-tuning & CodeT5 & C/C++ & Line & CVEFixes, and Big-Vul & 44\% \\ \hline
         \cite{168_contracttinker} & ContractTinker (2023) & Prompt Engineering & GPT-3.5 & Solidity & Function & Self-built & 48\%  \\ \hline
         \cite{170_secrepair} & SecRepair (2024) & Instruction Tuning, and Reinforcement Learning & CodeGen2 & C/C++ & Function & Self-built & -  \\ \hline
         \cite{171_vulmaster} & VulMaster (2024) & Continual Pre-training + Fine-tuning & CodeT5, and GPT-3.5 & C/C++ & Line & CVEFixes, and Big-Vul & 20\% \\ \hline
         \cite{172_vuladvisor} & VulAdvisor (2024) & Prompt Engineering, and Fine-tuning & GPT-3.5, and CodeT5 & C/C++ & Function & Self-built & - \\ \hline
         \cite{169_enhanced} & - (2024) & Fine-tuning & Code Llama, and Mistral & C/C++ & Line & CVEFixes, and Big-Vul & 57\%  \\ \hline
         \cite{173_dcodebert} & DCodeBERT (2025) & Continual Pre-training + Fine-tuning & CodeBERT & Python, Java, PHP, C++, JavaScript, and Ruby & Function & Self-built & - \\ \hline
         \cite{174_linejlocrepair} & LineJLocRepair (2025) & Continual Pre-training + Fine-tuning & CodeT5 & Python, Java, Ruby, JavaScript, Go, PHP, C, C++, and C\# & Function & CVEFixes, and Big-Vul & 47\% \\ \hline

\end{longtable}
}

\subsection{Incident Response}
Incident Response (IR) is a structured and well-planned approach an organization employs to manage and mitigate the impact of cybersecurity incidents. IR encompass any undesirable behavior involving a computer or network component and its primary goal is to minimize damage, limit losses, and swiftly restore normal operations after a security breach \cite{hassan2023role}. In the dynamic and fast-paced domain of cloud operations, incident management represents a critical challenge for large-scale cloud service providers such as Microsoft, Google, and Amazon. These incidents may vary in severity, spanning from minor service disruptions to major system outages, and can lead to significant consequences including financial losses, operational downtime, reputational damage, and potential legal liabilities \cite{180_nissist}. Consequently, recent research efforts have focused on reducing the duration of this process to enable faster responses to cybersecurity incidents \cite{181_autobnb, 183_autonomous, 185_ircopilot}.
For instance, the AutoBnB framework proposed in \cite{181_autobnb} is specifically designed to investigate LLM-based multi-agent collaboration in IR scenarios, employing the Backdoors \& Breaches tabletop game as a simulated environment for cybersecurity training. The methodology involves simulating realistic IR dynamics with various team structures, including centralized, decentralized, and hybrid configurations. By analyzing the interactions and performance of these LLM-based agents across different setups, the research provides valuable insights into optimizing multi-agent collaboration for incident response. Performance summaries from experiments highlight the importance of procedure selection and team coordination in these multi-agent IR scenarios. Moreover, several researchers introduce IRCopilot \cite{185_ircopilot}, a collaborative framework of four LLM-based components designed to coordinate strategy, execution, evaluation, and iterative refinement for incident response. This framework incorporates a variety of prompt engineering strategies alongside a deliberate segmentation of responsibilities. Evaluated on a comprehensive benchmark and real-world attack scenarios, IRCopilot demonstrates significantly better performance than general purpose LLMs.

Several studies have also explored the applications of LLMs in the domains of planning and mitigation. Planning involves the development and continuous refinement of structured procedures to ensure timely and coordinated actions during cybersecurity events. In \cite{29_employing}, they first outline challenges in creating and maintaining comprehensive Incident Response Plans (IRPs) and Standard Operating Procedures (SOPs) such as complexity of modern systems, high personnel turnover rates, insufficient documentation for legacy technologies, the difficulty of continuously updating plans for novel attack vectors, and the lack of composability of IRPs and SOPs. Then, they leverage LLMs to tackle these challenges and enhance different phases of Incident Response Planning. A recent study \cite{182_automated} explores the capabilities of LLMs from a reinforcement learning perspective to achieve automated tactics planning. This automation offers the advantages of increased efficiency by reducing response times, enhanced precision in detecting subtle patterns, continuous 24/7 monitoring and response capabilities, and optimized resource allocation by freeing up human personnel.

Mitigation models in incident response involve the strategies and actions taken to contain a cyber incident, prevent further damage, and restore normal operations. A work \cite{180_nissist} addresses the obstacles associated with incident management for enterprise-level cloud service providers. It emphasizes on complex incidents frequently require human intervention, hindered by Troubleshooting Guides (TSGs) that are often unstructured, inconsistent in quality, and outdated. They propose Nissist to overcome these obstacles by leveraging LLMs to reduce human intervention and provide proactive incident mitigation suggestions. Nissist's methodology involves building a comprehensive knowledge base by using LLMs to reformat unstructured TSGs into a structured, high-quality format, including sequential steps, and knowledge from incident mitigation history. This structured knowledge enables the multi-agent system design of Nissist to effectively interpret OCE intents, retrieve relevant knowledge nodes, and formulate actionable plans. The system's modules include an intent interpreter for understanding user queries, a node retriever and selector to find relevant information in the knowledge base, an action planner that recommends sequential steps, and finally, a Post Processor that uses a fine-tuned LLaMA2 model to correct potential LLM hallucinations. User experiments with OCEs demonstrated that Nissist significantly improved incident mitigation, achieving a Time-to-Mitigate (TTM) reduction of 98.93\% for simple incidents and 94.85\% for hard ones compared to manual mitigation. It also showed a 77.19\% full automation success rate for simple incidents and 52.63\% for hard ones, with only minor human intervention needed for remaining cases. Furthermore, ShieldGPT \cite{179_shieldgpt} has proposed  to mitigate the constantly evolving and complex DDoS attacks. It aims to overcome the limitations of previous AI-driven approaches by providing explainable predictions and actionable mitigation instructions. The ShieldGPT architecture consists of a DDoS classifier to categorize each flow as benign or a specific type of DDoS attack and GPT-4 to explain the attacking behaviors of a specific flow and generate actionable mitigation strategies for deployment on designated devices. ShieldGPT proved capable of generating pragmatic mitigation directives with actionable commands for defense devices such as Cisco routers and Snort IPS, including both stringent methods like IP blocking and less aggressive approaches like rate limiting.

There are also some more studies exploring application of LLMs in Log-based recommendation task \cite{177_logexpert,105_loglm, 178_logupdater}. This task focuses on automatically generating actionable solutions once anomalous logs have been detected. While existing log related methods concentrate on tasks like compression, parsing, and anomaly detection, engineers still face significant challenges and manual effort in generating actionable resolutions for detected anomalous logs. To overcome these limitations, the authors of \cite{177_logexpert} introduce LogExpert framework. LogExpert comprises three main components: a Log Recognizer that extracts anomalous logs and relevant information from noisy technical forum data into a structured format; an Extractive Summarization module that condenses resolutions to reduce noise and token consumption for the LLM; and a Resolution Generation module that combines the LLM's generative power with domain-specific knowledge, using few-shot prompting based on similar historical examples retrieved from a vector database, to produce customized, executable resolution steps. Furthermore, Human evaluations underscored the practical efficacy of the LogExpert framework. These assessments demonstrated a substantial improvement in the usefulness (by 16.10\%) and concreteness (by 26.67\%) of the generated resolutions when compared to outputs from baseline LLMs.

Despite the aforementioned advantages of employing LLMs, the volume of published research focusing on their application remains considerably lower compared to other domains within cybersecurity defense. This can be attributed to several significant obstacles. Firstly, the IR field suffers from a severe scarcity of real-world, detailed, and publicly accessible datasets, largely due to privacy and confidentiality concerns, which traditional anonymization methods often fail to adequately address, thereby hindering systematic research and robust model training \cite{galadima2024towards}. Secondly, existing log analysis and cybersecurity research has historically prioritized threat detection over automated resolution and repair, leaving a critical gap in generating actionable mitigation steps and requiring substantial manual effort from human experts \cite{178_logupdater}. Thirdly, IR compounded by the inherent complexity and dynamic nature, which demands rapid decision-making, adaptation to evolving threats, and deep domain knowledge, often proving challenging for conventional pre-trained models \cite{ahmed2023recommending}. Furthermore, LLMs themselves face specific limitations when applied to IR, including difficulties in handling heterogeneous network data, producing context-specific and accurate recommendations without hallucinations, and managing long-term conversational memory \cite{185_ircopilot}. Lastly, the absence of comprehensive, standardized benchmarks tailored for evaluating LLM performance across the multi-phased and intricate tasks of incident response impedes comparable and systematic advancements in this specialized area \cite{182_automated}.

Another observation that we can make in this section is the high adoption of LLM-based agents in this field, as shown in Table \ref{tab:incidentresponse}. Multi-agent systems offer a scalable and modular architecture that enables agents with specialized roles to operate autonomously while communicating and cooperating to achieve shared objectives. This approach enhances flexibility, supports parallel task execution, and allows for dynamic adaptation to evolving threat scenarios. Multi-agent architectures further enhance these benefits by mimicking human IR team structures with specialized, collaborative components, allowing for clear divisions of responsibility that mitigate issues such as factual inaccuracy and context loss common in monolithic LLMs \cite{185_ircopilot, 181_autobnb}. This collective intelligence, combined with their ability to build and refine comprehensive knowledge bases from diverse sources, makes LLM-based multi-agent systems a transformative approach for more efficient and effective incident response.

{
\scriptsize
\begin{longtable}{p{1.5cm} p{1.5cm} p{2.5cm} p{1.5cm} p{3cm}}
\caption{Overview of LLM-based incident response models} \label{tab:incidentresponse} \\
\hline
Ref. & Name & Method & LLM & Downstream Tasks \\
\hline
\endfirsthead

\multicolumn{5}{c}{{\bfseries Table \thetable\ (continued)}} \\
\hline
Ref. & Name & Method & LLM & Downstream Tasks \\
\hline
\endhead

\hline \multicolumn{5}{r}{{Continued on next page}} \\
\endfoot

\hline
\endlastfoot

         \cite{180_nissist} & Nissist (2024) & LLM-based agents & LLaMA-2 & Incident Mitigation \\ \hline
         \cite{29_employing} & - (2024) & Prompt Engineering & - & Incident Response Planning \\ \hline
         \cite{177_logexpert} & LogExpert (2024) & Prompt Engineering, and Fine-tuning & BERT, and GPT-4 & Log-based Recommendation \\ \hline
         \cite{179_shieldgpt} & ShieldGPT (2024) & Prompt Engineering & GPT-4 & DDOS Mitigation \\ \hline
         \cite{181_autobnb} & AutoBnB (2025) & LLM-based agents & GPT-4o & Incident Response \\ \hline
         \cite{178_logupdater} & LogUpdater (2025) & Prompt Engineering, and Fine-tuning & UniXcoder, and GPT-4o & Log-based Recommendation \\ \hline
         \cite{183_autonomous} & - (2025) & LLM-based Agents & - & Incident Response \\ \hline
         \cite{185_ircopilot} & IRCopilot (2025) & LLM-based Agents & - & Incident Response \\ \hline
         \cite{182_automated} & - (2025) & LLM-based Agents, and Reinforcement Learning & Various models & Incident Response Planning \\ \hline

\end{longtable}
}

\subsection{Answer of RQ5: }Within the three phases of the incident lifecycle, the Response phase has received the least research attention. This is largely due to its complex and mission-critical nature, which necessitates further investigation. Nonetheless, studies such as \cite{185_ircopilot} and \cite{180_nissist} have demonstrated models capable of achieving acceptable performance in real-world operations.

\section{RQ6: What are the challenges and future research directions for applying LLMs in SOC domain?}
\label{sec:rq6}

\subsection{Challenges}
Despite the significant advancements and promising potential of LLMs in transforming cybersecurity and SOCs, their application is met with a unique set of challenges and limitations. These hurdles span multiple dimensions, including the models themselves, the data they process, and the operational realities of SOC environments. Addressing these challenges is paramount for the safe, effective, and trustworthy deployment of LLMs in critical security operations.

\subsubsection{LLM related Challenges}
A primary category of limitations arises from the intrinsic characteristics of LLMs themselves. Their substantial computational overhead and resource intensity presents a significant hurdle, often precluding their efficient deployment in real-time cybersecurity operations, and posing a particular challenge for Small and Medium-sized Enterprises operating with limited IT infrastructure and budgets \cite{lilac}. This necessitates continuous advancements in model compression and more cost-effective inference strategies to make LLMs practically viable for high-volume, low-latency security tasks. Furthermore, LLMs frequently exhibit inconsistency and non-determinism in their outputs, occasionally generating hallucinations which severely undermines their reliability for critical security decision-making. These models also demonstrate limited understanding of the highly complex, nuanced, or ambiguous terminology and queries inherent to SOC tasks, often leading to an oversimplification of solutions or providing partial fixes that are not comprehensively robust. \cite{zibaeirad2024vulnllmeval}. Moreover, their current capabilities struggle to generalize effectively to zero-day threats not extensively represented in their training data, demanding constant updates and fine-tuning to maintain relevance against evolving attack landscapes \cite{145_securefalcon}.

Beyond foundational understanding, practical application of LLMs is further complicated by their sensitivity to prompt variation, meaning that the design of effective prompts is a non-trivial, iterative process, where even minor changes can drastically alter output quality . The "black box" nature of many state-of-the-art LLMs also presents a formidable challenge to explainability and interpretability. This lack of transparency makes it difficult for human analysts to validate AI-generated insights or comprehend the reasoning behind anomalous detections, which is crucial for building trust in high-stakes security environments. Additionally, current LLMs display limitations in dynamic decision-making and multi-step reasoning, struggling with tasks that require intricate, sequential thought processes or adaptive responses to unforeseen circumstances \cite{152_smartguard}. The phenomenon of catastrophic forgetting, where models lose previously acquired knowledge over time when exposed to new data, poses another serious risk to maintaining up-to-date threat intelligence and patch awareness \cite{zibaeirad2024vulnllmeval}. Finally, LLMs expand the attack surface of organizations and are vulnerable to adversarial techniques such as prompt injection, data poisoning, and model inversion. These attacks can be exploited by malicious actors to manipulate model behavior, extract sensitive information, or undermine existing security countermeasures \cite{akhtar2025llm}. As LLMs become increasingly integrated into SOC operations, developing robust defense mechanisms against such threats is essential.

\subsubsection{Data related Challenges}

A pervasive challenge in developing SOC related models is the limited availability of large volumes of high-quality, diverse, and real-world datasets \cite{182_automated, akhtar2025llm}. Existing datasets are often small, biased, contain duplications, or lack consistency, making robust model training difficult and often requiring labor-intensive manual labeling prone to human error \cite{alzu2025cyberattack}. The sensitive and proprietary nature of cybersecurity data, such as internal network logs and threat intelligence, raises significant privacy and security concerns when processed by external LLM services, leading organizations to be hesitant about data sharing. SOCs are also confronted with a massive volume of diverse and often unstructured data, making it challenging to collect, pre-process, validate, and ensure the completeness and consistency of information for LLM consumption \cite{khayat2025empowering}. Moreover, the dynamic and continuously evolving nature of cyber threats and associated data formats (e.g., log structures, malware variants) leads to "concept drift," rapidly rendering historical training data obsolete and necessitating constant retraining or adaptive mechanisms \cite{logprompt}. Additionally, many CTI reports and real-world security scenarios involve multimodal information, such as figures, charts, and raw network flow data, which current LLMs primarily trained on text struggle to process directly, limiting comprehensive analysis and requiring specialized interfaces to align modalities \cite{25_attackg+}. Lastly, the inherent non-linguistic structure of certain data modalities, such as time-ordered network flow data, presents a fundamental challenge for LLMs. Unlike natural language, which provides intrinsic contextual dependencies and grammatical structures that LLMs are designed to process, network flows often lack such inherent semantic organization. This modality mismatch consequently complicates the models' ability to effectively discern subtle patterns and mine inter-flow correlations within these datasets \cite{86_dollm}.

\subsubsection{SOC related Challenges}
Within the SOC environment, LLMs confront significant challenges stemming from operational realities. A primary practical hurdle involves the seamless and effective integration of LLMs with existing, complex SOC technology stacks, which are often heterogeneous and comprise a variety of legacy and modern systems, including SIEM platforms, SOAR platforms, and threat intelligence feeds. This integration is complicated by diverse data formats and interoperability issues across disparate systems, frequently necessitating custom development and careful orchestration. Moreover, the deployment of LLM-based solutions further adds a layer of complexity due to the need to ensure regulatory compliance with critical standards, especially when sensitive internal log data and user information are processed by external, third-party LLM services via public APIs, raising concerns about data leakage and unauthorized access. Finally, there is an ongoing need to find the optimal balance between human expertise and LLM-driven automation; while LLMs can augment human capabilities and automate routine tasks, critical SOC related decisions often require nuanced human judgment and cannot be fully delegated to autonomous AI agents. The lack of structured processes and heavy reliance on tacit knowledge among analysts also make it difficult to define clear objectives and measure performance improvements \cite{kersten2023give, hermann2024gpt}.

\subsection{Future Directions}
The application of LLMs within SOCs is a rapidly evolving area, holding significant promise for advancing its capabilities. While LLMs have demonstrated remarkable capabilities in NLP tasks and are increasingly being explored for various security functions, their widespread adoption in SOC environments necessitates further research to address existing limitations and unlock their full potential. Future research directions are focused on enhancing their accuracy, efficiency, and interpretability, as well as enabling more autonomous and integrated security operations. Here are key future research directions for applying LLMs in the SOC domain:

\begin{itemize}
    \item \textbf{Advancing Explainability and Trust in LLM-Driven SOCs: }A paramount challenge in integrating LLMs into SOCs is their "black box" nature, which can hinder trust among security analysts. Future research must focus on developing robust Explainable AI (XAI) techniques specifically tailored for LLMs in cybersecurity contexts. This includes methods to clearly articulate why an LLM classified an alert in a certain way, generated a specific recommendation, or identified a particular threat indicator. Techniques such as attention visualization, counterfactual explanations, and natural language justifications need further refinement to be practically useful in high-pressure SOC environments. Building trust also involves creating mechanisms for verifiability and human oversight, allowing analysts to easily validate LLM outputs and intervene when necessary. Research into human-AI collaboration frameworks, where the strengths of both are leveraged effectively, will be paramount \cite{desolda2025apollo, khayat2025empowering}.
    
    \item \textbf{Integration of LLMs and Federated Learning: }The integration of LLMs with FL is a promising future direction, particularly for enhancing security in distributed and privacy-sensitive environments. FL enables multiple clients, such as edge devices and base stations, to collaboratively learn a global model without exchanging raw data, aligning well with decentralized and latency-sensitive infrastructures like 5G \cite{hernandez2025intrusion}. FL inherently facilitates privacy-preserving, collaborative model training across diverse network nodes, which addresses critical challenges like data isolation and sensitive information exchange \cite{beuran2025fedmse, hendaoui2025fladen}. This decentralized paradigm significantly enhances scalability and fault tolerance for systems, enabling organizations to share CTI and collectively build more robust defense mechanisms \cite{el2023fedcti}. Additionally, This collaborative sharing of cybersecurity information is vital for vulnerability management, threat detection, real-time monitoring, and incident response across IoT ecosystems \cite{lin2025fedav, bierbrauer2025data}. While traditional FL-based anomaly detection frameworks often rely on manually designed numerical features, limiting their ability to grasp contextual and semantic patterns, LLMs can provide this crucial contextual understanding, thereby improving robustness against adversarial manipulations \cite{rezaei2025fedllmguard}. 
    
    \item \textbf{Robust Benchmarks and Datasets: }A significant challenge for LLM application in SOCs is the lack of standardized benchmarks and comprehensive evaluation methodologies. Future research needs to focus on creating more robust and representative datasets and benchmarks to accurately assess LLMs' performance in diverse SOC tasks. This includes generating large, varied datasets that reflect real-world attack scenarios, which would facilitate more reliable benchmarking of LLM performance \cite{10_triageprocess}. Moreover, There is a particular demand for simulation environments that can realistically mimic SOC operations and allow for the evaluation of LLM decision-making in different threat scenarios \cite{157_exploring}. These environments can help address data quality issues, mitigate biases introduced by existing datasets, and ensure models can adapt to evolving threat landscapes and previously unseen attacks \cite{bierbrauer2025data}.

    \item \textbf{Agent-Based RAG: }RAG frameworks address common LLM limitations such as hallucination and outdated knowledge by enabling models to retrieve relevant information from an external knowledge base before generating responses \cite{22_ragquestion}. Future research will increasingly focus on agent-based RAG systems, where LLM agents are endowed with the ability to dynamically query and integrate knowledge from various sources to perform complex tasks. These agents, capable of iterative interaction with their environment, reasoning about tasks, and planning actions, can significantly enhance functions like root cause analysis by accessing diverse information sources \cite{166_flow}. By combining the retrieval capabilities of RAG with the autonomous decision-making and tool-use of agents, such systems can provide more accurate, context-rich, and verifiable outputs for security professionals \cite{deng2025ai}.

    \item \textbf{Autonomous Security Decision-Making: }The future of SOCs involves a significant shift towards increased automation and orchestration, moving LLMs beyond mere "assistant" roles to more autonomous agents capable of handling complex decision-making tasks under human supervision. This entails developing LLMs that can autonomously perform tasks such as automated patch management, antivirus management, network reconfiguration, and port management \cite{73_securitybert}. Future research will focus on integrating LLMs for comprehensive automated incident response and threat mitigation, enabling systems to make immediate decisions and execute predefined actions to neutralize threats, thereby reducing the impact of attacks and improving overall efficiency. The goal is to streamline workflows and minimize manual intervention, enhancing SOC efficiency and enabling faster decision-making in response to dynamic cyber threats \cite{kshetri2025transforming}.

    \item \textbf{LLM-Integrated Architectures: } Combining LLMs with other AI approaches offers a promising direction to create more effective and efficient security solutions. Hybrid models leverage the strengths of different techniques to overcome individual limitations, leading to more lightweight, compact, and real-time accountable systems. This could involve traditional rule-based systems with LLMs, where LLMs provide contextual understanding and adapt rules to evolving threats. Rule-based systems ensure precision and deterministic outputs in known scenarios. Alternatively, integrating ML techniques with LLMs can create another powerful hybrid model. For instance, lightweight ML-based anomaly detection models can be guided by LLM knowledge distillation to improve performance and generalization ability while reducing computational overhead \cite{yang2024anomaly}. Similarly, hybrid NIDS can utilize LLMs like BERT to capture contextual semantic features from raw traffic data, enhancing accuracy and adaptability, even for resource-constrained environments like the Internet of Vehicles \cite{135_iovbertids}. Such hybrid approaches aim to achieve high accuracy and explainability while optimizing computational resources for real-time deployment.

\end{itemize}

\subsection{Answer of RQ6: }The most critical challenges of the proposed models in this domain include hallucinations, adversarial attacks against LLMs, the lack of standardized benchmarks, and the real-time nature of SOC tasks. Furthermore, integrating LLMs with advanced and emerging technologies such as Federated Learning, as well as combining it with traditional rule-based or ML-based approaches, represents a promising direction with significant potential for future research.

\section{Conclusion}
\label{sec:conclusion}
The rapid evolution of cyber threats has increased the demand for innovative solutions capable of enhancing the efficiency and resilience of SOCs. LLMs, with their ability to process unstructured data and support decision-making, have shown strong potential to address many of the limitations faced by SOCs today. In this survey, we examined the applications of LLMs within the literature as they apply to the SOC workflow. We believe that this survey offers employees, practitioners, and managers a comprehensive overview of the current state of LLM integration within SOCs, supporting informed decision-making and guiding future research directions. Ultimately, the effective integration of LLMs has the potential to transform SOC operations by augmenting human expertise and enabling more proactive cybersecurity defense.

\bibliographystyle{unsrt}  
\bibliography{references}


\end{document}